\documentclass[11pt,a4paper,oneside]{article}
\usepackage[top=3cm, bottom=3cm, left=2cm, right=2cm]{geometry}
\renewcommand*\arraystretch{.8}

\usepackage[english]{babel}
\usepackage[utf8x]{inputenc}
\usepackage[T1]{fontenc}

\usepackage{amsthm,amsmath,amssymb,mathrsfs,dsfont,mathtools}

\usepackage{bm}
\usepackage{bbm}
\usepackage{graphicx}
\usepackage[colorinlistoftodos]{todonotes}
\usepackage[colorlinks=true, allcolors=blue]{hyperref}
\usepackage{ulem}
\usepackage{booktabs,subcaption}
\usepackage[all]{nowidow}
\usepackage{setspace}
\usepackage{algpseudocode}
\usepackage{algorithm}
\usepackage{tikz} % For the plot
\usepackage[sf]{titlesec}
\usepackage{sectsty}
\allsectionsfont{\centering \normalfont\scshape} % Make all sections centered, the default font and small caps
\usepackage{lipsum}
\usepackage{adjustbox}
\usepackage[running]{lineno} 
\usepackage[round]{natbib}

\usepackage[all]{nowidow}
\usepackage{setspace}
\usepackage{algpseudocode}
\usepackage{algorithm}
\usetikzlibrary{decorations.pathreplacing,arrows.meta, positioning,chains,fit,shapes,calc}

\usepackage{flexisym}

\usepackage{titletoc}
\usepackage{diagbox}
\usepackage{tikz}
\usetikzlibrary{bayesnet}
\usepackage{longtable}
\usepackage{subcaption}
\usepackage{afterpage}
\usepackage{pifont}
\usepackage{bm}
\usepackage{amssymb}
\usepackage{booktabs}
\usepackage{caption}
\usepackage{slashbox}
\usepackage{lscape}
\usepackage{color, soul}
\usepackage{comment}
\usepackage{psfrag,epsf}
\usepackage{xurl} 
\usepackage{amsthm, amsfonts, tabu, enumitem, xcolor, authblk}
\usepackage{footnote}
\usepackage{animate}
\usepackage{xr-hyper}
\usepackage{multirow}
\usepackage{setspace}
\usepackage[title]{appendix}
                   
\algnewcommand{\Inputs}[1]{%
  \State \textbf{Inputs:}
  \Statex \hspace*{\algorithmicindent}\parbox[t]{.8\linewidth}{\raggedright #1}
}
\algnewcommand{\Initialize}[1]{%
  \State \textbf{Initialize:}
  \Statex \hspace*{\algorithmicindent}\parbox[t]{.8\linewidth}{\raggedright #1}
}

\usepackage{authblk}
\title{Bayesian modeling of multi-species labeling errors in ecological studies}
\date{}
\author[1]{Haoxuan Wang}
\author[2, 3]{Patrik Lauha} 
\author[1]{David B. Dunson}

\affil[1]{Department of Statistical Science, Duke University, Durham, NC, 27708, U.S.A.}
\affil[2]{Organismal and Evolutionary Biology Research Programme, Faculty of Biological and Environmental Sciences, University of Helsinki, Helsinki, 00014, Finland.}
\affil[3]{Department of Biological and Environmental Science, Faculty of Mathematics and Science, University of Jyväskylä, Jyväskylä, 40014, Finland}

\theoremstyle{definition}

\begin{document}

\maketitle
\textbf{Corresponding author}: Haoxuan Wang, H. Milton Stewart School of Industrial and Systems Engineering, Georgia Institute of Technology, 755 Ferst Drive NW, Atlanta, GA 30332-0205, U.S.A. Email: \href{mailto:hwang3111@gatech.edu}{hwang3111@gatech.edu}

\newpage

\begin{abstract}
1. Ecological and conservation studies rely on reliable species observations. In the case of birds, such data are typically produced through acoustic monitoring. Recently, machine learning algorithms have emerged that can accurately classify bird species from audio recordings, but such algorithms crucially depend on expert-labeled training data. Automated classification is particularly challenging when multiple species vocalize simultaneously, recordings contain background noise, or birds are far from the microphone. Passive acoustic  monitoring generates massive audio datasets, yet human experts can label only a tiny proportion of the available data, and experts may differ in their accuracy and breadth of knowledge across species.

2. We focus on combining sparse expert annotations to improve the quality of labels while providing uncertainty quantification. We propose a Bayesian hierarchical modeling framework tailored to multi-species identification that includes a family of model variants, aggregates sparse annotations, accounts for correlation among species vocalizations, and models heterogeneity in expert performance across species through an additional hierarchical structure. The framework produces posterior probabilities of species presence together with uncertainty measures and expert performance scores intended to support feedback, engagement, and improvement.

3. In comprehensive simulation studies, the proposed model variants outperform majority vote and simpler alternatives, especially in sparse and highly correlated annotation settings. We further evaluate the approach using data from a community science platform developed in Finland, comprising 3,997 audio clips with extremely sparse labels produced by 46 annotators. Using a subset of recordings annotated by a highly reliable ornithologist as a gold standard, our framework achieves substantially higher accuracy than majority vote and provides better-calibrated uncertainty estimates.

4. These results demonstrate that combining sparse expert annotations within our framework can substantially improve the quality of labels for ecological datasets. The framework offers a practical pathway for integrating limited expert effort into long-term ecological monitoring and conservation workflows, and enables scalable and uncertainty-aware analysis of the annotations. Even though we demonstrate our framework with labels for bird audio, the approach is applicable to any multi-label ecological data annotated by several experts or citizen scientists.
\end{abstract}

\textbf{Keywords}: Annotation; Bayesian modeling; Bio-monitoring; Crowd sourcing; Ecology; Measurement error; Multiple annotators

\newpage

\section{Introduction} \label{sec:intro}

Bio-monitoring is undergoing a remarkable technological revolution, driven by autonomous data collection devices, including passive acoustic recording units and camera traps. These innovations have ushered in an era of cost-effective, large-scale data acquisition, which has proven invaluable for ecological research \citep{%farina2011soundscape, 
shonfield2017autonomous},
%, frommolt2017information, krause2016using}, 
motivating development of deep learning based species identification \citep{%lebien2020pipeline, 
kahl2021birdnet}. However, creating accurate and reliable annotations for the massive amount of noisy unlabeled data that are now being routinely collected is a daunting challenge. We are motivated by the problem of identifying which bird species are present at a location based on audio recordings. Deep neural networks are promising, but require large training datasets consisting of audio recordings of known bird species \citep{kahl2021birdnet}. Although such data are available in online libraries, these are typically weakly labelled and might not match the data from bio-monitoring programs in terms of recording quality. It is critical to have strongly labeled training data under realistic conditions in the field for the region of interest to train a species classifier with sufficient accuracy \citep{lauha2022domain}. Unfortunately, annotation of recordings requires expert knowledge and is a laborious task \citep{%papadopoulos2015detecting,gibb2019emerging, 
lehikoinen2023successful}.

In response to this challenge, crowdsourcing has emerged as a popular and widely adopted technique for annotating large-scale datasets across various disciplines \citep{swanson2016general}. Crowdsourcing platforms such as MTurk provide task requesters with an online marketplace to post a batch of microtasks for some workers to complete for a small monetary compensation \citep{yin2021learning}. Likewise, citizen science has become more and more popular among scientists, and shows great potential for accomplishing large-scale tasks by sharing the workload \citep{%desherbinin2021critical, 
franzoni2022crowds}. Numerous bird hobbyists with extensive experience in identifying bird vocalizations are an ideal audience for crowdsourcing ornithological data \citep{sullivan2009ebird, lehikoinen2023successful}.

Our motivation is drawn from the Finnish Kerttu web portal \citep{lehikoinen2023successful}, which is a crowdsourcing project leveraging on birdwatcher expertise in annotating the species of birds vocalizing in audio recordings. The portal was designed to enable annotation of audio recordings collected with autonomous recorders. While the portal has subsequently been upgraded to cover global soundscapes (\url{https://bsg.laji.fi/identification/instructions}), the initial focus was on ten locations in Southern Finland. The users were asked to list all bird species vocalizing in the recording and to indicate if there are other bird species that they are not able to identify. The platform also asked users to rate their own bird sound identification skills, while providing them with feedback to improve their abilities and encourage their continued engagement with the site. In this paper, we use data produced in this project consisting of $3997$ $10$-second audio clips, which have been annotated by one or more bird experts. These data are described and referred to as "clips" in the paper by \cite{lehikoinen2023successful}. 

A critical question in this application, and other settings involving citizen scientist crowdsourcing, is how to account for the inevitable errors and variation in accuracy of the annotations provided by different users \citep{%ipeirotis2010quality,
aceves2017accuracy}. Even individuals with substantial expertise will face challenges in providing accurate classifications in cases with high noise, rare species, substantial distance from the recording, short vocalizations, species having similar vocalizations or multiple species having overlapping vocalizations. Errors through misclicks can also occur. Hence, in practice there are commonly inconsistencies in the annotations of different experts. To reduce error, one can assign each audio recording to multiple experts and then aggregate these annotations. Most existing aggregation methods focus on single-label scenarios \citep{johnson1996bayesian,
%johnson2006ordinal, 
raykar2010learning, ghosh2011moderates, dalvi2013aggregating, lin2018modeling, kim2021measuring}. However, in bird species annotation, we are faced with a multi-label crowdsourcing task.
% Each recording can contain vocalizations from multiple bird species. 
In this work we propose a solution for multi-label annotation aggregation. A more detailed description of this motivating application and the corresponding problem formulation is given in Section \ref{sec:motivation}.

\begin{figure}
\centering
\includegraphics[scale = 0.5]{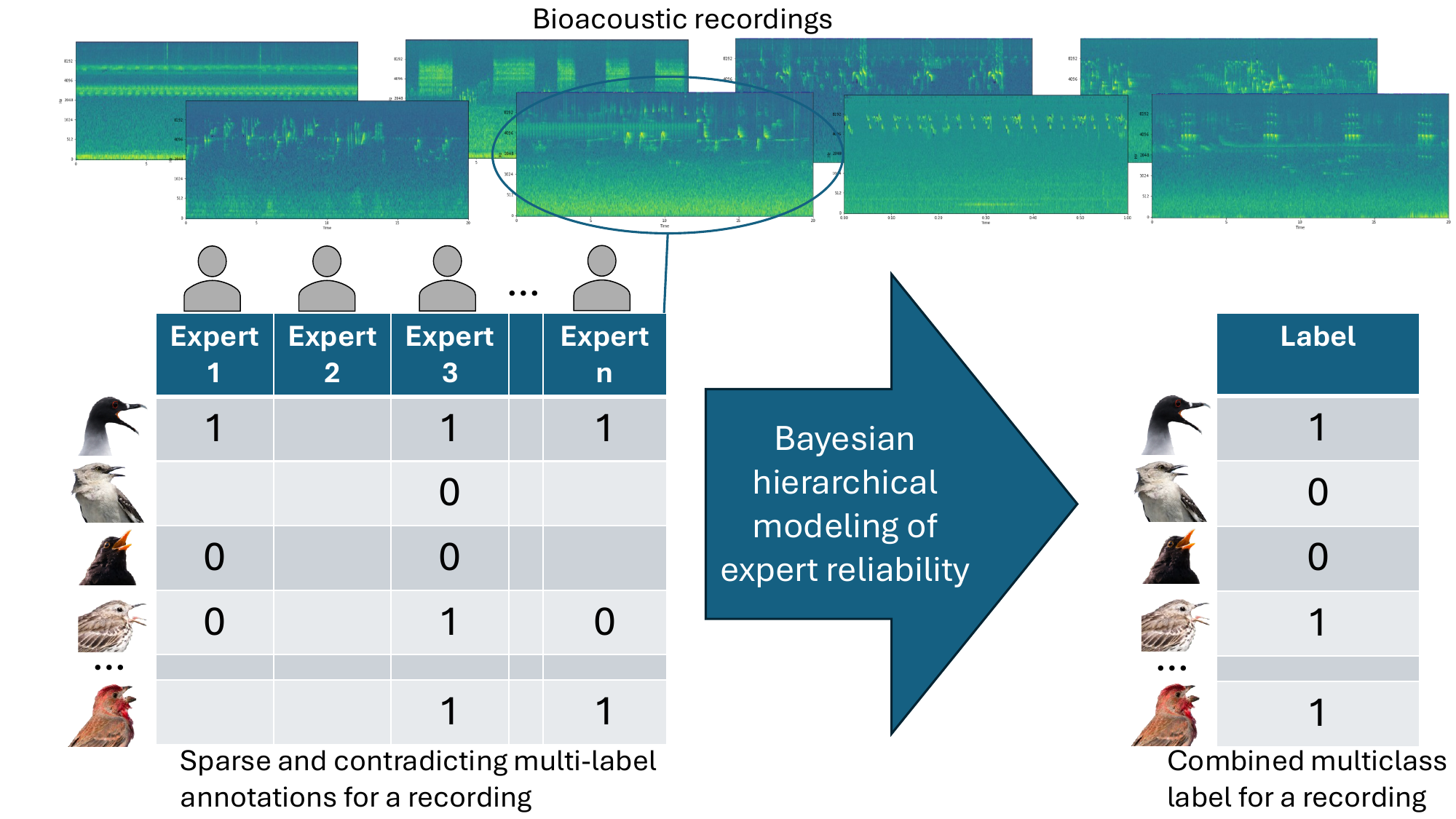}
\caption{Our framework combines sparse and possibly contradicting expert annotations into reliable multi-species labels through Bayesian modeling of expert reliability.}
\label{fig:workflow_figure}
\end{figure}

In this paper, we introduce a Bayesian hierarchical modeling framework tailored to the challenges of multi-species identification and apply it on bird sound annotation data from \cite{lehikoinen2023successful} (Fig. \ref{fig:workflow_figure}). We model the joint distribution of the different birds present in an audio recording in a flexible manner, while accommodating variability in quality of bird experts' annotations across different bird species. We employ informative priors for the model parameters to address the sparsity of the species annotation data, while using conjugacy and data augmentation to enable efficient posterior inference through collapsed Gibbs samplers in the more complex models falling in our framework.

The remainder of the paper is organized as follows. Section \ref{sec:models} presents the motivating application, the problem formulation, and the components of our Bayesian modeling framework. In Section \ref{sec:results}, we test the performance of several methods within our modeling framework on simulated data (Section \ref{sec:simulation}) and a dataset of Finnish bird vocalizations which has been annotated by bird experts through a crowdsourcing project \citep{lehikoinen2023successful} (Section \ref{sec:application}). We compare our methods to Majority Vote approach (MV), which was originally used for annotation aggregation by \cite{lehikoinen2023successful}. Lastly, potential avenues of future research are discussed in Section \ref{sec:discussion}.

\section{Materials and methods} \label{sec:models}
In this section, we first describe the motivating bird-sound annotation problem and introduce the notation used throughout the paper. We then present several Bayesian hierarchical models for this multi-species identification task. We begin the modeling part with a straightforward baseline model in Section \ref{sec:baseline}. Building from this baseline, we propose a nonparametric BMM to induce dependence in species occurrence in Section \ref{sec:BMM} and \ref{sec:DP}. We further extend the measurement error model to account for variation in annotation performance across species through a hierarchical structure described in Section \ref{sec:varying_ability}. There are four sub-models falling in our general framework, as summarized in Table \ref{table:model_sum}.

\begin{table}[ht]
\centering
\caption{Summary of key features of different models}
\begin{tabular}{lcc}
\toprule
\textbf{Model} & \textbf{Species Dependence} & \textbf{Varying Expertise} \\
\midrule
\textbf{Base} & \textbf{\ding{55}} & \textbf{\ding{55}} \\
\textbf{Base-Hierarchical} & \textbf{\ding{55}} & \textbf{\checkmark} \\
\textbf{DP-BMM} & \textbf{\checkmark} & \textbf{\ding{55}} \\
\textbf{DP-BMM-Hierarchical} & \textbf{\checkmark} & \textbf{\checkmark} \\
\bottomrule
\end{tabular}
\label{table:model_sum}
\end{table}

\subsection{Motivating application and problem formulation} \label{sec:motivation}
Our motivating application is bird-sound annotation on the Finnish Kerttu web portal \citep{lehikoinen2023successful}. Suppose that $N_2$ bird experts provide bird-song identification results for $N_1$ audio recordings, and let $N_3$ denote the number of candidate bird species. We represent the collected annotations by a three-dimensional array $\mathcal{T}$, where $\mathcal{T}_{i,j,k}=1$ means annotator $j$ reports that species $k$ is present in recording $i$, $\mathcal{T}_{i,j,k}=0$ means annotator $j$ reports that species $k$ is absent, and $\mathcal{T}_{i,j,k}=\texttt{NA}$ means that this entry is missing. Missingness is substantial because not every annotator reviews every recording, and annotators may skip species when they are not sufficiently confident.

Each recording can contain vocalizations from multiple bird species, so this is naturally a multi-label annotation aggregation problem \citep{bragg2013crowdsourcing, li2016conditional, li2018multi, zhang2018multi, zhang2019multi, shi2021gaussian, yin2021learning}. For recording $i \in \{1,2,\dots,N_1\}$, let $\bm{y}_{i} = (y_{i,1}, y_{i,2}, \dots, y_{i,N_3})^T$ denote the latent binary vector indicating which species are truly present. For annotator $j \in \{1,2,\dots,N_2\}$, let $\mathcal{S}_j \subseteq \{(i,k): i=1,2,\dots,N_1;\ k=1,2,\dots,N_3\}$ denote the recording--species pairs annotated by annotator $j$. If $(i,k)\notin \mathcal{S}_j$, then $\mathcal{T}_{i,j,k}=\texttt{NA}$; otherwise, $\mathcal{T}_{i,j,k}\in\{0,1\}$. Our first goal is to jointly infer $\{\bm{y}_1,\bm{y}_2,\dots,\bm{y}_{N_1}\}$ from the noisy and incomplete annotations $\mathcal{T}$, while accurately characterizing uncertainty.

A second goal is to model annotator expertise. In binary settings, annotator models usually rely on either one-coin or two-coin formulations \citep{raykar2010learning, ghosh2011moderates, dalvi2013aggregating}. Since sensitivity and specificity can differ substantially in bird-sound identification, we use a two-coin formulation and model annotator-specific true positive rates $\lambda_j = \Pr(\mathcal{T}_{i,j,k}=1 \mid y_{i,k}=1)$ and false positive rates $\psi_j = \Pr(\mathcal{T}_{i,j,k}=1 \mid y_{i,k}=0)$ for $(i,k)\in\mathcal{S}_j$. It is also unrealistic to assume that annotators have the same expertise across all species, which motivates the hierarchical extension introduced later in this section \citep{bragg2013crowdsourcing, padmanabhan2016topic, yin2021learning}.

The motivating dataset is extremely sparse. On average, each bird expert annotates about 129 recordings, whereas the dataset contains more than 3900 recordings in total, leading to a missing rate of $97.126\%$ in $\mathcal{T}$. At the same time, we have useful prior knowledge about annotator performance and species occurrence in the study region. For example, the average agreement score is approximately $85\%$ for species identified by at least one annotator, and false positive rates are expected to be very low \citep{lehikoinen2023successful}. These features motivate our use of informative priors for the model parameters.

\subsection{Baseline model} \label{sec:baseline}
We start with introducing a simple baseline model corresponding to row 1 of Table \ref{table:model_sum}. The latent binary species occurrence data are assumed to follow:
\begin{align}
    y_{i, k}  \sim \text{Bernoulli}(o_k),  \label{eq:baseline_indicator}
\end{align}
where $o_k$ represents the occurrence probability of species $k$ in a random audio segment. Then, we allow each annotator to have their own TPR $\lambda_j$ and FPR $\psi_j$ as follows:
\begin{equation}
    \text{Pr}(\mathcal{T}_{i,j,k} = 1 \mid y_{i, k} = 1) = \lambda_j,\qquad 
    \text{Pr}(\mathcal{T}_{i,j,k} = 1 \mid y_{i, k} = 0) = \psi_j,
\label{eq:baseline_tpr&fpr}
\end{equation}
defined for $(i, k) \in \mathcal{S}_j$. Equation \eqref{eq:baseline_tpr&fpr} implies:
\begin{align}
    \mathcal{T}_{i, j, k} \mid y_{i, k}, \lambda_j, \psi_j \sim \text{Bernoulli}\left(\lambda_j^{y_{i, k}} \psi_j^{1 - y_{i, k}}\right), \ j = 1, 2, \dots, N_2. 
\end{align}

Introducing conjugate priors for parameters $\bm{o} = (o_1, o_2, \dots, o_{N_3})^T$, $\bm{\lambda} = (\lambda_1, \lambda_2, \dots, \lambda_{N_2})^T$ and $\bm{\psi} = (\psi_1, \psi_2, \dots, \psi_{N_2})^T$, we let 
\begin{equation}
o_k \sim \text{Beta}(a_{o}, b_{o}),\qquad 
\lambda_j \sim \text{Beta}(a_{\lambda}, b_{\lambda}),\qquad
\psi_j \sim \text{Beta}(a_{\psi}, b_{\psi}).
\label{eq:three_priors}
\end{equation}
Here $a_{o}$ and $b_{o}$ characterize variation in abundance across different bird species in the region where recordings are collected. In addition, $a_{\lambda}$ and $b_{\lambda}$ characterize the distribution of TPRs across annotators, while $a_{\psi}$ and $b_{\psi}$ characterize the distribution of FPRs. We provide a detailed discussion on choosing appropriate values for the hyperparameters in Section \ref{sec:application}.

\subsection{Bernoulli mixture models} \label{sec:BMM}
Next, we account for correlations among different bird species by modeling the distribution of $\bm{y}_i = (y_{i, 1}, y_{i, 2}, \dots, y_{i, N_3})^T$ as a mixture of Bernoulli distributions \citep{bishop2006pattern}. In Section \ref{sec:DP}, we introduce a Bayesian nonparametric extension to infer the number of mixture components $R$. As we add additional components, the model can accurately characterize arbitrarily complex joint distributions of the species occurrence indicators \citep{dunson2009nonparametric}. For now, assume $R$ is pre-specified. The joint distribution of $\bm{y}_i$ is:
\begin{align}
    p(\bm{y}_i \mid \bm{\pi}, \bm{O}) = \sum_{r = 1}^{R} \pi_r \prod_{k = 1}^{N_3} o_{r, k}^{y_{i, k}} \left(1 - o_{r, k}^{1 - y_{i, k}} \right), 
\end{align}
where $o_{r, k}$ represents the occurrence probability of bird species $k$ specific to mixture component $r$, $\pi_r$ is the probability weight on component $r$, $\bm{\pi} = (\pi_1, \pi_2, \dots, \pi_R)^T$ and $\bm{O} = \{ o_{r,k} \}$ is an $R \times N_3$ matrix of the occurrence probabilities. For $R=1$, the model assumes independent occurrences of the different species, and as $R$ increases more complex dependence structures are characterized. Unlike previous methods, such as \cite{bragg2013crowdsourcing, duan2014separate, hung2017computing}, BMM flexibly captures both positive and negative correlations among the $N_3$ bird species.

We specify standard conjugate priors for component probabilities $\bm{\pi} = (\pi_1, \pi_2, \dots, \pi_{R})^T$ and occurrence probabilities within each component $o_{r, k}$ as follows:
\begin{equation}
    \bm{\pi} \sim \text{Dirichlet}(\bm{\alpha}),\qquad
    o_{r, k} \sim \text{Beta}(a_o, b_o), 
\label{eq:BMM_add_prior}
\end{equation}
where $ \bm{\alpha} = (\alpha_1, \alpha_2, \dots, \alpha_R)$, $\alpha_r > 0$, $r = 1, 2, \dots, R$, is the parameter of the Dirichlet distribution, encoding our prior beliefs about each mixture component's weight $\pi_r$. Commonly, $R$ is chosen as an upper bound on the number of components with $\alpha_r = 1/R$ or some other small number to favor setting unnecessary mixture component weights close to zero; this is sometimes referred to as an over-fitted mixture model \citep{rousseau2011asymptotic, van2015overfitting}.

\subsection{Dirichlet process extension} \label{sec:DP}
\cite{dunson2009nonparametric} proposed a nonparametric Bayes modeling approach for multivariate unordered categorical data, which is flexible enough to accurately approximate any possible joint probability mass function. Following their model, we assumed that each latent binary variable $\bm{y}_i = (y_{i, 1}, y_{i, 2}, \dots, y_{i, N_3})^T$ conforms to a DP mixture with countably infinite Bernoulli mixture components. In accordance, we have mixing weights $\bm{\pi} = \left\{\pi_r\right\}_{r = 1}^{\infty}$, and the matrix of the occurrence probabilities in Section \ref{sec:BMM} becomes $\bm{O} = \left\{\bm{o}_r \right\}_{r = 1}^{\infty}$, where $\bm{o}_r = \left(o_{r, 1}, o_{r, 2}, \dots, o_{r, N_3}\right)^T$. We introduce $\theta_i$ and $z_i$ for audio recording $i \in \left\{1, 2, \dots, N_1\right\}$, where each $\theta_i$ designates the parameter selected from the sample path $G$ of the DP that generates the corresponding latent binary variable $\bm{y}_i = (y_{i, 1}, y_{i, 2}, \dots, y_{i, N_3})^T$ and $z_i$ indicates which mixture component the audio recording $i$ is assigned to. Consequently, we can portray the infinite mixture model for $\bm{y}_i = (y_{i, 1}, y_{i, 2}, \dots, y_{i, N_3})^T$ alongside the associated prior as follows:
\begin{equation}
    G \sim \text{DP}(\gamma, H), \quad 
    \theta_i \mid G \sim G, \ i = 1, 2, \dots, N_1, \quad 
    \bm{y}_i \mid \theta_i \sim f(\bm{y}_i \mid \theta_i), \ i = 1, 2, \dots, N_1,
\label{eq:dp_bmm_hierary}
\end{equation}
where $f(\bm{y}_i \mid \theta_i)$ signifies a distribution parameterized by $\theta_i$, and the base measure $H$ outlines the prior for the occurrence probabilities of the $N_3$ bird species within each mixture component. Specifically, the base measure $H$ can then be expressed as:
\begin{align}
    H(\bm{o}) = \prod_{k = 1}^{N_3} \text{Beta}(o_k \mid a_o, b_o),
\end{align}
where $\bm{o} = (o_1, o_2, \dots, o_{N_3})^T$ and $a_o$ and $b_o$ have the same meaning as we indicate in Section \ref{sec:baseline}, encapsulating our prior beliefs about the occurrence probabilities of the $N_3$ bird species. For audio recording $i$, if $\theta_i = \bm{o}_r$, we can express the generative process of $\bm{y}_i = (y_{1, 1}, y_{1, 2}, \dots, y_{i, N_3})^T$ in Equation \eqref{eq:dp_bmm_hierary} as:
\begin{equation}
    f(\bm{y}_i \mid \theta_i) = p(\bm{y}_i \mid \bm{O}, z_i = r) = \prod_{k = 1}^{N_3} \text{Bernoulli}\left(y_{i, k} \mid o_{r, k}\right).
\end{equation}
Finally, we chose a $\text{Gamma}(u_1, u_2)$ prior for the concentration parameter $\gamma$ in the DP \citep{escobar1995bayesian, blei2006variational}, where $u_1$ and $u_2$ are hyper-parameters.

\subsection{Hierarchical modeling of annotator expertise} \label{sec:varying_ability}
In previous models, we assume the same levels of bird song identification expertise of annotators across the $N_3$ bird species. Since this is very unlikely to hold in practice, we generalize Equation \eqref{eq:baseline_tpr&fpr} to allow experts to vary in their ability levels, both overall and in terms of their skills with particular species, via the following hierarchical model: 
\begin{equation}
\begin{aligned}
    \lambda_j \mid \mu_{\lambda}, \phi_{\lambda} &\sim \mathcal{N}(\mu_{\lambda}, \phi_{\lambda}^2), \quad \lambda_{j, k} \mid \lambda_j, \phi_{\lambda}^{*} \sim \mathcal{N}\left(\lambda_j, \left(\phi_{\lambda}^{*}\right)^2\right), \\
    \psi_j \mid \mu_{\psi}, \phi_{\psi} &\sim \mathcal{N}(\mu_{\psi}, \phi_{\psi}^2), \quad
    \psi_{j, k} \mid \psi_j, \phi_{\psi}^{*} \sim\mathcal{N}\left(\psi_j, \left(\phi_{\psi}^{*}\right)^2\right), 
\end{aligned}
\label{eq:lambda_psi_hierary}
\end{equation}
where $\left(\lambda_j, \psi_j\right)$ have different meanings compared to their usage in \textbf{Base} and \textbf{DP-BMM} and represent annotator $j$'s overall bird identification expertise, $\left(\lambda_{j, k}, \psi_{j, k}\right)$ represent annotator $j$'s identification expertise for species $k$, and $(\phi_{\lambda}^{*}, \phi_{\psi}^{*})$ control the variability of annotators' identification expertise across different species. Incorporating this hierarchical structure in Equation \eqref{eq:lambda_psi_hierary}, the measurement error process for each bird expert is as follows: 
\begin{equation}
    \text{Pr}(\mathcal{T}_{i,j,k} = 1 \mid y_{i, k} = 1) = \sigma\left(\lambda_{j, k}\right), \qquad
    \text{Pr}(\mathcal{T}_{i,j,k} = 1 \mid y_{i, k} = 0) = \sigma\left(\psi_{j, k}\right),
\label{eq:varying_baseline_tpr&fpr}
\end{equation}
for $(i, k) \in \mathcal{S}_j$, where $\sigma(z) = \frac{1}{1 + \exp(-z)}$ is the logistic transformation function. For $(i, k) \in \mathcal{S}_j$, a more compact form of Equation \eqref{eq:varying_baseline_tpr&fpr} is:
\begin{align}
    \mathcal{T}_{i, j, k} \mid y_{i, k}, \lambda_{j,k}, \psi_{j,k} \sim \text{Bernoulli}\left(\sigma(\lambda_{j,k})^{y_{i, k}} \sigma(\psi_{j,k})^{1 - y_{i, k}}\right), \ j = 1, 2, \dots, N_2.
    \label{eq:baseline_combine_annotate}
\end{align}
If annotator $j$ does not have expertise in species $k$ then we do not include any data from that annotator on that species in our analysis.

\subsection{Posterior computation}
For the models summarized in Table \ref{table:model_sum}, we developed corresponding easy to implement Gibbs samplers. To improve efficiency in posterior sampling for our DP mixture models, we marginalized out $\bm{o}_{r} = (o_{r, 1}, o_{r, 2}, \dots, o_{r, N_3})^T$ and $\pi_r$ for each mixture component, and derive a collapsed Gibbs sampler \citep{liu1994collapsed}, which samples the assignments of the $N_1$ audio recordings. For models taking the variability of annotators' sound identification expertise across species into consideration, we adopted the Pólya-Gamma (PG) augmentation technique widely used in logistic regression-type models \citep{polson2013bayesian}. Details are provided in Appendix 1.

\subsection{Simulation setup} \label{sec:simulation_setup}
We assessed the proposed framework through simulation studies under a range of data-generating mechanisms, sparsity levels, and prior settings. In all experiments, we generated $1000$ audio recordings, $20$ annotators, and $25$ bird species.

For posterior inference, we ran three chains of $2000$ iterations each for \textbf{Base} and \textbf{DP-BMM}. We discarded the first $1000$ draws in each chain as burn-in, and collected sufficient samples to ensure a minimum of $100$ effective sample size (ESS) for each parameter. For the more complex \textbf{Base-Hierarchical} and \textbf{DP-BMM-Hierarchical} models, to achieve the minimum ESS requirement, each chain was run for $5000$ iterations, with the initial $2000$ draws discarded as burn-in. In addition to estimating ESS values for different parameters, we assessed MCMC convergence and mixing via Gelman--Rubin statistics, which were below $1.1$ for all parameters in all experiments.

We considered scenarios that vary according to whether the occurrences of different species are independent or correlated, whether annotators' expertise varies across species, the average number of annotations per recording ($0.8$, $1.6$, $2.4$, $3.2$, $4.0$), and the prior specification used in the analysis. We used the default prior (see Section \ref{sec:prior_specify}) and also performed a sensitivity analysis on the choice of priors.

When occurrences of species are independent, we drew $o_k \sim \text{Beta}(2, 98)$, $k = 1, 2, \dots, 25$, and then generated the indicators $y_{i, k} \mid o_k \sim \text{Bernoulli}(o_k)$, $i = 1, 2, \dots, 1000$, $k = 1, 2, \dots, 25$, for all species in each recording. When occurrences of species are correlated, we drew the indicators $y_{i,k}$ for the first $15$ species as described above, and then directly set $y_{i, k+15} = y_{i, k}$, $i = 1, 2, \dots, 1000$, $k = 1, 2, \dots, 10$. This represents an extreme scenario in which species $k$ and species $k+15$ always co-occur.

We generated three types of annotators: random ($10\%$), normal ($70\%$), and excellent ($20\%$). The average across-species TPRs for individuals in these three groups were drawn independently from $\text{Uniform}(0.60, 0.70)$, $\text{Uniform}(0.75, 0.85)$, and $\text{Uniform}(0.90, 0.95)$, respectively. As birders tend not to record a bird species as being present unless they are almost certain, FPRs tend to be very low; hence, we generated average across-species FPRs for all individuals independently from $\text{Uniform}(0.001, 0.01)$. For cases in which expertise does not vary across species, we generated $\mathcal{T}_{i, j, k}$ from $\text{Bernoulli}\left(\lambda_{j}^{y_{i,k}} \psi_{j}^{1-y_{i,k}}\right)$. For cases in which expertise does vary, we characterized this variability and generated annotators' species-specific TPRs $\lambda_{j, k}$ and FPRs $\psi_{j,k}$, as well as annotations $\mathcal{T}_{i, j, k}$, through Equations \eqref{eq:lambda_psi_hierary} -- \eqref{eq:baseline_combine_annotate}, where $\phi_{\lambda}^{*} = 2$ and $\phi_{\psi}^{*} = 1$.

Since the goal of our simulation studies was to assess the performance of our approach in realistic settings, we started with a default prior as in the real data analysis and then conducted a sensitivity analysis. We considered different combinations of priors for the occurrence probabilities of species and annotators' TPRs. Our default prior for the occurrence probability was $\text{Beta}(0.02 \times 100, (1 - 0.02) \times 100)$. Then, we varied the prior mean within the plausible set $\{0.01, 0.015, 0.025, 0.30\}$. Our default prior for annotators' $\lambda_j$ in \textbf{Base} and \textbf{DP-BMM} was $\text{Beta}(0.81 \times 20, (1 - 0.81) \times 20)$; we varied the prior mean within $\{0.75, 0.78, 0.84, 0.87\}$. For \textbf{Base-Hierarchical} and \textbf{DP-BMM-Hierarchical}, the default prior for $\lambda_j$ was $\mathcal{N}(\log(0.81/(1-0.81)), 0.58^2)$; the other choices of priors for $\lambda_j$ were $\mathcal{N}(\log(0.75/(1-0.75)), 0.53^2)$, $\mathcal{N}(\log(0.78/(1-0.78)), 0.55^2)$, $\mathcal{N}(\log(0.84/(1-0.84)), 0.63^2)$, and $\mathcal{N}(\log(0.875/(1-0.875)), 0.72^2)$. As for annotators' FPRs, we used $\text{Beta}(0.005 \times 1200, (1 - 0.005) \times 1200)$ as the prior in \textbf{Base} and \textbf{DP-BMM}, and used $\mathcal{N}(0.005/(1 - 0.005), 0.41^2)$ as the prior in \textbf{Base-Hierarchical} and \textbf{DP-BMM-Hierarchical}. For \textbf{DP-BMM} and \textbf{DP-BMM-Hierarchical}, we placed a $\text{Gamma}(0.5, 0.5)$ prior on concentration parameter $\gamma$; for \textbf{Base-Hierarchical} and \textbf{DP-BMM-Hierarchical}, we employed an empirical Bayes approach to estimate the hyperparameters for annotators' species-specific identification expertise. Specifically, we maximized the likelihood with respect to $\phi_{\lambda}^{*}$ and $\phi_{\psi}^{*}$, and updated them as follows:
\begin{equation}
    \left(\phi_{\lambda}^{*}\right)^2 = \frac{\sum_{j = 1}^{20} \sum_{k = 1}^{25} \mathbb{I}\{k \in l_j\}(\lambda_{j, k} - \lambda_j)^2}{\sum_{j = 1}^{20} \sum_{k = 1}^{25} \mathbb{I}\{k \in l_j\}},
    \left(\phi_{\psi}^{*}\right)^2 = \frac{\sum_{j = 1}^{20} \sum_{k = 1}^{25} \mathbb{I}\{k \in l_j\}(\psi_{j, k} - \psi_j)^2}{\sum_{j = 1}^{20} \sum_{k = 1}^{25} \mathbb{I}\{k \in l_j\}},
\end{equation}
where $l_j$ is the pre-specified list of bird species for annotator $j$, which is mentioned in Section \ref{sec:varying_ability}, and coded as $1, 2, \dots, 25$.

\subsection{Application-specific prior specification and evaluation setup} \label{sec:prior_specify}

For the real-data analysis, we used bird species annotation data from the Finnish Kerttu project introduced in Section \ref{sec:motivation}, available at \url{https://zenodo.org/record/7030863#.Y6GgtIRBwuU}. The data are highly sparse, so careful prior elicitation is important.

Similar to the simulation studies, we ran three chains of $3000$ iterations each for \textbf{Base} and \textbf{DP-BMM}. We discarded the first $1500$ draws in each chain as burn-in. For \textbf{Base-Hierarchical} and \textbf{DP-BMM-Hierarchical}, each chain was run for $7000$ iterations, with the initial $2000$ draws discarded as burn-in for both models. For the stickiest $\psi_{j, k}$'s in both \textbf{Base-Hierarchical} and \textbf{DP-BMM-Hierarchical}, we obtained approximately $20$ ESS, but such low ESS values were only observed for a small fraction of the $\psi_{j, k}$'s. The overwhelming majority of parameters had ESS over $100$. Further diagnostics of MCMC convergence and mixing are provided in Appendix \ref{sec:diagnostic_application}.

We first specified priors for \textbf{Base} and \textbf{DP-BMM}, assuming uniform identification expertise across species. Based on the domain knowledge described in Section \ref{sec:motivation}, we expect (i) the average TPR among our annotators to be $\sim 0.9$, (ii) nearly $95\%$ of annotators achieve TPRs exceeding $85\%$, and (iii) the majority exhibit FPRs around $0.005$. Taking into consideration factors such as multiple bird species vocalizing simultaneously and potential background noise, we set the hyperparameters for $\lambda_j$ and $\psi_j$ to $a_{\lambda} = 45$, $b_{\lambda} = 5$, $a_{\psi} = 5$, and $b_{\psi} = 995$, resulting in the equal-tailed $95\%$ prior credible interval for $\lambda_j$ to be $(0.804, 0.966)$ and for $\psi_j$ to be $(0.00163, 0.0102)$. For the bird species occurrence probabilities, we know that there are usually only two or three species present in a recording. If we ignore correlations among species and consider the occurrence of one species as a Bernoulli trial, the prior for $o_k$ should satisfy $a_o / (a_o + b_o) \approx 2 / N_3$, where $N_3 = 117$ in our problem. Moreover, as $95\%$ of recordings contain at most $4$ or $5$ species, the hyperparameters for $o_k$ in \textbf{Base} and $o_{r, k}$ in \textbf{DP-BMM} are chosen to be $a_o = 2$ and $b_o = 98$, leading to equal-tailed $95\%$ prior credible intervals for both $o_k$ and $o_{r, k}$ of $(0.00246, 0.0550)$.

We next specified priors for \textbf{Base-Hierarchical} and \textbf{DP-BMM-Hierarchical}, incorporating the hierarchical structure of annotators' identification expertise. For the overall expertise of annotator $j$, we set the hyperparameters as $\mu_{\lambda} = \log\left(0.9 / (1 - 0.9)\right)$, $\phi_{\lambda} = 0.48$, $\mu_{\psi} = \log\left(0.005 / (1 - 0.005)\right)$, and $\phi_{\psi} = 0.45$. These result in equal-tailed $95\%$ prior credible intervals for $\sigma(\lambda_j)$ of $(0.778, 0.959)$ and for $\sigma(\psi_j)$ of $(0.00207, 0.0120)$. Similar to the simulation studies, we employed an empirical Bayes approach to estimate $\phi_{\lambda}^{*}$ and $\phi_{\psi}^{*}$.

For evaluation of species identification, based on the results of \textbf{Base}, we identified $115$ recordings with the highest levels of uncertainty, which potentially contain bird species that are difficult to identify. We then asked an experienced and reliable ornithologist to provide annotations for these recordings. These annotations were used as the gold standard for evaluating recovery of $\{\bm{y}_1, \bm{y}_2, \dots, \bm{y}_{N_1}\}$ from noisy $\mathcal{T}$.

\section{Results} \label{sec:results}

In this Section \ref{sec:results}, we test the performance of several methods within our modeling framework on simulated data (Section \ref{sec:simulation}) and a dataset of Finnish bird vocalizations which has been annotated by bird experts through a crowdsourcing project \citep{lehikoinen2023successful} (Section \ref{sec:application}). We compare our methods to Majority Vote approach (MV), which was originally used for annotation aggregation by \cite{lehikoinen2023successful}.

\subsection{Simulation results} \label{sec:simulation}
We present results under the simulation settings described in Section \ref{sec:simulation_setup}. We evaluate the proposed framework in terms of species identification and estimation of annotators' bird-song identification expertise.

\subsubsection{Species identification}
\label{sec:simulation_annotation_aggregation}
We assess the performance of different annotation aggregation methods under different generative mechanisms in terms of the Area under the Curve (AUC). Under the default priors for annotators' TPRs and the occurrence probabilities, performance is summarized in Table \ref{tab:simu_default_auc} for the four scenarios -- \textbf{Scenario 1}: independent species and no variability of annotators' identification expertise, \textbf{Scenario 2}: correlated species and no variability, \textbf{Scenario 3}: independent species and variability and \textbf{Scenario 4}: correlated species and variability. Sensitivity analyses are provided in Appendix \ref{sec:_additional_simulation_agg}.

\begin{table}[!t]
\centering
\caption{AUCs of different methods under four different scenarios with default priors. \#Anns represents the number of annotations for each recording.}
\resizebox{0.8\textwidth}{!}{
\renewcommand{\arraystretch}{1} % ensure single-spacing
\begin{tabular}{lcrrrrr}
  \toprule
  & \backslashbox{\#Anns}{Method} & \textbf{MV} & \textbf{Base} & \textbf{Base-Hierarchical} & \textbf{DP-BMM} & \textbf{DP-BMM-Hierarchical} \\ 
\midrule
Scenario 1 & \hspace{30pt} & & & & & \\
\midrule
& 0.8 & 0.834 & 0.863 & 0.866 & 0.863 & \textbf{0.867} \\ 
& 1.6 & 0.910 & 0.938 & \textbf{0.940} & 0.938 & \textbf{0.940} \\ 
& 2.4 & 0.951 & 0.976 & \textbf{0.977} & 0.976 & 0.976 \\ 
& 3.2 & 0.978 & \textbf{0.993} & \textbf{0.993} & \textbf{0.993} & \textbf{0.993} \\ 
& 4.0 & 0.989 & \textbf{0.997} & \textbf{0.997} & \textbf{0.997} & 0.996 \\ 
\midrule
Scenario 2 & \hspace{30pt} & & & & & \\
\midrule
& 0.8 & 0.832 & 0.873 & 0.869 & \textbf{0.875} & 0.872 \\
& 1.6 & 0.920 & 0.952 & 0.953 & \textbf{0.956} & \textbf{0.956} \\
& 2.4 & 0.954 & 0.978 & 0.978 & \textbf{0.981} & 0.980 \\
& 3.2 & 0.977 & 0.993 & 0.993 & \textbf{0.996} & 0.995 \\
& 4.0 & 0.987 & 0.998 & 0.997 & \textbf{0.999} & 0.998 \\
\midrule
Scenario 3 & \hspace{30pt} & & & & & \\
\midrule
& 0.8 & 0.804 & 0.827 & \textbf{0.837} & 0.828 & 0.834 \\ 
& 1.6 & 0.861 & 0.896 & \textbf{0.906} & 0.897 & 0.905 \\ 
& 2.4 & 0.917 & 0.951 & \textbf{0.959} & 0.951 & 0.957 \\ 
& 3.2 & 0.953 & 0.976 & \textbf{0.984} & 0.976 & 0.980 \\ 
& 4.0 & 0.969 & 0.984 & \textbf{0.989} & 0.984 & \textbf{0.989} \\ 
\midrule
Scenario 4 & \hspace{30pt} & & & & & \\
\midrule
& 0.8 & 0.790 & 0.800 & 0.842 & 0.803 & \textbf{0.845} \\ 
& 1.6 & 0.887 & 0.914 & 0.931 & 0.915 & \textbf{0.933} \\ 
& 2.4 & 0.924 & 0.958 & 0.965 & 0.961 & \textbf{0.967} \\ 
& 3.2 & 0.953 & 0.973 & 0.979 & 0.977 & \textbf{0.980} \\ 
& 4.0 & 0.961 & 0.983 & 0.983 & 0.985 & \textbf{0.988} \\ 
\bottomrule
\end{tabular}}
\label{tab:simu_default_auc}
\end{table}

Generally speaking, performance of all methods in terms of AUCs improves significantly with an increasing number of annotations per recording for all scenarios. All of our models consistently outperform \textbf{MV} across scenarios, which demonstrates the necessity of taking annotators' identification expertise into account when aggregating annotations. As expected, \textbf{DP-BMM}/\textbf{DP-BMM-Hierarchical} perform slightly better than \textbf{Base}/\textbf{Base-Hierarchical} when the occurrences of species are correlated. When the number of annotations per recording is low, the performance gap is even larger, suggesting that combining BMMs with measurement error with multiple annotators is effective in our settings. In comparing \textbf{Base}/\textbf{DP-BMM} and \textbf{Base-Hierarchical}/\textbf{DP-BMM-Hierarchical}, the latter  performs much better in the presence of variability across species in annotator expertise. Based on the sensitivity analyses in Appendix \ref{sec:_additional_simulation_agg}, we find the performance of the models doesn't change significantly under reasonable changes to the prior.

\subsubsection{Assessment of species identification expertise}
\label{sec:simulation_assess_skills}
In addition to species identification, we evaluate the accuracy of the estimated annotators' TPRs for methods except \textbf{MV} under different generative mechanisms in terms of coverages of $95\%$ credible intervals (CIs) and mean squared errors (MSEs). As shown in Section \ref{sec:simulation_annotation_aggregation}, we summarize the coverages for annotators' TPRs in Table \ref{tab:simu_default_coverage} and the MSEs in Table \ref{tab:simu_default_mse} under default priors. Sensitivity analyses are provided in Appendix \ref{sec:_additional_simulation_ability}.

\begin{table}[!t]
\centering
\caption{Coverage of 95\% CIs for TPRs of different methods under four different scenarios with default priors. \#Anns represents the number of annotations for each recording.}
\resizebox{0.8\textwidth}{!}{
\renewcommand{\arraystretch}{1} % ensure single-spacing
\begin{tabular}{lcrrrr}
  \toprule
  & \backslashbox{\#Anns}{Method} & \textbf{Base} & \textbf{Base-Hierarchical} & \textbf{DP-BMM} & \textbf{DP-BMM-Hierarchical} \\ 
  \midrule
  \textbf{Scenario 1} & \hspace{30pt} & & & & \\
  \midrule
  & 0.8 & 0.50 & \textbf{0.85} & 0.50 & \textbf{0.85} \\ 
  & 1.6 & 0.60 & 0.85 & 0.65 & \textbf{0.95} \\ 
  & 2.4 & 0.75 & 0.85 & 0.75 & \textbf{0.90} \\ 
  & 3.2 & \textbf{1.00} & 0.95 & 0.95 & 0.95 \\ 
  & 4.0 & 0.95 & \textbf{1.00} & 0.95 & 0.90 \\ 
  \midrule
  \textbf{Scenario 2} & \hspace{30pt} & & & & \\
  \midrule
  & 0.8 & 0.55 & \textbf{0.85} & 0.65 & 0.80 \\ 
  & 1.6 & 0.85 & 0.90 & \textbf{0.95} & \textbf{0.95} \\ 
  & 2.4 & 0.85 & \textbf{0.90} & 0.85 & \textbf{0.90} \\ 
  & 3.2 & \textbf{0.95} & \textbf{0.95} & 0.90 & 0.90 \\ 
  & 4.0 & \textbf{1.00} & 0.90 & \textbf{1.00} & 0.85 \\ 
  \midrule
  \textbf{Scenario 3} & \hspace{30pt} & & & & \\
  \midrule
  & 0.8 & 0.25 & \textbf{0.85} & 0.25 & \textbf{0.85} \\ 
  & 1.6 & 0.15 & 0.90 & 0.15 & \textbf{1.00} \\ 
  & 2.4 & 0.25 & \textbf{0.90} & 0.30 & \textbf{0.90} \\ 
  & 3.2 & 0.40 & 0.90 & 0.35 & \textbf{0.95} \\ 
  & 4.0 & 0.45 & \textbf{0.95} & 0.40 & \textbf{0.95} \\ 
  \midrule
  \textbf{Scenario 4} & \hspace{30pt} & & & & \\
  \midrule
  & 0.8 & 0.25 & \textbf{0.95} & 0.25 & 0.90 \\ 
  & 1.6 & 0.15 & \textbf{0.90} & 0.15 & \textbf{0.90} \\ 
  & 2.4 & 0.45 & 0.85 & 0.50 & \textbf{0.90} \\ 
  & 3.2 & 0.45 & \textbf{0.90} & 0.55 & \textbf{0.90} \\ 
  & 4.0 & 0.60 & \textbf{0.85} & 0.60 & \textbf{0.85} \\ 
  \bottomrule
\end{tabular}}
\label{tab:simu_default_coverage}
\end{table}

\begin{table}[!t]
\centering
\caption{The MSEs for TPRs of different methods under four different scenarios with default priors (All values in units of $10^{-3}$). \#Anns represents the number of annotations for each recording.}
\resizebox{0.8\textwidth}{!}{
\renewcommand{\arraystretch}{1} % ensure single-spacing
\begin{tabular}{lcrrrr}
  \toprule
  & \backslashbox{\#Anns}{Method} & \textbf{Base} & \textbf{Base-Hierarchical} & \textbf{DP-BMM} & \textbf{DP-BMM-Hierarchical} \\
  \midrule
  \textbf{Scenario 1} & \hspace{30pt} & & & & \\
  \midrule
  & 0.8 & 23.7 & \textbf{6.95} & 24.1 & 10.2 \\
  & 1.6 & 14.0 & 4.60 & 11.9 & \textbf{3.54} \\
  & 2.4 & 7.46 & \textbf{3.62} & 7.24 & 3.71 \\
  & 3.2 & \textbf{1.44} & 1.71 & 1.49 & 1.50 \\
  & 4.0 & 2.09 & \textbf{1.46} & 1.98 & 4.13 \\
  \midrule
  \textbf{Scenario 2} & \hspace{30pt} & & & & \\
  \midrule
  & 0.8 & 38.2 & \textbf{8.82} & 38.2 & 12.9 \\ 
  & 1.6 & 9.58 & 3.93 & 8.46 & \textbf{3.61} \\
  & 2.4 & 4.81 & 3.68 & 4.64 & \textbf{3.49} \\
  & 3.2 & \textbf{2.31} & 2.53 & \textbf{2.31} & 3.46 \\
  & 4.0 & 2.27 & \textbf{1.98} & 2.34 & 3.07 \\
  \midrule
  \textbf{Scenario 3} & \hspace{30pt} & & & & \\
  \midrule
  & 0.8 & 86.3 & 5.85 & 86.3 & \textbf{5.70} \\ 
  & 1.6 & 74.3 & \textbf{3.73} & 70.4 & 4.41 \\ 
  & 2.4 & 38.6 & \textbf{3.59} & 37.7 & 3.69 \\ 
  & 3.2 & 20.0 & \textbf{2.74} & 20.0 & 4.04 \\ 
  & 4.0 & 11.2 & 2.65 & 11.8 & \textbf{2.55} \\ 
  \midrule
  \textbf{Scenario 4} & \hspace{30pt} & & & & \\
  \midrule
  & 0.8 & 83.6 & \textbf{4.12} & 81.0 & 4.43 \\ 
  & 1.6 & 68.1 & \textbf{5.86} & 66.9 & 6.53 \\ 
  & 2.4 & 33.8 & 4.58 & 28.6 & \textbf{3.69} \\ 
  & 3.2 & 19.4 & \textbf{4.82} & 18.4 & 4.97 \\ 
  & 4.0 & 13.2 & 5.13 & 13.1 & \textbf{4.56} \\ 
  \bottomrule
\end{tabular}}
\label{tab:simu_default_mse}
\end{table}

In general, all of the models' performances improve in terms of both coverage and MSEs of annotators' TPRs as the number of annotations per recording increases. Comparing \textbf{Base} and \textbf{DP-BMM}, the latter has higher coverages and lower MSEs when the occurrences of bird species are correlated, indicating the DP Bernoulli mixture model's advantage in capturing and exploiting correlations among species. Comparing \textbf{Base-Hierarchical} and \textbf{DP-BMM-Hierarchical}, the simpler model performs better when the number of annotations is low even when occurrences of species are correlated; hence, for inferring annotator ability scores, an overly simple model may be preferred when training data are extremely sparse. However, from Tables \ref{tab:simu_default_coverage} and \ref{tab:simu_default_mse}, \textbf{Base-Hierarchical}/\textbf{DP-BMM-Hierarchical} have much better performance than \textbf{Base}/\textbf{DP-BMM} in terms of both coverage and MSE across the scenarios especially when the number of annotations is low. The latter ones could be regarded as the limiting case of the former ones, where $\phi_{\lambda}^{*} = \phi_{\psi}^{*} = 0$. From the sampling algorithms listed in Appendix \ref{sec:posterior_dpbmm_h}, when the annotations are sparse and reasonable priors are adopted, the additional hierarchical structure help pull $\lambda_j$ towards the prior mean, thus constraining the posterior samples of $\sigma\left(\lambda_j\right)$ within a reasonable range. Additionally, the gap in the performances between the two types of models in estimating annotators' identification expertise decreases as more annotations are collected.

According to the additional results in Appendix \ref{sec:_additional_simulation_ability}, \textbf{Base-Hierarchical} and \textbf{DP-BMM-Hierarchical} generally have significantly better performances under different priors for annotators' TPRs and occurrence probabilities when few annotations are obtained, whereas \textbf{Base} and \textbf{DP-BMM} are more sensitive to the prior specification in sparse data cases. If somewhat unreasonable priors are chosen, performance of the non-hierarchical models suffers even for moderately large numbers of annotations.

\subsection{Results for bird species annotations} \label{sec:application}
We now present results for the Finnish Kerttu data using the prior specification and evaluation setup described in Section \ref{sec:prior_specify}.

\subsubsection{Species identification} \label{sec:annoattion_aggregation}
\begin{figure}
\centering
\includegraphics[scale = 0.70]{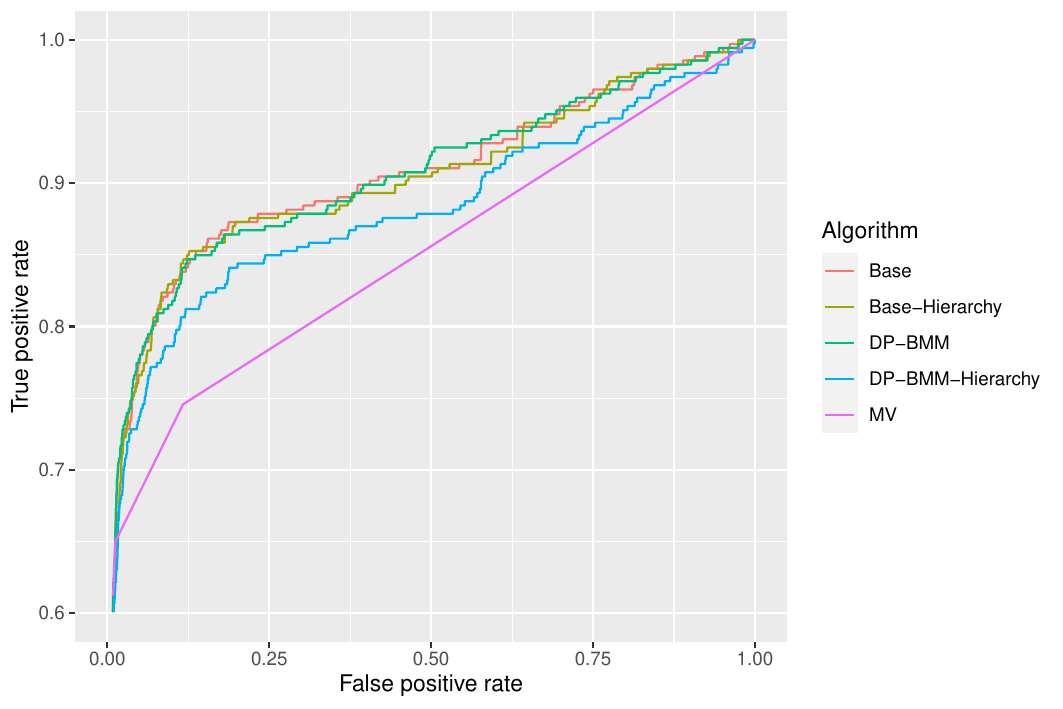}
\caption{ROC curves for five annotation aggregation methods. Colors correspond to different aggregation methods. The AUCs of \textbf{MV}, \textbf{Base}, \textbf{Base-Hierarchical}, \textbf{DP-BMM} and \textbf{DP-BMM-Hierarchical} are 0.849, 0.905, 0.903, 0.905 and 0.881, respectively.}
\label{fig:app_aggregation_roc}
\end{figure}

% Based on the results of \textbf{Base}, we identified $115$ recordings which exhibited the highest level of uncertainty, potentially containing bird species that are challenging to identify. Subsequently, we asked an experienced and reliable ornithologist to provide annotations for these recordings. These annotations are considered as the gold standard, enabling us to evaluate our models in recovering $\{\bm{y}_1, \bm{y}_2, \dots, \bm{y}_{N_1}\}$ from noisy $\mathcal{T}$. Their performances are visualized through Receiver Operating Characteristic (ROC) curves (see Figure \ref{fig:app_aggregation_roc}). The results presented below are derived by averaging samples after burn-in from three Markov chains, each initialized with a different random seed.
We first assess species identification performance. Figure \ref{fig:app_aggregation_roc} displays the ROC curves for the five annotation aggregation methods.

As shown in Figure \ref{fig:app_aggregation_roc}, all of our models demonstrate significantly superior performances compared with \textbf{MV} in terms of AUC. This remarkable improvement can be primarily attributed to accounting for heterogeneous bird song identification expertise among annotators. Upon closer examination of Figure \ref{fig:app_aggregation_roc}, at a given FPR, \textbf{Base}, \textbf{Base-Hierarchical} and \textbf{DP-BMM} exhibit similar TPRs, which are slightly higher than that of \textbf{DP-BMM-Hierarchical}. In our problem, most annotators correctly indicate the absence of a bird species in a recording. Then what we care about is whether annotators can correctly identify the species when it vocalizes. In our current data, \textbf{Base}/\textbf{DP-BMM} outperform the other three methods in terms of TPR. \textbf{Base-Hierarchical} and \textbf{MV} have similar performances in terms of sensitivity, whereas \textbf{DP-BMM-Hierarchical} seems to be the worst. When annotations are sparse it is difficult to accurately estimate annotators' species-specific identification expertise under a highly flexible model. However, we expect based on our simulations that \textbf{DP-BMM-Hierarchical}’s relative performance will improve as additional annotations become available.

In addition to predictive performance in inferring the species in terms of AUC, we care about how well the models fit the data. For Bayesian hierarchical models, we assess predictive loss using the Watanabe-Akaike information criterion (WAIC) \citep{watanabe2010asymptotic},  $\text{WAIC} = -2 \left(\text{lppd} - p_{\text{WAIC}}\right)$, where lppd represents the log point-wise predictive density and $p_{\text{WAIC}}$ corresponds to a correction for effective number of parameters to adjust for over-fitting; see  \cite{gelman2013bayesian, gelman2014understanding}. Sometimes, an overly simple model will do just as well in terms of prediction error but then predictive uncertainty is underestimated. Therefore, we are interested in the accuracy of probabilistic predictions, and we choose to use the Brier score (BS) to measure the model calibration, with lower values indicating better calibration. All models' performances in terms of WAIC and BS are summarized in Table \ref{tab:model_comparison}. Combining these metrics, we see a trade-off between model complexity and predictive calibration. \textbf{DP-BMM} excels in fitting the data but falls short in predictive calibration; conversely, \textbf{Base-Hierarchical}, while not the best in terms of WAIC, offers the most reliable predictive calibration. For our problem, \textbf{Base-Hierarchical} may be the most appropriate model overall, balancing acceptable fit to the data with superior calibration of uncertainty.
\begin{table}[ht]
    \centering
    \caption{WAICs and Brier scores of four Bayesian hierarchical models, which are computed based on posterior draws from Gibbs samplers.}
    \renewcommand{\arraystretch}{1} % ensure single-spacing
    \begin{tabular}{lcccc}
        \toprule
        & \textbf{Base} & \textbf{Base-Hierarchical} & \textbf{DP-BMM} & \textbf{DP-BMM-Hierarchical} \\
        \midrule
        lppd & -8120.59 & -8274.19 & \textbf{-7719.85} & -11431.57 \\
        \( p_{\text{WAIC}} \) & 15693.27 & 18464.95 & \textbf{14354.50} & 18979.46 \\
        WAIC & 47627.73 & 53478.30 & \textbf{44148.72} & 60822.06 \\
        \midrule
        BS & 0.0172 & \textbf{0.0163} & 0.0174 & 0.0165 \\
        \bottomrule
    \end{tabular}
    \label{tab:model_comparison}
\end{table}

\subsubsection{Assessment of sound identification expertise} \label{sec:assess_skills}
\begin{figure}
    \centering
    \begin{subfigure}{1.05\textwidth}
        \includegraphics[width=1.00\textwidth]{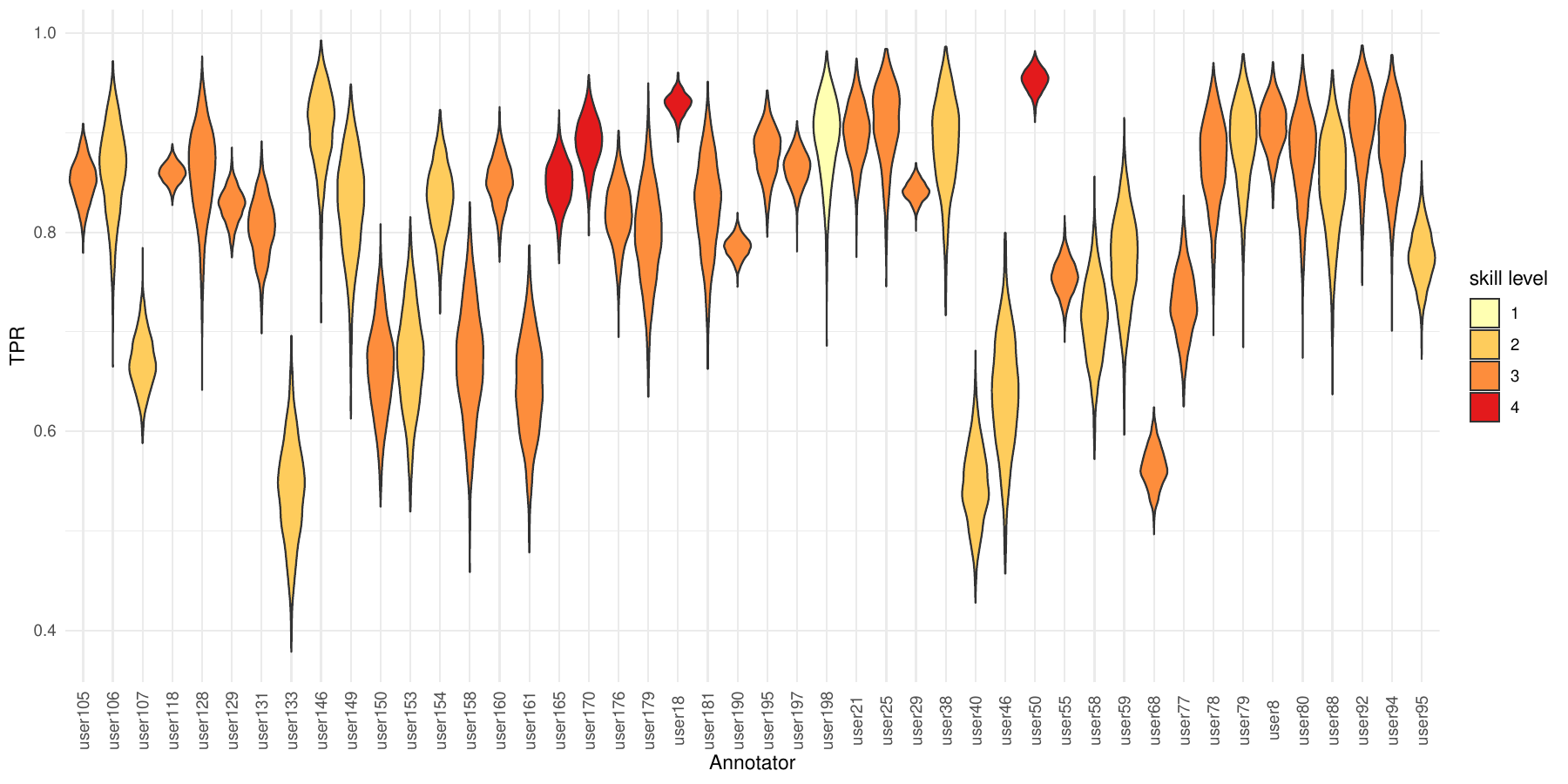}
        \caption{The posterior distributions of annotators' TPRs in \textbf{Base}.}
    \end{subfigure}

    \begin{subfigure}{1.05\textwidth}
        \includegraphics[width=1.00\textwidth]{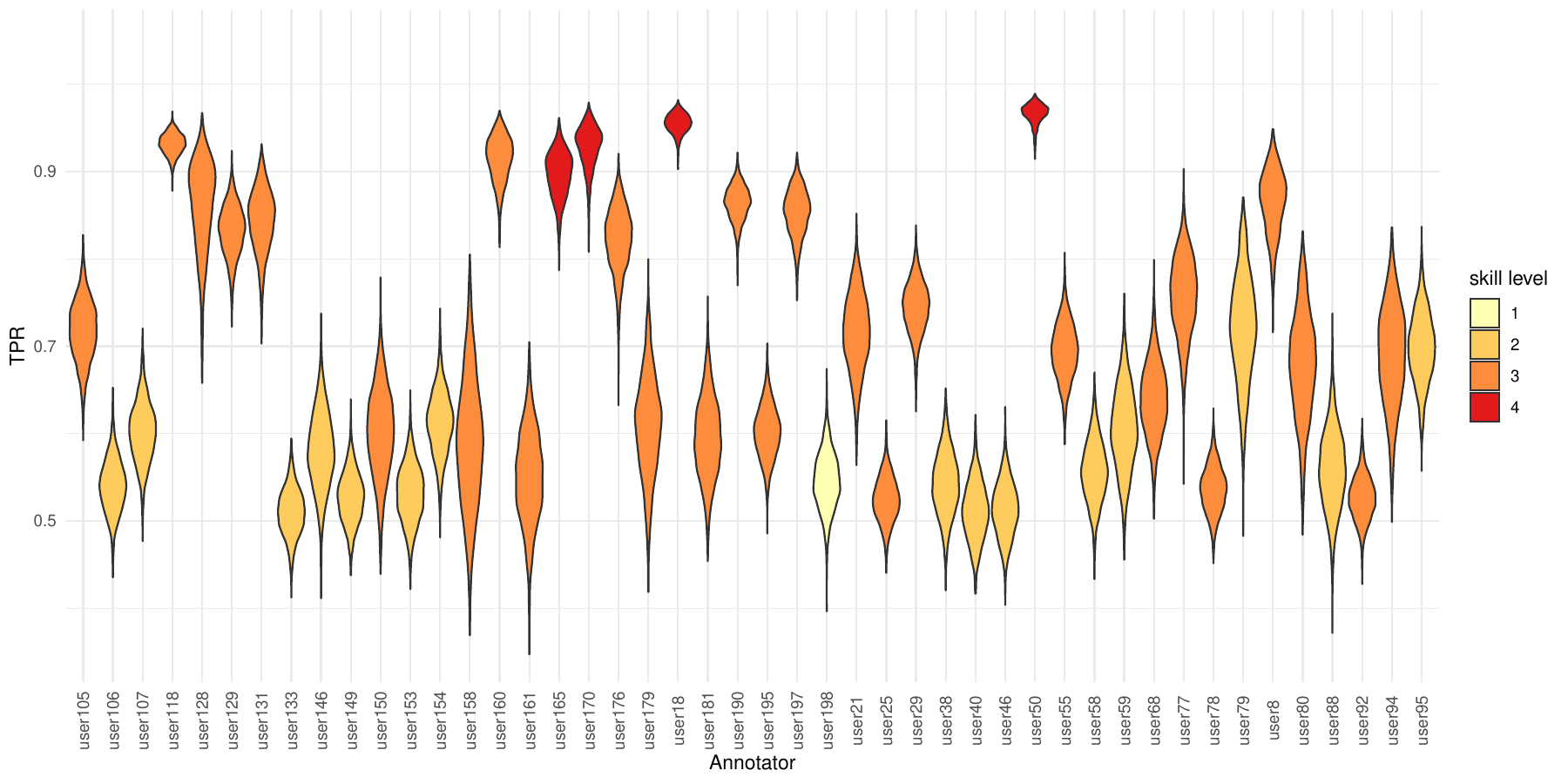}
        \caption{The posterior distributions of annotators' TPRs in \textbf{Base-Hierarchical}.}
    \end{subfigure}
    \caption{The posterior distributions of annotators' TPRs in two models that do not take into account correlations among bird species. Levels $1, 2, 3, 4$ represent the different levels of annotators' bird song identification expertise. Level $4$ corresponds to the highest skill level, followed by decreasing levels of 3, 2, and 1 in that order.}
    \label{fig:app_ability_base}
\end{figure}

\begin{figure}
    \centering
    \begin{subfigure}{1.05\textwidth}
        \includegraphics[width=1.00\textwidth]{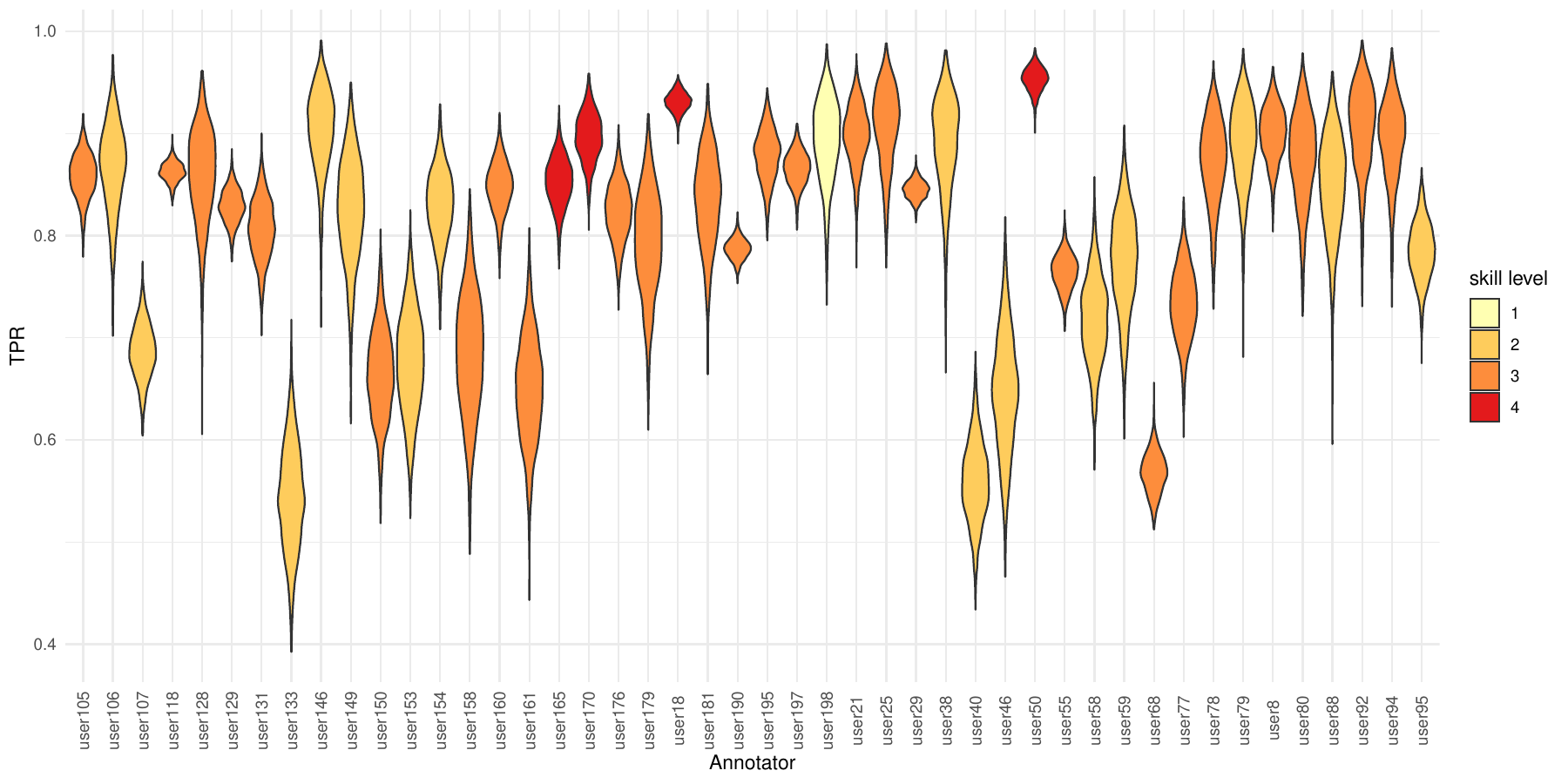}
        \caption{The posterior distributions of annotators' TPRs in \textbf{DP-BMM}.}
    \end{subfigure}

    \begin{subfigure}{1.05\textwidth}
        \includegraphics[width=1.00\textwidth]{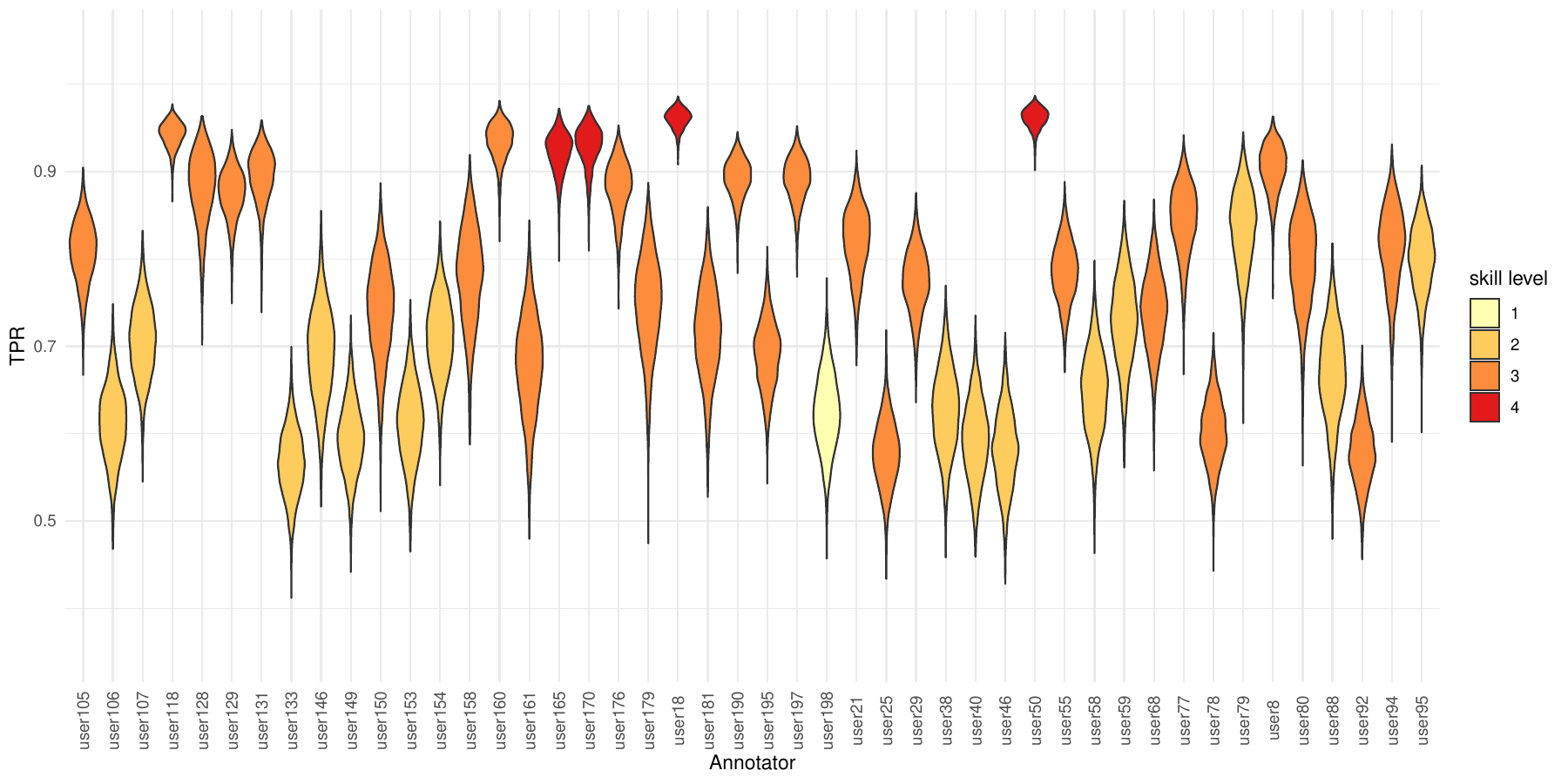}
        \caption{The posterior distributions of annotators' TPRs in \textbf{DP-BMM-Hierarchical}.}
    \end{subfigure}
    \caption{The posterior distributions of annotators' TPRs in two models that take into account correlations among bird species. Levels $1, 2, 3, 4$ represent the different levels of annotators' bird song identification expertise. Level $4$ corresponds to the highest skill level, followed by decreasing levels of 3, 2, and 1 in that order.}
    \label{fig:app_ability_dpbmm}
\end{figure}

In our problem, annotators are more likely to omit species that are actually present in the recordings than to falsely identify species that are not vocalizing in recordings. Consequently, given that our annotators exhibit low FPRs, our primary focus is to assess their bird song identification expertise based on their TPRs $\lambda_{j}$'s. The posterior distributions of each annotator's $\psi_j$ are provided in Appendix \ref{suppsec:additional_app}. Our Markov chains appear to have converged given that the posterior distributions obtained from three chains with widely different starting points are quite similar for each model. We visualize the posterior distribution of each annotator's $\lambda_{j}$ in Figures \ref{fig:app_ability_base} and \ref{fig:app_ability_dpbmm}. In \textbf{Base-Hierarchical} and \textbf{DP-BMM-Hierarchical}, we do not directly model annotators' overall TPRs or FPRs. Instead, we model their species-specific TPRs $\sigma(\lambda_{j, k})$'s and FPRs $\sigma(\psi_{j, k})$'s. However, here we use $\sigma(\lambda_j)$ and $\sigma(\psi_j)$ to roughly represent their overall TPRs and FPRs.

While we do not have access to the ground truth of annotators' identification expertise, we can utilize their self-reported Finnish bird song identification levels as a reference to assess our approach. To analyze the posterior distributions of annotators' TPRs and FPRs, we used distinct colors to represent different skill levels and create violin plots. Figures \ref{fig:app_ability_base} and \ref{fig:app_ability_dpbmm} reveal that annotators with higher levels generally exhibit higher estimated TPRs across all four models. This alignment with annotators' self-assessments of their expertise is reassuring. Comparing Figures \ref{fig:app_ability_base} and \ref{fig:app_ability_dpbmm}, we observed that the posterior distributions of different annotators' TPRs in \textbf{Base} and \textbf{DP-BMM} are quite similar, while the posterior distributions of annotators' TPRs from \textbf{Base-Hierarchical} and \textbf{DP-BMM-Hierarchical} are also similar to each other. Accurately characterizing species correlation seems to have little impact on inferences on annotators' identification expertise for these data. 

Comparing two subplots in Figure \ref{fig:app_ability_base} or \ref{fig:app_ability_dpbmm}, we notice that the estimated TPRs in \textbf{Base-Hierarchical}/\textbf{DP-BMM-Hierarchical} exhibit more variability compared with those in \textbf{Base}/\textbf{DP-BMM}. Specifically, some annotators with high TPRs in \textbf{Base}/\textbf{DP-BMM} would get significantly lower TPRs in \textbf{Base-Hierarchical}/\textbf{DP-BMM-Hierarchical}. Upon closer examination, we find annotators who are knowledgeable about many species and have provided limited annotations tend to have lower TPRs in the models with the hierarchical structure of identification expertise. In the models without the hierarchical structure, the posterior distributions of their TPRs are dominated by informative priors because they contributed fewer annotations compared with others. However, for the models with the hierarchical structure, we noticed the estimated $\phi_{\lambda}^{*}$ is considerably larger than $\phi_{\lambda}$, indicating that species-specific TPRs are primarily influenced by the data likelihood. Consequently, annotators who provided a higher proportion of $0$ will have much lower species-specific TPRs or $\lambda_{j, k} \in (-\infty, +\infty)$ compared with other annotators. Furthermore, different from Section \ref{sec:simulation}, where we only estimate annotators' species-specific TPRs for species annotators gave annotations to, here we also consider the species-specific TPRs for species in their expertise sets mentioned in Section \ref{sec:varying_ability}. For annotators whose pre-specified sets contain extensive species, the presence of numerous relatively lower $\lambda_{j, k}$'s will lead to low $\lambda_j$'s, and limited annotations are not rich enough to reliably estimate their expertise in the models with the complex hierarchical structure.

\section{Discussion} \label{sec:discussion}
Our research has provided significant insights into the complexities and challenges inherent in combining annotations for bird songs, offering valuable contributions to citizen science-based avian acoustics. We explored Bayesian hierarchical models for species annotation, aiming to improve accuracy of aggregated annotations by addressing the variability in annotators' identification expertise and the dependence among species occurrences. Our main results show that i) annotation aggregation through Bayesian hierarchical models outperforms the widely used Majority Vote method and ii) the most flexible models are not necessarily the best when applied on very sparse datasets which are typical in citizen science projects.

One characteristic of our modeling framework is the consideration of annotators' varying expertise across different species. This variability is crucial as it mirrors real-world scenarios where annotators may excel in identifying certain species but struggle with others due to multiple species having similar vocalizations or the expert being specialized in a specific group of species. However, due to limited annotations, the results in Section \ref{sec:application} demonstrate that the additional modeling flexibility didn't lead to improved predictive performance or better fit to the data. Contrary to our initial expectations, the most complex model \textbf{DP-BMM-Hierarchical} did not consistently outperform simpler alternatives. Specifically, while \textbf{DP-BMM} exhibited superior performance in fitting the data, its predictive calibration suffered compared to \textbf{Base-Hierarchical}, which offered a more reliable estimation of uncertainty and better computational efficiency. This discrepancy highlights the need to balance model complexity with practical utility in biomonitoring applications.

Moving forward, our research opens up several promising avenues for further exploration and application:
\\
$\triangleright \ \textbf{Incorporating more information and structure:}$ One promising direction is to enrich our models by incorporating additional information and structure. This could involve integrating data from other sources, such as environmental variables or bird migration patterns, to further enhance the accuracy of species annotation aggregation methods.
\\
$\triangleright \ \textbf{Scaling up to larger datasets:}$ While our current research is based on a specific dataset, expanding our models to handle larger datasets covering a wider geographic region is a natural progression. This scalability will be crucial for real-world applications, such as monitoring bird populations on a global scale.
\\
$\triangleright \ \textbf{Combining modeling with active learning:}$ An innovative approach is to integrate our models into a website or mobile app for bird enthusiasts. By collecting new data through active learning, we can automatically identify and prioritize challenging cases for annotation based on our model results. This dynamic feedback loop will not only improve the accuracy of our annotations but also engage citizen scientists in species identification research.
\\ 
$\triangleright \ \textbf{Applications to other types of ecological data:}$ Although we focus in this article on bird sound annotations, the same models could generally be applied to any kind of multi-label crowd sourcing scenario. For example eBird, iNaturalist and Zooniverse are examples of global platforms collecting both audio and image observations of birds and other animals or different environmental subjects, which are annotated and confirmed by other users. The data quality questions can be a major challenge in citizen science projects, but when addressed correctly, citizen science projects can produce reliable data with high scientific importance \citep{munson2010method, bonney2014next, balazs2021data}. By improving the aggregation of partially disagreeing annotations, our method can thus help to produce meaningful ecological data in large scale.

In conclusion, our study has advanced the state-of-the-art in bird species annotation aggregation by leveraging Bayesian hierarchical models and addressing the complexities of annotator expertise and species dependencies. We provided a robust framework for improving the accuracy of citizen science-based bird sound identification, with implications for ecological studies and conservation efforts. Looking ahead, our ongoing efforts will focus on enhancing model robustness, expanding applicability to diverse datasets, and integrating real-time learning strategies to empower stakeholders in avian research.

\section*{Authors' contribution}
Haoxuan Wang, Patrik Lauha, and David Dunson jointly designed the method.
Haoxuan Wang implemented the methodology and analyses and took the lead in writing the paper.
Patrik Lauha recognized the need for the proposed methodology for addressing measurement errors in species classification, and gave substantial feedback on the methods and results. 
David Dunson took the lead in conceptualizing the modeling framework and providing feedback on initial results and the writing.
All authors contributed critically to the drafts and approved the final version for publication.

\section*{Acknowledgements} \label{sec:acknowledgements}
We would like to thank expert ornithologist Sebastian Andrejeff for annotating the test dataset of Finnish birds. This project has received funding from the European Research Council (ERC) under the European Union’s Horizon 2020 research and innovation programme through the LIFEPLAN project (grant agreement No 856506). This work was also supported by the National Science Foundation under Grant No. IIS-2426762.

\section*{Conflict of interest}
The authors declare no conflict of interest.

\section*{Data availability}
 Data and code to reproduce the tables and figures are also available at \href{https://github.com/Master-Savitar/Bayes-Species-Identification}{Github repository}.

\clearpage
% Bibliography
\bibliographystyle{plainnat}
\bibliography{references}

\clearpage
\appendix
\section*{Supplementary Material}

%%%%% Section A1 %%%%%
\section{Posterior computation} \label{suppsec:posterior}
For simplicity, in this section, we assume that there are no missing values in our identification data $\mathcal{T} \in \left\{0, 1\right\}^{N_1 \times N_2 \times N_3}$, as we can conveniently handle any missing values in R by setting \texttt{na.rm = TRUE}.
\subsection{Posterior inference for Base Model}
\subsubsection{A list of model parameters to be sampled}
In \textbf{Base}, model parameters to be sampled include:
\begin{itemize}
    \item $\bm{o} = (o_1, o_2, \dots, o_{N_3})^T$, where $o_k$ is the occurrence probability of bird species $k$;
    \item $\bm{\lambda} = (\lambda_1, \lambda_2, \dots, \lambda_{N_2})^T$, where $\lambda_j$ is the True Positive Rate (TPR) of annotator $j$;
    \item $\bm{\psi} = (\psi_1, \psi_2, \dots, \psi_{N_2})^T$,
    where $\psi_j$ is the False Positive Rate (FPR) of annotator $j$;
    \item $\bm{Y} = \left(\bm{y}_1, \bm{y}_2, \dots, \bm{y}_{N_1}\right)$, where $\bm{y}_i = \left(y_{i, 1}, y_{i, 2}, \dots, y_{i, N_3}\right)^T$ is the latent binary variable for audio recording $i$.
\end{itemize}

\subsubsection{The posterior distribution}
The posterior distribution of model parameters to be updated is as follows when assuming no missing values in $\mathcal{T}$:
\begin{equation}
    \begin{aligned}
        & p\left(\bm{o}, \bm{\lambda}, \bm{\psi}, \bm{Y} \mid \mathcal{T}; a_o, b_o, a_{\lambda}, b_{\lambda}, a_{\psi}, b_{\psi} \right) \\
       \propto & \prod_{k = 1}^{N_3} p(o_k \mid a_o, b_o) \times \left\{\prod_{j = 1}^{N_2} p(\lambda_j \mid a_{\lambda}, b_{\lambda}) p(\psi_j \mid a_{\psi}, b_{\psi}) \right\} \\
       & \ \ \ \times \left\{\prod_{k = 1}^{N_3} \prod_{i = 1}^{N_1} p(y_{i, k} \mid o_k) \right\} \times \left\{\prod_{k = 1}^{N_3} \prod_{i = 1}^{N_1} \prod_{j = 1}^{N_2} p(\mathcal{T}_{i, j, k} \mid y_{i, k}, \lambda_j, \psi_j)\right\}.
    \end{aligned}
\end{equation}

\subsubsection{The sampling algorithm} \label{sec:base_algorithm}
$\triangleright \ \textbf{Sampling } \bm{o}:$
\\
\\
The full conditional of $\bm{o}$ is:
\begin{equation}
    \begin{aligned}
    & p\left(\bm{o} \mid \bm{\lambda}, \bm{\psi}, \bm{Y}, \mathcal{T}; a_o, b_o, a_{\lambda}, b_{\lambda}, a_{\psi}, b_{\psi} \right) \\
    \propto \ & \prod_{k = 1}^{N_3} p(o_k \mid a_o, b_o) \times \left\{\prod_{k = 1}^{N_3} \prod_{i = 1}^{N_1} p(y_{i, k} \mid o_k) \right\}.
    \end{aligned}
\end{equation}
The full conditional of $o_{k}$ is:
\begin{equation}
    \begin{aligned}
        & p\left(o_k \mid \bm{o}_{-k}, \bm{\lambda}, \bm{\psi}, \bm{Y}, \mathcal{T}; a_o, b_o, a_{\lambda}, b_{\lambda}, a_{\psi}, b_{\psi} \right) \\
        \propto \ & p(o_k \mid a_o, b_o) \prod_{i = 1}^{N_1} p(y_{i, k} \mid o_k) \\
        \propto \ & o_k^{a_o - 1} (1 - o_k)^{b_o - 1} \prod_{i = 1}^{N_1} o_k^{y_{i, k}} (1 - o_k)^{1 - y_{i, k}} \\
        \propto \ & o_k^{a_o + \sum_{i = 1}^{N_1} y_{i, k} - 1} (1 - o_k)^{b_o + \sum_{i = 1}^{N_1}(1 - y_{i, k}) - 1},
    \end{aligned}
\end{equation}
where $\bm{o}_{-k} = (o_1, \dots, o_{k-1}, o_{k+1}, \dots, o_{N_3})^T$. Therefore, we sampled each $o_k$, $k = 1, 2, \dots, N_3$ from its full conditional:
\begin{equation}
    o_k \mid \bm{o}_{-k}, \bm{\lambda}, \bm{\psi}, \bm{Y}, \mathcal{T}; a_o, b_o, a_{\lambda}, b_{\lambda}, a_{\psi}, b_{\psi} \sim \text{Beta}\left(a_o + \sum_{i = 1}^{N_1} y_{i, k}, b_o + \sum_{i = 1}^{N_1}(1 - y_{i, k})\right).
    \label{eq:base_gibbs_p}
\end{equation}
\\
$\triangleright \ \textbf{Sampling } \bm{\lambda}:$
\\
\\
The full conditional of $\bm{\lambda}$ is:
\begin{equation}
    \begin{aligned}
        & p(\bm{\lambda} \mid \bm{o}, \bm{\psi}, \bm{Y}, \mathcal{T}; a_o, b_o, a_{\lambda}, b_{\lambda}, a_{\psi}, b_{\psi}) \\
        \propto \ & \left\{\prod_{j = 1}^{N_2} p(\lambda_j \mid a_{\lambda}, b_{\lambda}) \right\} \times \left\{\prod_{k = 1}^{N_3} \prod_{i = 1}^{N_1} \prod_{j = 1}^{N_2} p(\mathcal{T}_{i, j, k} \mid y_{i, k}, \lambda_j, \psi_j)\right\}.
    \end{aligned}
\end{equation}
The full conditional of $\lambda_j$ is:
\begin{equation}
    \begin{aligned}
    & p(\lambda_j \mid \bm{\lambda}_{-j}, \bm{o}, \bm{\psi}, \bm{Y}, \mathcal{T}; a_o, b_o, a_{\lambda}, b_{\lambda}, a_{\psi}, b_{\psi}) \\
    \propto \ & p(\lambda_j \mid a_{\lambda}, b_{\lambda}) \times \left\{\prod_{k = 1}^{N_3} \prod_{i = 1}^{N_1} p(\mathcal{T}_{i, j, k} \mid y_{i, k}, \lambda_j, \psi_j)\right\} \\
    \propto \ & \lambda_j^{a_{\lambda} - 1} (1 - \lambda_j)^{b_{\lambda} - 1} \times \left\{\prod_{k = 1}^{N_3} \prod_{i = 1}^{N_1} \lambda_j^{y_{i, k} \mathcal{T}_{i, j, k}} (1 - \lambda_j)^{y_{i, k} (1 - \mathcal{T}_{i, j, k})}\right\} \\
    \propto \ & \lambda_j^{a_{\lambda} + \sum_{k = 1}^{N_3} \sum_{i = 1}^{N_1} y_{i, k} \mathcal{T}_{i, j, k} - 1} (1 - \lambda_j)^{b_{\lambda} + \sum_{k = 1}^{N_3} \sum_{i = 1}^{N_1} y_{i, k} (1 - \mathcal{T}_{i, j, k}) - 1},
\end{aligned}
\end{equation}
where $\bm{\lambda}_{-j} = (\lambda_1, \dots, \lambda_{j-1}, \lambda_{j+1}, \lambda_{N_2})^T$. Therefore, we sampled each $\lambda_j$, $j = 1, 2, \dots, N_2$ from its full conditional:
\begin{equation}
    \lambda_j \mid \bm{\lambda}_{-j}, \bm{o}, \bm{\psi}, \bm{Y}, \mathcal{T}; a_o, b_o, a_{\lambda}, b_{\lambda}, a_{\psi}, b_{\psi} \sim \text{Beta}\left(a_{\lambda} + \sum_{k = 1}^{N_3} \sum_{i = 1}^{N_1} y_{i, k} \mathcal{T}_{i, j, k}, b_{\lambda} + \sum_{k = 1}^{N_3} \sum_{i = 1}^{N_1} y_{i, k} (1 - \mathcal{T}_{i, j, k}) \right).
    \label{eq:base_gibbs_lambda}
\end{equation}
\\
$\triangleright \ \textbf{Sampling } \bm{\psi}:$
\\
\\
The full conditional of $\bm{\psi}$ is:
\begin{equation}
    \begin{aligned}
        & p(\bm{\psi} \mid \bm{o}, \bm{\lambda}, \bm{Y}, \mathcal{T}; a_o, b_o, a_{\lambda}, b_{\lambda}, a_{\psi}, b_{\psi}) \\
        \propto \ & \left\{\prod_{j = 1}^{N_2} p(\psi_j \mid a_{\psi}, b_{\psi}) \right\} \times \left\{\prod_{k = 1}^{N_3} \prod_{i = 1}^{N_1} \prod_{j = 1}^{N_2} p(\mathcal{T}_{i, j, k} \mid y_{i, k}, \lambda_j, \psi_j)\right\}.
    \end{aligned}
\end{equation}
The full conditional of $\psi_j$ is:
\begin{equation}
    \begin{aligned}
    & p(\psi_j \mid \bm{\psi}_{-j}, \bm{o}, \bm{\lambda}, \bm{Y}, \mathcal{T}; a_o, b_o, a_{\lambda}, b_{\lambda}, a_{\psi}, b_{\psi}) \\
    \propto \ & p(\psi_j \mid a_{\psi}, b_{\psi}) \times \left\{\prod_{k = 1}^{N_3} \prod_{i = 1}^{N_1} p(\mathcal{T}_{i, j, k} \mid y_{i, k}, \lambda_j, \psi_j)\right\} \\
    \propto \ & \psi_j^{a_{\psi} - 1} (1 - \psi_j)^{b_{\psi} - 1} \times \left\{\prod_{k = 1}^{N_3} \prod_{i = 1}^{N_1} \psi_j^{(1 - y_{i, k}) \mathcal{T}_{i, j, k}} (1 - \psi_j)^{(1 - y_{i, k}) (1 - \mathcal{T}_{i, j, k})}\right\} \\
    \propto \ & \psi_j^{a_{\psi} + \sum_{k = 1}^{N_3} \sum_{i = 1}^{N_1} (1 - y_{i, k}) \mathcal{T}_{i, j, k} - 1} (1 - \psi_j)^{b_{\psi} + \sum_{k = 1}^{N_3} \sum_{i = 1}^{N_1} (1 - y_{i, k}) (1 - \mathcal{T}_{i, j, k}) - 1},
\end{aligned}
\end{equation}
where $\bm{\psi}_{-j} = (\psi_1, \dots, \psi_{j-1}, \psi_{j+1}, \dots, \psi_{N_2})^T$. Therefore, we sampled each $\lambda_j$, $j = 1, 2, \dots, N_2$ from its full conditional:
\begin{equation}
\begin{aligned}
    & \psi_j \mid \bm{\psi}_{-j}, \bm{o}, \bm{\psi}, \bm{Y}, \mathcal{T}; a_o, b_o, a_{\psi}, b_{\lambda}, a_{\psi}, b_{\psi} \\ 
    & \sim \text{Beta}\left(a_{\psi} + \sum_{k = 1}^{N_3} \sum_{i = 1}^{N_1} (1 - y_{i, k}) \mathcal{T}_{i, j, k}, b_{\psi} + \sum_{k = 1}^{N_3} \sum_{i = 1}^{N_1} (1 - y_{i, k}) (1 - \mathcal{T}_{i, j, k}) \right).
\end{aligned}
    \label{eq:base_gibbs_psi}
\end{equation}
\\
$\triangleright \ \textbf{Sampling } \bm{Y}:$
\\
\\
The full conditional of $\bm{Y}$ is:
\begin{equation}
    \begin{aligned}
        & p(\bm{Y} \mid \bm{o}, \bm{\lambda}, \bm{\psi}, \mathcal{T}; a_o, b_o, a_{\lambda}, b_{\lambda}, a_{\psi}, b_{\psi}) \\
        \propto \ & \left\{\prod_{k = 1}^{N_3} \prod_{i = 1}^{N_1} p(y_{i, k} \mid o_k) \right\} \times \left\{\prod_{k = 1}^{N_3} \prod_{i = 1}^{N_1} \prod_{j = 1}^{N_2} p(\mathcal{T}_{i, j, k} \mid y_{i, k}, \lambda_j, \psi_j)\right\}.
    \end{aligned}
\end{equation}
The full conditional of $y_{i, k}$ is:
\begin{equation}
    \begin{aligned}
        & p(y_{i, k} \mid \bm{Y}_{-(i, k)}, \bm{o}, \bm{\lambda}, \bm{\psi}, \mathcal{T}; a_o, b_o, a_{\lambda}, b_{\lambda}, a_{\psi}, b_{\psi}) \\
        \propto \ & p(y_{i, k} \mid o_k) \times \left\{\prod_{j = 1}^{N_2} p(\mathcal{T}_{i, j, k} \mid y_{i, k}, \lambda_j, \psi_j)\right\} \\
        \propto \ & o_k^{y_{i, k}} (1 - o_k)^{1 - y_{i, k}} \times \prod_{j = 1}^{N_2} \left\{\left(\lambda_j^{y_{i, k}} \psi_j^{1 - y_{i, k}}\right)^{\mathcal{T}_{i, j, k}} \left( (1 - \lambda_j)^{y_{i, k}} (1 - \psi_j)^{1 - y_{i, k}} \right)^{1 - \mathcal{T}_{i, j, k}}\right\} \\
        \propto \ & \left[o_k \times \prod_{j=1}^{N_2} \lambda_j^{\mathcal{T}_{i, j, k}} (1 - \lambda_j)^{1 - \mathcal{T}_{i, j, k}} \right]^{y_{i, k}} \left[(1 - o_k) \times \prod_{j=1}^{N_2} \psi_j^{\mathcal{T}_{i, j, k}} (1 - \psi_j)^{1 - \mathcal{T}_{i, j, k}}\right]^{1 - y_{i, k}} \\
        \propto \ & \left(\hat{o}_{i, k}\right)^{y_{i, k}} \left(1 - \hat{o}_{i, k}\right)^{1 - y_{i, k}},
    \end{aligned}
\end{equation}
where 
$$\hat{o}_{i, k} = \frac{o_k \times \prod_{j=1}^{N_2} \lambda_j^{\mathcal{T}_{i, j, k}} (1 - \lambda_j)^{1 - \mathcal{T}_{i, j, k}}}{o_k \times \prod_{j=1}^{N_2} \lambda_j^{\mathcal{T}_{i, j, k}} (1 - \lambda_j)^{1 - \mathcal{T}_{i, j, k}} + (1 - o_k) \times \prod_{j=1}^{N_2} \psi_j^{\mathcal{T}_{i, j, k}} (1 - \psi_j)^{1 - \mathcal{T}_{i, j, k}}}
$$ is the parameter of $y_{i,k}$'s full conditional distribution, and $\bm{Y}_{-(i, k)}$ contains all elements in $\bm{Y}$ except $y_{i, k}$. Therefore, we sampled each $y_{i, k}$, $i = 1, 2, \dots, N_1$, $k = 1, 2, \dots, N_3$ from its full conditional:
\begin{equation}
    \begin{aligned}
        y_{i, k} \mid \bm{Y}_{-(i, k)}, \bm{o}, \bm{\lambda}, \bm{\psi}, \mathcal{T}; a_o, b_o, a_{\lambda}, b_{\lambda}, a_{\psi}, b_{\psi} \sim \text{Bernoulli}(\hat{o}_{i, k}).
    \end{aligned}
\end{equation}

\subsection{Posterior inference for DP-BMM} \label{sec:dpbmm_posterior}
\subsubsection{A list of model parameters to be sampled}
In \textbf{DP-BMM} introduced in Section \ref{sec:BMM} and \ref{sec:DP}, model parameters to be sampled include:
\begin{itemize}
    \item $\gamma$ is the concentration parameter in the Dirichlet process (DP);
    \item $\bm{z} = (z_1, z_2, \dots, z_{N_1})^T$ is the collection of the $N_1$ audio recordings' assignments, where $z_i$ is the assignment of recording $i$;
    \item $\bm{\lambda} = (\lambda_1, \lambda_2, \dots, \lambda_{N_2})^T$, where $\lambda_j$ is the True Positive Rate (TPR) of annotator $j$;
    \item $\bm{\psi} = (\psi_1, \psi_2, \dots, \psi_{N_2})^T$,
    where $\psi_j$ is the False Positive Rate (FPR) of annotator $j$;
    \item $\bm{Y} = \left(\bm{y}_1, \bm{y}_2, \dots, \bm{y}_{N_1}\right)$, where $\bm{y}_i = \left(y_{i, 1}, y_{i, 2}, \dots, y_{i, N_3}\right)^T$ is the latent binary variable for audio recording $i$.
\end{itemize}

\subsubsection{The posterior distribution}
As we have mentioned in Section \ref{sec:models}, we integrated out mixing coefficients $\bm{\pi} = \left\{\pi_r\right\}_{r = 1}^{\infty}$. Additionally, we integrated out parameters $\bm{O} = \left\{\bm{o}_r\right\}_{r = 1}^{\infty}$. Here $\bm{o}_r = (o_{r, 1}, o_{r, 2}, \dots, o_{r, N_3})^T$ represents the occurrence probabilities of the $N_3$ bird species within the $r^{\text{th}}$ mixture component, and only sample the remaining model parameters listed above. This technique not only simplifies the model but also results in faster convergence of the Markov chain to its stationary distribution in most cases \citep{blei2006variational}.

Then, the posterior distribution is as follows:
\begin{equation}
    \begin{aligned}
        & p(\gamma, \bm{z}, \bm{Y}, \bm{\lambda}, \bm{\psi} \mid \mathcal{T}; u_1, u_2, a_o, b_o, a_{\lambda}, b_{\lambda}, a_{\psi}, b_{\psi}) \\
        = \ & \int_{\mathcal{O}} p(\bm{O}, \gamma, \bm{z}, \bm{Y}, \bm{\lambda}, \bm{\psi} \mid \mathcal{T}; u_1, u_2, a_o, b_o, a_{\lambda}, b_{\lambda}, a_{\psi}, b_{\psi}) d \bm{O} \\
        \propto \ & \int_{\mathcal{O}} p(\bm{O} \mid a_o, b_o) p(\gamma \mid u_1, u_2) p(\bm{z} \mid \gamma) p(\bm{Y} \mid \bm{O}, \bm{z}) \\
        & \ \ \ \times p(\bm{\lambda} \mid a_{\lambda}, b_{\lambda}) p(\bm{\psi} \mid a_{\psi}, b_{\psi}) p(\mathcal{T} \mid \bm{Y}, \bm{\lambda}, \bm{\psi}) d\bm{O} \\
        = \ & \left(\int_{\mathcal{O}} p(\bm{O} \mid a_o, b_o) p(\bm{Y} \mid \bm{O}, \bm{z}) d\bm{O} \right) \\
        & \ \ \ \times p(\gamma \mid u_1, u_2) p(\bm{z} \mid \gamma) p(\bm{\lambda} \mid a_{\lambda}, b_{\lambda}) p(\bm{\psi} \mid a_{\psi}, b_{\psi}) p(\mathcal{T} \mid \bm{Y}, \bm{\lambda}, \bm{\psi}) \\
        = \ & \left(\int_{\mathcal{O}} p(\bm{O} \mid a_o, b_o) p(\bm{Y} \mid \bm{O}, \bm{z}) d\bm{O} \right) \times p(\gamma \mid u_1, u_2) \times p(\bm{z} \mid \gamma) \\
        & \ \ \ \times \left\{\prod_{j = 1}^{N_2} p(\lambda_j \mid a_{\lambda}, b_{\lambda}) p(\psi_j \mid a_{\psi}, b_{\psi}) \right\} \times \left\{\prod_{k = 1}^{N_3} \prod_{i = 1}^{N_1} \prod_{j = 1}^{N_2} p(\mathcal{T}_{i, j, k} \mid y_{i, k}, \lambda_j, \psi_j)\right\}.
    \end{aligned}
    \label{eq:dpbmm_posterior}
\end{equation}
As we can see from Equation \eqref{eq:dpbmm_posterior}, we do not give the explicit form of the integration, which we will address while sampling $\bm{z} = (z_1, z_2, \dots, z_{N_1})^T$ and $\bm{Y} = (\bm{y}_1, \bm{y}_2, \dots, \bm{y}_{N_1})$.

\subsubsection{The sampling algorithm} \label{sec:dpbmm_algorithm}
$\triangleright \ \textbf{Sampling } \gamma:$
\\
\\
The full conditional of $\gamma$ is:
\begin{equation}
    \begin{aligned}
        & p(\gamma \mid \bm{z}, \bm{Y}, \bm{\lambda}, \bm{\psi}, \mathcal{T}; u_1, u_2, a_o, b_o, a_{\lambda},b_{\lambda}, a_{\psi}, b_{\psi}) \\
        \propto \ & \underbrace{p(\bm{z} \mid \gamma)}_{\text{the first term}} \times \underbrace{p(\gamma \mid u_1, u_2)}_{\text{the second term}}.
    \end{aligned}
    \label{eq:dpbmm_gamma}
\end{equation}
As seen in Equation \eqref{eq:dpbmm_gamma}, the primary challenge in deriving the full conditional of $\gamma$ lies in obtaining the explicit form of the first term.

We have formally defined the DP in Equation \eqref{sec:DP}, but there's still the matter of constructing and representing it. One commonly used constructive definition for the DP is the stick-breaking process \citep{sethuraman1994constructive}. For more comprehensive insights, please refer to \cite{murphy2012machine, gelman2013bayesian}. 

However, it's problematic to deal with a countably infinite number of sticks in practice. Consequently, we will delve into an alternative approach, namely the Chinese restaurant process (CRP) \citep{aldous1985exchangeability, griffiths2003hierarchical}. The CRP offers us an effective way to construct a DP.

Assuming we have the assignments of the $N_1$ audio recordings $\bm{z} = (z_1, z_2, \dots, z_{N_1})^T$, which correspond to $R$ distinct Bernoulli mixture components. As presented in Equation \eqref{eq:dp_bmm_hierary}, if $\theta_i \mid G \sim G$ represents $N_1$ observations from $G \sim \text{DP}(\gamma, H)$, generated by $R$ distinct parameters $\bm{o}_r$, then by the definition of the DP, the predictive distribution can be expressed as:
\begin{equation}
    \begin{aligned}
        p(\theta_{n + 1} = \bm{o} \mid \theta_1, \dots, \theta_n, \gamma, H) = \frac{\gamma}{\gamma + n} H(\bm{o}) + \frac{1}{\gamma + n} \sum_{r = 1}^{R} n_r \delta_{\bm{o}_r}(\bm{o}),
    \end{aligned}
\end{equation}
where $n_r$ is the number of audio recordings generated by the parameter $\bm{o}_r$ or assigned to the $r^{\text{th}}$ mixture component. This predictive distribution is also known as Blackwell-MacQueen sampling scheme \citep{blackwell1973ferguson}. Utilizing it, we could easily obtain:
\begin{equation}
    \begin{aligned}
        p(z_{n + 1} = \tilde{r} \mid z_1, \dots, z_n, \gamma) &= \frac{\gamma}{\gamma + n} \mathbb{I}\left(\tilde{r} = r^{*}\right) + \frac{1}{\gamma + n} \sum_{r = 1}^{R} n_r \mathbb{I}\left(\tilde{r} = r\right) \\
        &= \left\{
\begin{aligned}
    & \frac{n_{\tilde{r}}}{\gamma + n}, \ \ \ \text{if} \ \tilde{r} \ \text{exists}, \\ 
    & \frac{\gamma}{\gamma + n}, \ \ \ \text{if} \ \tilde{r} \ \text{is new}, \\
\end{aligned}
\right.
    \end{aligned}
    \label{eq:crp}
\end{equation}
where $\mathbb{I}(\cdot)$ represents the indicator function, and $r^{*}$ denotes a new Bernoulli component that is outside the existing $R$ mixture components. This predictive distribution, expressed in terms of assignments $\bm{z} = (z_1, z_2, \dots, z_{N_1})^T$, is the Chinese restaurant process. It draws an analogy to the seemingly infinite supply of tables at certain Chinese restaurants \citep{murphy2012machine}.

Based on the predictive distribution presented in Equation \eqref{eq:crp}, we obtain the explicit form of the first term in Equation \eqref{eq:dpbmm_posterior} as follows:
\begin{equation}
    \begin{aligned}
        p(\bm{z} \mid \gamma) &= \prod_{i = 1}^{N_1} p(z_{i} \mid \bm{z}_{1:i-1}, \gamma) \\
        &= \gamma^{R} \frac{\Gamma(\gamma)}{\Gamma(\gamma + n)} \prod_{r = 1}^{R} (n_r - 1)! \\
        &= \gamma^{R} \frac{\prod_{r = 1}^{R} (n_r - 1)!}{\prod_{i = 1}^{N_1} (i - 1 + \gamma)},
    \end{aligned}
    \label{eq:dpbmm_gamma_first}
\end{equation}
where $\bm{z}_{1:i-1} = (z_1, z_2, \dots, z_{i-1})^T$, and $\Gamma(\cdot)$ is the Gamma function. It is worth mentioning that this probability distribution over sequences of assignments of mixture components doesn't depend on the specific order. This further implies that the assignments $z_1, z_2, \dots, z_{N_1}$ is exchangeable by de Finetti's theorem.

Now we can put the first term and the second term in Equation \eqref{eq:dpbmm_gamma} together, and explicitly express the concentration parameter $\gamma$'s full conditional as follows:
\begin{equation}
    \begin{aligned}
        & p(\gamma \mid \bm{z}, \bm{Y}, \bm{\lambda}, \bm{\psi}, \mathcal{T}; u_1, u_2, a_o, b_o, a_{\lambda},b_{\lambda}, a_{\psi}, b_{\psi}) \\
        \propto \ & p(\bm{z} \mid \gamma) p(\gamma \mid u_1, u_2) \\
        \propto \ & \frac{\gamma^R}{\prod_{i = 1}^{N_1} (i - 1 + \gamma)} \times \gamma^{u_1 - 1} \exp(-u_2 \gamma) \\
        \propto \ & \frac{\gamma^{R + u_1 - 1}}{\prod_{i = 1}^{N_1} (i - 1 + \gamma)} \exp(-u_2 \gamma).
    \end{aligned}
    \label{eq:dpbmm_gamma_final}
\end{equation}

As direct sampling of $\gamma$ from its full conditional is not feasible, we opted for an adaptive Metropolis algorithm to update it. Adaptive Metropolis was firstly proposed in \cite{haario2001adaptive}. The covariance matrix of the proposal distribution has a fixed scaling parameters:
\begin{equation}
    s = \frac{2.38^2}{d},
\end{equation}
where $d$ is the the dimension of the parameter. 

Specifically, the proposal distribution for $\gamma$ is:
\begin{equation}
    q(\gamma, \gamma^{*}) = \mathcal{N}(\gamma, s_{\gamma}^2).
\end{equation}
In our context, where $d = 1$, the optimal acceptance rate is relatively higher, approximately around $0.44$. Consequently, in order to update $\gamma$ at each iteration, we adopted a more direct adaptive approach targeting the optimal acceptance rate of $0.44$. 

In practice, we can consider $50$ iterations as a batch, and increases or decreases $s_{\gamma}$ according to the proportion of accepted proposals within these $50$ iterations. To explore the space of suitable values more effectively, it's convenient to work within the logarithmic scale. As a result, if the proportion of accepted values for $\gamma$ exceeds $0.44$, we will increment $\log(s_{\gamma})$ by $\min\left\{0.01, \frac{1}{\sqrt{t}}\right\}$, where $t$ represents the number of iterations. Conversely, if the proportion of accepted values for $\gamma$ falls below $0.44$, we will decrement $\log(s_{\gamma})$ by $\min\left\{0.01, \frac{1}{\sqrt{t}}\right\}$.

Based on the full conditional as well as the proposal distribution, the acceptance rate for $\gamma$ is:
\begin{equation}
    \begin{aligned}
        \alpha(\gamma, \gamma^{*}) &= \min\left\{1 , \frac{p(\gamma^{*} \mid \bm{z}, \bm{Y}, \bm{\lambda}, \bm{\psi}, \mathcal{T}; u_1, u_2, a_o, b_o, a_{\lambda},b_{\lambda}, a_{\psi}, b_{\psi}) q(\gamma^{*}, \gamma)}{p(\gamma \mid \bm{z}, \bm{Y}, \bm{\lambda}, \bm{\psi}, \mathcal{T}; u_1, u_2, a_o, b_o, a_{\lambda},b_{\lambda}, a_{\psi}, b_{\psi}) q(\gamma, \gamma^{*})}\right\} \\
        &= \min\left\{1 , \frac{p(\gamma^{*} \mid \bm{z}, \bm{Y}, \bm{\lambda}, \bm{\psi}, \mathcal{T}; u_1, u_2, a_o, b_o, a_{\lambda},b_{\lambda}, a_{\psi}, b_{\psi})}{p(\gamma \mid \bm{z}, \bm{Y}, \bm{\lambda}, \bm{\psi}, \mathcal{T}; u_1, u_2, a_o, b_o, a_{\lambda},b_{\lambda}, a_{\psi}, b_{\psi})}\right\} \\
        &= \min\left\{1, \left(\frac{\gamma^{*}}{\gamma}\right)^{R + u_1 - 1} \times \exp\left(-u_2(\gamma^{*} - \gamma)\right) \times \prod_{i = 1}^{N_1} \left(\frac{i - 1 + \gamma}{i - 1 + \gamma^{*}}\right) \right\}.
    \end{aligned}
\end{equation}

It should be noted that in practice, it's better calculate the factorial term as follow to avoid numerical issues:
\begin{equation}
    \begin{aligned}
        \prod_{i = 1}^{N_1} \left(\frac{i - 1 + \gamma}{i - 1 + \gamma^{*}}\right) = \exp\left\{\sum_{i = 1}^{N_1} \log(i - 1 + \gamma) - \sum_{i = 1}^{N_1} \log(i - 1 + \gamma^{*})\right\}.
    \end{aligned}
\end{equation}
\\
$\triangleright \ \textbf{Sampling } \bm{z}:$
\\
\\
In this part, we demonstrate the sequential update process for each $z_i$, $ i = 1, 2, \dots, N_1$. 

The full conditional of $z_i$ is:
\begin{equation}
    \begin{aligned}
        & p(z_i = \tilde{r} \mid \bm{z}_{-i}, \gamma, \bm{Y}, \bm{\lambda}, \bm{\psi}, \mathcal{T}; u_1, u_2, a_o, b_o, a_{\lambda},b_{\lambda}, a_{\psi}, b_{\psi}) \\
        \propto \ & p(z_i = \tilde{r}, \bm{z}_{-i} \mid \gamma) \left(\int_{\mathcal{O}} p(\bm{O} \mid a_o, b_o) p(\bm{Y} \mid \bm{O}, z_i = \tilde{r}, \bm{z}_{-i}) d \bm{O} \right) \\
        \propto \ & \underbrace{p(z_i = \tilde{r} \mid \bm{z}_{-i}, \gamma)}_{\text{the first term}} \times \underbrace{\left(\int_{\mathcal{O}} p(\bm{O} \mid a_o, b_o) p(\bm{Y} \mid \bm{O}, z_i = \tilde{r}, \bm{z}_{-i}) d \bm{O} \right)}_{\text{the second term}}
    \end{aligned}
    \label{eq:dpbmm_z}
\end{equation}

For the first term in Equation \eqref{eq:dpbmm_z}, we can leverage the exchangeability of $\bm{z} = (z_1, z_2, \dots, z_{N_1})^T$ implied by Equation \eqref{eq:dpbmm_gamma_first}. This will lead to the following expression:
\begin{equation}
    \begin{aligned}
        p(z_i = \tilde{r} \mid \bm{z}_{-i}, \gamma) = 
        \left\{
        \begin{aligned}
            & \frac{n_{\tilde{r}, -i}}{\gamma + N_1 - 1}, \ \ \ \text{if} \ \tilde{r} \ \text{exists}, \\ 
            & \frac{\gamma}{\gamma + N_1 - 1}, \ \ \ \text{if} \ \tilde{r} \ \text{is new}, \\
        \end{aligned}
        \right.
    \end{aligned}
    \label{eq:dpbmm_z_first}
\end{equation}
where $n_{\tilde{r}, -i}$ represents the number of audio recordings assigned to the $\tilde{r}^{\text{th}}$ component when we exclude recording $i$ from consideration.

For the second term, $\bm{z}_{-i}$ and $\bm{Y}$ are given, since we want to derive the full conditional distribution of $z_i$. This leads to:
\begin{equation}
    \begin{aligned}
        & \int_{\mathcal{O}} p(\bm{O} \mid a_o, b_o) p(\bm{Y} \mid \bm{O}, z_i = \tilde{r}, \bm{z}_{-i}) d \bm{O} \\
        \propto \ & \int_{\mathcal{O}} p(\bm{O} \mid a_o, b_o) p(\bm{Y}_{-i} \mid \bm{O}, \bm{z}_{-i}) p(\bm{y}_i \mid \bm{O}, z_i = \tilde{r}) d \bm{O} \\
        \propto \ & \int_{\bm{o}_{\tilde{r}}} p(\bm{o}_{\tilde{r}} \mid a_o, b_o) \left\{\prod_{i^{'} \neq i, z_{i^{'}} = \tilde{r}} p(\bm{y}_{i^{'}} \mid \bm{o}_{\tilde{r}})\right\} p(\bm{y}_{i} \mid \bm{o}_{\tilde{r}}) d \bm{o}_{\tilde{r}} \\
        \propto \ & \underbrace{\int_{0}^{1} \dots \int_{0}^{1}}_{N_3} \left\{\prod_{k = 1}^{N_3} \text{Beta}(o_{\tilde{r}, k} \mid a_o + n^{(+)}_{\tilde{r}, k, -i}, b_o + n^{(-)}_{\tilde{r}, k, -i})\right\} \left\{\prod_{k = 1}^{N_3} \text{Bernoulli}(y_{i, k} \mid o_{\tilde{r}, k})\right\} do_{\tilde{r}, 1} \dots do_{\tilde{r}, N_3} \\
        \propto \ & \prod_{k = 1}^{N_3} \left(\int_{0}^{1} \text{Bernoulli}(y_{i, k} \mid o_{\tilde{r}, k}) \text{Beta}(o_{\tilde{r}, k} \mid a_o + n^{(+)}_{\tilde{r}, k, -i}, b_o + n^{(-)}_{\tilde{r}, k, -i}) d o_{\tilde{r}, k} \right) \\
        \propto \ & \prod_{k = 1}^{N_3} \left[\frac{B(a_o + n^{(+)}_{\tilde{r}, k, -i} + y_{i, k}, b_o + n^{(-)}_{\tilde{r}, k, -i} + (1 - y_{i, k}))}{B(a_o + n^{(+)}_{\tilde{r}, k, -i}, b_o + n^{(-)}_{\tilde{r}, k, -i})}\right],
    \end{aligned}
    \label{eq:dpbmm_z_second}
\end{equation}
where $\bm{Y}_{-i} = (\bm{y}_1, \dots, \bm{y}_{i - 1}, \bm{y}_{i + 1}, \dots, \bm{y}_{N_1})$, $n^{(+)}_{\tilde{r}, k, -i}$ represents the cardinality of the set $\{i^{'}: i^{'} \neq i, z_{i^{'}} = \tilde{r}, y_{i^{'}, k} = 1\}$ and $n^{(-)}_{\tilde{r}, k, -i}$ represents the cardinality of the set $\{i^{'}: i^{'} \neq i, z_{i^{'}} = \tilde{r}, y_{i^{'}, k} = 0\}$. Here $B(\cdot, \cdot)$ is defined as follows:
\begin{equation}
    B(\alpha, \beta)  = \frac{\Gamma(\alpha) \Gamma(\beta)}{\Gamma(\alpha + \beta)},
\end{equation}
where $\Gamma(\cdot)$ is the Gamma function we've used when deriving the full conditional of the concentration parameter $\gamma$.

Now, combining the first term in Equation \eqref{eq:dpbmm_z_first} with the second term in Equation \eqref{eq:dpbmm_z_second}, we obtain the explicit form of $z_i$'s full condition as follows:
\begin{equation}
    \begin{aligned}
        & p(z_i = \tilde{r} \mid \bm{z}_{-i}, \gamma, \bm{Y}, \bm{\lambda}, \bm{\psi}, \mathcal{T}; u_1, u_2, a_o, b_o, a_{\lambda},b_{\lambda}, a_{\psi}, b_{\psi}) \\
        \propto \ & p(z_i = \tilde{r}, \bm{z}_{-i} \mid \gamma) \left(\int_{\mathcal{O}} p(\bm{O} \mid a_o, b_o) p(\bm{Y} \mid \bm{O}, z_i = \tilde{r}, \bm{z}_{-i}) d \bm{O} \right) \\
        \propto \ & p(z_i = \tilde{r} \mid \bm{z}_{-i}, \gamma) \left(\int_{\mathcal{O}} p(\bm{O} \mid a_o, b_o) p(\bm{Y} \mid \bm{O}, z_i = \tilde{r}, \bm{z}_{-i}) d \bm{O} \right) \\
        \propto \ & 
        \left\{
        \begin{aligned}
            & \frac{n_{\tilde{r}, -i}}{\gamma + N_1 - 1} \times \left\{\prod_{k = 1}^{N_3} \frac{B(a_o + n^{(+)}_{\tilde{r}, k, -i} + y_{i, k}, b_o + n^{(-)}_{\tilde{r}, k, -i} + (1 - y_{i, k}))}{B(a_o + n^{(+)}_{\tilde{r}, k, -i}, b_o + n^{(-)}_{\tilde{r}, k, -i})}\right\}, \ \ \ \text{if} \ \tilde{r} \ \text{exists}, \\ 
            & \frac{\gamma}{\gamma + N_1 - 1} \times \left\{\prod_{k = 1}^{N_3} \frac{B(a_o + y_{i, k}, b_o + (1 - y_{i, k}))}{B(a_o, b_o)}\right\}, \ \ \ \text{if} \ \tilde{r} \ \text{is new}. \\
        \end{aligned}
        \right.
    \end{aligned}
    \label{eq:dpbmm_z_final}
\end{equation}
Following this, we proceeded to calculate the product in Equation \eqref{eq:dpbmm_z_final} for each $\tilde{r} \in \{1, 2, \dots, R\}$ as well as for the scenario where $\tilde{r}$ corresponds to a new mixture component. We stored these results in a vector and normalized them, denoted as $\bm{q}_i \in (0, 1)^{R+1}$. Consequently, we can sample each $z_i$, $i = 1, 2, \dots, N_1$ from its full conditional:
\begin{equation}
    z_i \mid \bm{z}_{-i}, \gamma, \bm{Y}, \bm{\lambda}, \bm{\psi}, \mathcal{T}; u_1, u_2, a_o, b_o, a_{\lambda},b_{\lambda}, a_{\psi}, b_{\psi} \sim \text{Categorical}(\bm{q}_i).
\end{equation}
\\
$\triangleright \ \textbf{Sampling } \bm{Y}:$
\\
\\
In this part, we show how to sample each $y_{i, k}$, $i = 1, 2, \dots, N_1$, $k = 1, 2, \dots, N_3$.

The full conditional of $y_{i, k}$ is:
\begin{equation}
    \begin{aligned}
        & p(y_{i, k} \mid \bm{Y}_{-(i, k)}, \gamma, \bm{z}, \bm{\lambda}, \bm{\psi}, \mathcal{T}; u_1, u_2, a_o, b_o, a_{\lambda}, b_{\lambda}, a_{\psi}, b_{\psi}) \\
        \propto \ & \left(\int_{\mathcal{O}} p(\bm{O} \mid a_o, b_o) p(\bm{Y} \mid \bm{O}, \bm{z}) d\bm{O} \right) \times \left\{\prod_{j = 1}^{N_2} p(\mathcal{T}_{i, j, k} \mid y_{i, k}, \lambda_j, \psi_j)\right\} \\
        \propto \ & \underbrace{B\left(a_o + n_{z_i, k, -i}^{(+)} + y_{i, k}, b_o + n_{z_i, k, -i}^{(-)} + (1 - y_{i, k})\right)}_{\text{the first term}} \\
        & \times \underbrace{\left\{\prod_{j = 1}^{N_2} \left[\lambda_{j}^{y_{i, k}} \psi_j^{1 - y_{i, k}}\right]^{\mathcal{T}_{i, j, k}} \left[(1 - \lambda_{j})^{y_{i, k}} (1 - \psi_j)^{1 - y_{i, k}}\right]^{1 - \mathcal{T}_{i, j, k}}\right\}}_{\text{the second term}}.
    \end{aligned}
    \label{eq:dpbmm_y}
\end{equation}
It should be noted that the first term can be directly obtained from Equation \eqref{eq:dpbmm_z_second}. Here, we only needed to remove the denominator and replace $\tilde{r}$ with $z_i$, as $\bm{z} = (z_1, z_2, \dots, z_{N_1})^T$ is given.

The next step is to calculate the product in Equation \eqref{eq:dpbmm_y} for both $y_{i, k} = 0$ and $y_{i, k} = 1$. Afterwards, we normalized them, and obtained the probability of $y_{i, k} = 1$, denoted as $\eta_{i, k}$, $i = 1, 2, \dots, N_1$, $k = 1, 2, \dots, N_3$. 

Finally, we sampled each $y_{i, k}$ from its full conditional:
\begin{equation}
    y_{i, k} \mid \bm{Y}_{-(i, k)}, \gamma, \bm{z}, \bm{\lambda}, \bm{\psi}, \mathcal{T}; u_1, u_2, a_o, b_o, a_{\lambda}, b_{\lambda}, a_{\psi}, b_{\psi} \sim \text{Bernoulli}(\eta_{i, k}).
\end{equation}
\\
$\triangleright \ \textbf{Sampling } \bm{\lambda}:$
\\
The full conditional of $\bm{\lambda}$ is:
\begin{equation}
    \begin{aligned}
        & p(\bm{\lambda} \mid \gamma, \bm{z}, \bm{Y}, \bm{\psi}, \mathcal{T}; u_1, u_2, a_o, b_o, a_{\lambda}, b_{\lambda}, a_{\psi}, b_{\psi}) \\
        \propto \ & \left\{\prod_{j = 1}^{N_2} p(\lambda_j \mid a_{\lambda}, b_{\lambda}) \right\} \times \left\{\prod_{k = 1}^{N_3} \prod_{i = 1}^{N_1} \prod_{j = 1}^{N_2} p(\mathcal{T}_{i, j, k} \mid y_{i, k}, \lambda_j, \psi_j)\right\}.
    \end{aligned}
\end{equation}

The full conditional of $\lambda_j$ is:
\begin{equation}
    \begin{aligned}
    & p(\lambda_j \mid \bm{\lambda}_{-j}, \gamma, \bm{z}, \bm{Y}, \bm{\psi}, \mathcal{T}; u_1, u_2, a_o, b_o, a_{\lambda}, b_{\lambda}, a_{\psi}, b_{\psi}) \\
    \propto \ & p(\lambda_j \mid a_{\lambda}, b_{\lambda}) \times \left\{\prod_{k = 1}^{N_3} \prod_{i = 1}^{N_1} p(\mathcal{T}_{i, j, k} \mid y_{i, k}, \lambda_j, \psi_j)\right\} \\
    \propto \ & \lambda_j^{a_{\lambda} - 1} (1 - \lambda_j)^{b_{\lambda} - 1} \times \left\{\prod_{k = 1}^{N_3} \prod_{i = 1}^{N_1} \lambda_j^{y_{i, k} \mathcal{T}_{i, j, k}} (1 - \lambda_j)^{y_{i, k} (1 - \mathcal{T}_{i, j, k})}\right\} \\
    \propto \ & \lambda_j^{a_{\lambda} + \sum_{k = 1}^{N_3} \sum_{i = 1}^{N_1} y_{i, k} \mathcal{T}_{i, j, k} - 1} (1 - \lambda_j)^{b_{\lambda} + \sum_{k = 1}^{N_3} \sum_{i = 1}^{N_1} y_{i, k} (1 - \mathcal{T}_{i, j, k}) - 1},
\end{aligned}
\end{equation}
where $\bm{\lambda}_{-j} = (\lambda_1, \dots, \lambda_{j-1}, \lambda_{j+1}, \lambda_{N_2})^T$. Therefore, we sampled each $\lambda_j$, $j = 1, 2, \dots, N_2$ from its full conditional:
\begin{equation}
    \begin{aligned}
        & \lambda_j \mid \bm{\lambda}_{-j}, \gamma, \bm{z}, \bm{Y}, \bm{\psi}, \mathcal{T}; u_1, u_2, a_o, b_o, a_{\lambda}, b_{\lambda}, a_{\psi}, b_{\psi} \\
        & \sim \text{Beta}\left(a_{\lambda} + \sum_{k = 1}^{N_3} \sum_{i = 1}^{N_1} y_{i, k} \mathcal{T}_{i, j, k}, b_{\lambda} + \sum_{k = 1}^{N_3} \sum_{i = 1}^{N_1} y_{i, k} (1 - \mathcal{T}_{i, j, k}) \right),
    \end{aligned}
    \label{eq:dpbmm_lambda}
\end{equation}
which is the same as Equation \eqref{eq:base_gibbs_lambda}.
\\
\\
$\triangleright \ \textbf{Sampling } \bm{\psi}:$
\\
\\
The full conditional of $\bm{\psi}$ is:
\begin{equation}
    \begin{aligned}
        & p(\bm{\psi} \mid \gamma, \bm{z}, \bm{Y}, \bm{\lambda}, \mathcal{T}; u_1, u_2, a_o, b_o, a_{\lambda}, b_{\lambda}, a_{\psi}, b_{\psi}) \\
        \propto \ & \left\{\prod_{j = 1}^{N_2} p(\psi_j \mid a_{\psi}, b_{\psi}) \right\} \times \left\{\prod_{k = 1}^{N_3} \prod_{i = 1}^{N_1} \prod_{j = 1}^{N_2} p(\mathcal{T}_{i, j, k} \mid y_{i, k}, \lambda_j, \psi_j)\right\}.
    \end{aligned}
\end{equation}
The full conditional of $\psi_j$ is:
\begin{equation}
    \begin{aligned}
    & p(\psi_j \mid \bm{\psi}_{-j}, \gamma, \bm{z}, \bm{Y}, \bm{\lambda}, \mathcal{T}; u_1, u_2, a_o, b_o, a_{\lambda}, b_{\lambda}, a_{\psi}, b_{\psi}) \\
    \propto \ & p(\psi_j \mid a_{\psi}, b_{\psi}) \times \left\{\prod_{k = 1}^{N_3} \prod_{i = 1}^{N_1} p(\mathcal{T}_{i, j, k} \mid y_{i, k}, \lambda_j, \psi_j)\right\} \\
    \propto \ & \psi_j^{a_{\psi} - 1} (1 - \psi_j)^{b_{\psi} - 1} \times \left\{\prod_{k = 1}^{N_3} \prod_{i = 1}^{N_1} \psi_j^{(1 - y_{i, k}) \mathcal{T}_{i, j, k}} (1 - \psi_j)^{(1 - y_{i, k}) (1 - \mathcal{T}_{i, j, k})}\right\} \\
    \propto \ & \psi_j^{a_{\psi} + \sum_{k = 1}^{N_3} \sum_{i = 1}^{N_1} (1 - y_{i, k}) \mathcal{T}_{i, j, k} - 1} (1 - \psi_j)^{b_{\psi} + \sum_{k = 1}^{N_3} \sum_{i = 1}^{N_1} (1 - y_{i, k}) (1 - \mathcal{T}_{i, j, k}) - 1},
\end{aligned}
\end{equation}
where $\bm{\psi}_{-j} = (\psi_1, \dots, \psi_{j-1}, \psi_{j+1}, \dots, \psi_{N_2})^T$. Therefore, we sampled each $\lambda_j$, $j = 1, 2, \dots, N_2$ from its full conditional:
\begin{equation}
    \begin{aligned}
        & \psi_j \mid \bm{\psi}_{-j}, \gamma, \bm{z}, \bm{Y}, \bm{\lambda}, \mathcal{T}; u_1, u_2, a_o, b_o, a_{\lambda}, b_{\lambda}, a_{\psi}, b_{\psi} \\ 
        & \sim \text{Beta}\left(a_{\psi} + \sum_{k = 1}^{N_3} \sum_{i = 1}^{N_1} (1 - y_{i, k}) \mathcal{T}_{i, j, k}, b_{\psi} + \sum_{k = 1}^{N_3} \sum_{i = 1}^{N_1} (1 - y_{i, k}) (1 - \mathcal{T}_{i, j, k}) \right),
    \end{aligned}
    \label{eq:dpbmm_psi}
\end{equation}
which is also the same as Equation \eqref{eq:base_gibbs_psi}.

\subsection{Posterior inference for DP-BMM-Hierarchical} \label{sec:posterior_dpbmm_h}
Please note that we will not provide a detailed derivation of the posterior computation for \textbf{Base-Hierarchical}. Instead, we will focus on demonstrating how to sample model parameters in \textbf{DP-BMM-Hierarchical}. The key distinction between these two models lies in whether we assume independence in bird species occurrence. Therefore, to sample parameters from their joint posterior distribution in \textbf{Base-Hierarchical}, you can simply combine the sampling algorithm outlined in this section with the one presented in Section \ref{sec:base_algorithm}.

Also, for the sake of clarification, within this subsection, we are exclusively considering a scenario in which each annotator has the capacity to identify all $N_2$ bird species at least partly by their vocalizations. In practical application, for $\lambda_{j, k}$'s and $\psi_{j, k}$'s that fall outside the lists mentioned in Section \ref{sec:varying_ability}, we can designate them as \texttt{NA}. Handling these \texttt{NA} values in R is straightforward, as we can simply set \texttt{na.rm = TRUE}.

\subsubsection{A list of model parameters to be sampled}
In \textbf{DP-BMM-Hierarchical} introduced in Section \ref{sec:varying_ability}, model parameters to be sampled include:
\begin{itemize}
    \item $\gamma$ is the concentration parameter in the DP;
    \item $\bm{z} = (z_1, z_2, \dots, z_{N_1})^T$ is the collection of the $N_1$ audio recordings' assignments, where $z_i$ is the assignment of recording $i$;
    \item $\bm{Y} = (\bm{y}_1, \bm{y}_2, \dots, \bm{y}_{N_1})$, where $\bm{y}_{i} = (y_{i, 1}, y_{i, 2}, \dots, y_{i, N_3})^T$ is the latent binary variable for recording $i$;
    \item $\bm{\lambda} = (\lambda_1, \lambda_2, \dots, \lambda_{N_2})^T$, where $\lambda_j$ represents annotator $j$'s bird song identification expertise when birds are actually present in recordings;
    \item $\bm{\Lambda} = \left(\lambda_{j, k}\right) \in \mathbb{R}^{N_2 \times N_3}$, where $\lambda_{j, k}$ represents annotator $j$'s identification expertise for species $k$ when birds are actually present in recordings;
    \item For $j = 1, 2, \dots, N_2$, $k = 1, 2, \dots, N_3$, \\
    $W_{j, k}^{(\lambda)} = \left\{w_{j, k}^{(\lambda, 1)}, w_{j, k}^{(\lambda, 2)}, \dots, w_{j, k}^{\left(\lambda, \sum_{i = 1}^{N_1} y_{i, k} \mathcal{T}_{i, j, k}\right)}, \bar{w}_{j, k}^{(\lambda, 1)}, \bar{w}_{j, k}^{(\lambda, 2)}, \dots, \bar{w}_{j, k}^{\left(\lambda, \sum_{i = 1}^{N_1} y_{i, k} (1 - \mathcal{T}_{i, j, k})\right)}\right\}$ is the collection of Pólya-Gamma variables introduced due to the presence of the two terms: $\sigma\left(\lambda_{j, k}\right)$ and $\sigma\left(-\lambda_{j, k}\right)$, in the annotation process of \textbf{DP-BMM-Hierarchical};
    \item $\bm{\psi} = (\psi_1, \psi_2, \dots, \psi_{N_2})^T$, where $\psi_j$ represents annotator $j$'s bird song identification expertise when birds are not present in recordings;
    \item $\bm{\Psi} = \left(\psi_{j, k}\right) \in \mathbb{R}^{N_2 \times N_3}$, where $\psi_{j, k}$ represents annotator $j$'s identification expertise for species $k$ when birds are not present in recordings;
    \item For $j = 1, 2, \dots, N_2$, $k = 1, 2, \dots, N_3$, \\
    $W_{j, k}^{(\psi)} = \left\{w_{j, k}^{(\psi, 1)}, w_{j, k}^{(\psi, 2)}, \dots, w_{j, k}^{\left(\psi, \sum_{i = 1}^{N_1} y_{i, k} \mathcal{T}_{i, j, k}\right)}, \bar{w}_{j, k}^{(\psi, 1)}, \bar{w}_{j, k}^{(\psi, 2)}, \dots, \bar{w}_{j, k}^{\left(\psi, \sum_{i = 1}^{N_1} y_{i, k} (1 - \mathcal{T}_{i, j, k})\right)}\right\}$ is the collection of Pólya-Gamma variables introduced due to the presence of the two terms: $\sigma\left(\psi_{j, k}\right)$ and $\sigma\left(-\psi_{j, k}\right)$, in the annotation process of \textbf{DP-BMM-Hierarchical};
\end{itemize}

\subsubsection{The posterior distribution}
As demonstrated in Section \ref{sec:dpbmm_posterior}, for \textbf{DP-BMM-Hierarchical}, we integrated out mixing coefficients $\bm{\pi} = \left\{\pi_r\right\}_{r = 1}^{\infty}$ as well as $\bm{O} = \left\{\bm{o}_r\right\}_{r = 1}^{\infty}$, and only sample the remaining parameters listed above.

Without introduced Pólya-Gamma variables, the posterior distribution is as follows:
\begin{equation}
    \begin{aligned}
        & p\left(\gamma, \bm{z}, \bm{Y}, \bm{\lambda}, \bm{\Lambda}, \bm{\psi}, \bm{\Psi} \mid \mathcal{T}; u_1, u_2, a_o, b_o, \mu_{\lambda}, \phi_{\lambda}, \phi_{\lambda}^{*}, \mu_{\psi}, \phi_{\psi}, \phi_{\psi}^{*}\right) \\
        = \ & \int_{\mathcal{O}} p\left(\bm{O}, \gamma, \bm{z}, \bm{Y}, \bm{\lambda}, \bm{\Lambda}, \bm{\psi}, \bm{\Psi} \mid \mathcal{T}; u_1, u_2, a_o, b_o, \mu_{\lambda}, \phi_{\lambda}, \phi_{\lambda}^{*}, \mu_{\psi}, \phi_{\psi}, \phi_{\psi}^{*}\right) d\bm{O} \\
        \propto \ & \left(\int_{\mathcal{O}} p(\bm{O} \mid a_o, b_o) p(\bm{Y} \mid \bm{O}, \bm{z}) d\bm{O}\right) \times p(\gamma \mid u_1, u_2) \times p(\bm{z} \mid \gamma) \\
        & \ \ \ \times p(\bm{\lambda} \mid \mu_{\lambda}, \phi_{\lambda}) p(\bm{\Lambda} \mid \bm{\lambda}, \phi_{\lambda}^{*}) \times p(\bm{\psi} \mid \mu_{\psi}, \phi_{\psi}) p(\bm{\Psi} \mid \bm{\psi}, \phi_{\psi}^{*}) \times p(\mathcal{T} \mid \bm{Y}, \bm{\Lambda}, \bm{\Psi}) \\
        \propto \ & \left(\int_{\mathcal{O}} p(\bm{O} \mid a_o, b_o) p(\bm{Y} \mid \bm{O}, \bm{z}) d\bm{O}\right) \times p(\gamma \mid u_1, u_2) \times p(\bm{z} \mid \gamma) \\
        & \ \ \ \times \left\{\prod_{j = 1}^{N_2} p(\lambda_j \mid \mu_{\lambda}, \phi_{\lambda}) p(\psi_j \mid \mu_{\psi}, \phi_{\psi})\right\} \times \left\{\prod_{j = 1}^{N_2} \prod_{k = 1}^{N_3} p(\lambda_{j, k} \mid \lambda_j, \phi_{\lambda}^{*}) p(\psi_{j, k} \mid \psi_j, \phi_{\psi}^{*}) \right\} \\
        & \ \ \ \times \left\{\prod_{k = 1}^{n_3} \prod_{i = 1}^{N_1} \prod_{j = 1}^{N_2} p(\mathcal{T}_{i, j, k} \mid y_{i, k}, \lambda_{j, k}, \psi_{j, k}) \right\}.
    \end{aligned}
\end{equation}

We will introduce Pólya-Gamma variables when sampling $\Lambda = (\lambda_{j, k}) \in \mathbb{R}^{N_2 \times N_3}$ as well as $\Psi = (\psi_{j, k}) \in \mathbb{R}^{N_2 \times N_3}$.

\subsubsection{The sampling algorithm} \label{sec:dpbmm_hierarchy_algorithm}
$\triangleright \ \textbf{Sampling } \gamma:$
\\
\\
As we've discussed in Section \ref{sec:dpbmm_algorithm}, the full conditional of $\gamma$ is:
\begin{equation}
    \begin{aligned}
        & p(\gamma \mid \bm{z}, \bm{Y}, \bm{\lambda}, \bm{\Lambda}, \bm{\psi}, \bm{\Psi}, \mathcal{T}; u_1, u_2, a_o, b_o, \mu_{\lambda}, \phi_{\lambda}, \phi_{\lambda}^{*}, \mu_{\psi}, \phi_{\psi}, \phi_{\psi}^{*}) \\
        \propto \ & p(\bm{z} \mid \gamma) p(\gamma \mid u_1, u_2) \\
        \propto \ & \frac{\gamma^R}{\prod_{i = 1}^{N_1} (i - 1 + \gamma)} \times \gamma^{u_1 - 1} \exp(-u_2 \gamma) \\
        \propto \ & \frac{\gamma^{R + u_1 - 1}}{\prod_{i = 1}^{N_1} (i - 1 + \gamma)} \exp(-u_2 \gamma).
    \end{aligned}
    \label{eq:dpbmm_hierarchy_gamma}
\end{equation}

Still, we adopted an adaptive Metropolis algorithm to update $\gamma$. And the proposal distribution for $\gamma$ is:
\begin{equation}
    q(\gamma, \gamma^{*}) = \mathcal{N}(\gamma, s_{\gamma}^2).
\end{equation}

Based on the full conditional as well as the proposal distribution, the acceptance rate for $\gamma$ is:
\begin{equation}
    \begin{aligned}
        \alpha(\gamma, \gamma^{*}) &= \min\left\{1 , \frac{p(\gamma^{*} \mid \bm{z}, \bm{Y}, \bm{\lambda}, \bm{\Lambda}, \bm{\psi}, \bm{\Psi}, \mathcal{T}; u_1, u_2, a_o, b_o, \mu_{\lambda}, \phi_{\lambda}, \phi_{\lambda}^{*}, \mu_{\psi}, \phi_{\psi}, \phi_{\psi}^{*}) q(\gamma^{*}, \gamma)}{p(\gamma \mid \bm{z}, \bm{Y}, \bm{\lambda}, \bm{\Lambda}, \bm{\psi}, \bm{\Psi}, \mathcal{T}; u_1, u_2, a_o, b_o, \mu_{\lambda}, \phi_{\lambda}, \phi_{\lambda}^{*}, \mu_{\psi}, \phi_{\psi}, \phi_{\psi}^{*}) q(\gamma, \gamma^{*})}\right\} \\
        &= \min\left\{1 , \frac{p(\gamma^{*} \mid \bm{z}, \bm{Y}, \bm{\lambda}, \bm{\Lambda}, \bm{\psi}, \bm{\Psi}, \mathcal{T}; u_1, u_2, a_o, b_o, \mu_{\lambda}, \phi_{\lambda}, \phi_{\lambda}^{*}, \mu_{\psi}, \phi_{\psi}, \phi_{\psi}^{*})}{p(\gamma \mid \bm{z}, \bm{Y}, \bm{\lambda}, \bm{\Lambda}, \bm{\psi}, \bm{\Psi}, \mathcal{T}; u_1, u_2, a_o, b_o, \mu_{\lambda}, \phi_{\lambda}, \phi_{\lambda}^{*}, \mu_{\psi}, \phi_{\psi}, \phi_{\psi}^{*})}\right\} \\
        &= \min\left\{1, \left(\frac{\gamma^{*}}{\gamma}\right)^{R + u_1 - 1} \times \exp\left(-u_2(\gamma^{*} - \gamma)\right) \times \prod_{i = 1}^{N_1} \left(\frac{i - 1 + \gamma}{i - 1 + \gamma^{*}}\right) \right\}.
    \end{aligned}
\end{equation}
\\
$\triangleright \ \textbf{Sampling } \bm{z}:$
\\
\\
As shown in Section \ref{sec:dpbmm_algorithm}, the full conditional of $z_i$ is:
\begin{equation}
    \begin{aligned}
        & p(z_i = \tilde{r} \mid \bm{z}_{-i}, \gamma, \bm{Y}, \bm{\lambda}, \bm{\Lambda}, \bm{\psi}, \bm{\Psi}, \mathcal{T}; u_1, u_2, a_o, b_o, \mu_{\lambda}, \phi_{\lambda}, \phi_{\lambda}^{*}, \mu_{\psi}, \phi_{\psi}, \phi_{\psi}^{*}) \\
        \propto \ & p(z_i = \tilde{r}, \bm{z}_{-i} \mid \gamma) \left(\int_{\mathcal{O}} p(\bm{O} \mid a_o, b_o) p(\bm{Y} \mid \bm{O}, z_i = \tilde{r}, \bm{z}_{-i}) d \bm{O} \right) \\
        \propto \ & p(z_i = \tilde{r} \mid \bm{z}_{-i}, \gamma) \left(\int_{\mathcal{O}} p(\bm{O} \mid a_o, b_o) p(\bm{Y} \mid \bm{O}, z_i = \tilde{r}, \bm{z}_{-i}) d \bm{O} \right) \\
        \propto \ & 
        \left\{
        \begin{aligned}
            & \frac{n_{\tilde{r}, -i}}{\gamma + N_1 - 1} \times \left\{\prod_{k = 1}^{N_3} \frac{B(a_o + n^{(+)}_{\tilde{r}, k, -i} + y_{i, k}, b_o + n^{(-)}_{\tilde{r}, k, -i} + (1 - y_{i, k}))}{B(a_o + n^{(+)}_{\tilde{r}, k, -i}, b_o + n^{(-)}_{\tilde{r}, k, -i})}\right\}, \ \ \ \text{if} \ \tilde{r} \ \text{exists}, \\ 
            & \frac{\gamma}{\gamma + N_1 - 1} \times \left\{\prod_{k = 1}^{N_3} \frac{B(a_o + y_{i, k}, b_o + (1 - y_{i, k}))}{B(a_o, b_o)}\right\}, \ \ \ \text{if} \ \tilde{r} \ \text{is new}. \\
        \end{aligned}
        \right.
    \end{aligned}
    \label{eq:dpbmm_hierarchy_z}
\end{equation}
Similar to what we've done for \textbf{DP-BMM}, we proceeded to calculate the product in Equation \eqref{eq:dpbmm_hierarchy_z} for each $\tilde{r} \in \{1, 2, \dots, R\}$ as well as for the scenario where $\tilde{r}$ corresponds to a new mixture component. We stored these results in a vector and normalized them, denoted as $\bm{q}_i \in (0, 1)^{R+1}$. Consequently, we can sample each $z_i$, $i = 1, 2, \dots, N_1$ from its full conditional:
\begin{equation}
    z_i \mid \bm{z}_{-i}, \gamma, \bm{Y}, \bm{\lambda}, \bm{\Lambda}, \bm{\psi}, \bm{\Psi}, \mathcal{T}; u_1, u_2, a_o, b_o, \mu_{\lambda}, \phi_{\lambda}, \phi_{\lambda}^{*}, \mu_{\psi}, \phi_{\psi}, \phi_{\psi}^{*} \sim \text{Categorical}(\bm{q}_i).
\end{equation}
\\
$\triangleright \ \textbf{Sampling } \bm{Y}:$
\\
\\
Similar to our discussions in Section \ref{sec:dpbmm_algorithm}, we sampled each $y_{i, k}$, $i = 1, 2, \dots, N_1$, $k = 1, 2, \dots, N_3$. The only difference lies in our modeling approach for each annotator. Consequently, we only needed to replace $\lambda_j$ or $\psi_j$ with $\sigma(\lambda_{j, k})$ or $\sigma(\psi_{j, k})$. And the full conditional of $y_{i, k}$ is:
\begin{equation}
    \begin{aligned}
        & p(y_{i, k} \mid \bm{Y}_{-(i, k)}, \gamma, \bm{z}, \bm{\lambda}, \bm{\Lambda}, \bm{\psi}, \bm{\Psi}, \mathcal{T}; u_1, u_2, a_o, b_o, \mu_{\lambda}, \phi_{\lambda}, \phi_{\lambda}^{*}, \mu_{\psi}, \phi_{\psi}, \phi_{\psi}^{*}) \\
        \propto \ & \left(\int_{\mathcal{O}} p(\bm{O} \mid a_o, b_o) p(\bm{Y} \mid \bm{O}, \bm{z}) d\bm{O} \right) \times \left\{\prod_{j = 1}^{N_2} p(\mathcal{T}_{i, j, k} \mid y_{i, k}, \lambda_{j, k}, \psi_{j, k})\right\} \\
        \propto \ & \underbrace{B\left(a_o + n_{z_i, k, -i}^{(+)} + y_{i, k}, b_o + n_{z_i, k, -i}^{(-)} + (1 - y_{i, k})\right)}_{\text{the first term}} \\
        & \times \underbrace{\left\{\prod_{j = 1}^{N_2} \left[\left(\sigma(\lambda_{j, k})\right)^{y_{i, k}} \left(\sigma(\psi_{j, k})\right)^{1 - y_{i, k}}\right]^{\mathcal{T}_{i, j, k}} \left[(\sigma(-\lambda_{j, k}))^{y_{i, k}} (\sigma(-\psi_{j, k}))^{1 - y_{i, k}}\right]^{1 - \mathcal{T}_{i, j, k}}\right\}}_{\text{the second term}}.
    \end{aligned}
    \label{eq:dpbmm_hierarchy_y}
\end{equation}
Similar to the procedure for \textbf{DP-BMM}, the next step is to calculate the product in Equation \eqref{eq:dpbmm_hierarchy_y} for both $y_{i, k} = 0$ and $y_{i, k} = 1$. Afterwards, we normalized them, and obtained the probability of $y_{i, k} = 1$, denoted as $\eta_{i, k}$, $i = 1, 2, \dots, N_1$, $k = 1, 2, \dots, N_3$. 

Finally, we sampled each $y_{i, k}$ from its full conditional:
\begin{equation}
    y_{i, k} \mid \bm{Y}_{-(i, k)}, \gamma, \bm{z}, \bm{\lambda}, \bm{\Lambda}, \bm{\psi}, \bm{\Psi}, \mathcal{T}; u_1, u_2, a_o, b_o, \mu_{\lambda}, \phi_{\lambda}, \phi_{\lambda}^{*}, \mu_{\psi}, \phi_{\psi}, \phi_{\psi}^{*} \sim \text{Bernoulli}(\eta_{i, k}).
\end{equation}
\\
$\triangleright \ \textbf{Sampling } \bm{\lambda}:$
\\
\\
As mentioned in Section \ref{sec:varying_ability} and outlined in our list of model parameters for \textbf{DP-BMM-Hierarchical}, it should be noted that both $\lambda_j$ and $\psi_j$ have different meanings compared to their usage in \textbf{DP-BMM} and \textbf{Base}. In the following parts, we will provide a detailed explanation of how to sample $\bm{\lambda}$, $\bm{\Lambda}$, $\bm{\psi}$ and $\bm{\Psi}$ from their full conditional distributions. We will particularly focus on how the introduction of Pólya-Gamma variables allows us to implement a straightforward Gibbs sampler, even in the presence of the challenging logistic term.

The full conditional of $\lambda_j$, $j = 1, 2, \dots, N_2$ is:
\begin{equation}
    \begin{aligned}
        & p(\lambda_j \mid \bm{\lambda}_{-j}, \gamma, \bm{z}, \bm{Y}, \bm{\Lambda}, \bm{\psi}, \bm{\Psi}, \mathcal{T}; u_1, u_2, a_o, b_o, \mu_{\lambda}, \phi_{\lambda}, \phi_{\lambda}^{*}, \mu_{\psi}, \phi_{\psi}, \phi_{\psi}^{*}) \\
        \propto \ & p(\lambda_j \mid \mu_{\lambda}, \phi_{\lambda}) \times \left\{\prod_{k = 1}^{N_3} p(\lambda_{j, k} \mid \lambda_j, \phi_{\lambda}^{*})\right\} \\
        \propto \ & \exp\left(-\frac{(\lambda_j - \mu_{\lambda})^2}{2 \phi_{\lambda}^2}\right) \times \exp\left\{-\frac{\sum_{k = 1}^{N_3} (\lambda_{j, k} - \lambda_j)^2}{2\left(\phi_{\lambda}^{*}\right)^2}\right\} \\
        \propto \ & \exp\left(-\frac{\left(\lambda_j - \mu^{(\lambda)}_{N_3, j}\right)^2}{2\left(\tau_{N_3, j}^{(\lambda)}\right)^2}\right),
    \end{aligned}
\end{equation}
\begin{equation}
    \mu^{(\lambda)}_{N_3, j} = \frac{\mu_{\lambda} / \phi_{\lambda}^2 + \left(\sum_{k = 1}^{N_3} \lambda_{j, k}\right) / \left(\phi_{\lambda}^{*}\right)^2}{1 / \phi_{\lambda}^2 + N_3 / \left(\phi_{\lambda}^{*}\right)^2}, \ \ \ \left(\tau_{N_3, j}^{(\lambda)}\right)^2 = \frac{1}{1 / \phi_{\lambda}^2 + N_3 / \left(\phi_{\lambda}^{*}\right)^2}.
\end{equation}
\\
Therefore, we sampled each $\lambda_j$, $j = 1, 2, \dots, N_2$ from its full conditional:
\begin{equation}
    \lambda_j \mid \bm{\lambda}_{-j}, \gamma, \bm{z}, \bm{Y}, \bm{\Lambda}, \bm{\psi}, \bm{\Psi}, \mathcal{T}; u_1, u_2, a_o, b_o, \mu_{\lambda}, \phi_{\lambda}, \phi_{\lambda}^{*}, \mu_{\psi}, \phi_{\psi}, \phi_{\psi}^{*} \sim \mathcal{N}\left(\mu^{(\lambda)}_{N_3, j}, \left(\tau_{N_3, j}^{(\lambda)}\right)^2\right).
\end{equation}
\\
$\triangleright \ \textbf{Sampling } \bm{\psi}:$
\\
\\
The full conditional of $\psi_j$, $j = 1, 2, \dots, N_2$ is:
\begin{equation}
    \begin{aligned}
        & p(\psi_j \mid \bm{\psi}_{-j}, \gamma, \bm{z}, \bm{Y}, \bm{\lambda}, \bm{\Lambda}, \bm{\Psi}, \mathcal{T}; u_1, u_2, a_o, b_o, \mu_{\lambda}, \phi_{\lambda}, \phi_{\lambda}^{*}, \mu_{\psi}, \phi_{\psi}, \phi_{\psi}^{*}) \\
        \propto \ & p(\psi_j \mid \mu_{\psi}, \phi_{\psi}) \times \left\{\prod_{k = 1}^{N_3} p(\psi_{j, k} \mid \psi_j, \phi_{\psi}^{*})\right\} \\
        \propto \ & \exp\left(-\frac{(\psi_j - \mu_{\psi})^2}{2 \phi_{\psi}^2}\right) \times \exp\left\{-\frac{\sum_{k = 1}^{N_3} (\psi_{j, k} - \psi_j)^2}{2\left(\phi_{\psi}^{*}\right)^2}\right\} \\
        \propto \ & \exp\left(-\frac{\left(\psi_j - \mu^{(\psi)}_{N_3, j}\right)^2}{2\left(\tau_{N_3, j}^{(\psi)}\right)^2}\right),
    \end{aligned}
\end{equation}
\begin{equation}
    \mu^{(\psi)}_{N_3, j} = \frac{\mu_{\psi} / \phi_{\psi}^2 + \left(\sum_{k = 1}^{N_3} \psi_{j, k}\right) / \left(\phi_{\psi}^{*}\right)^2}{1 / \phi_{\psi}^2 + N_3 / \left(\phi_{\psi}^{*}\right)^2}, \ \ \ \left(\tau_{N_3, j}^{(\psi)}\right)^2 = \frac{1}{1 / \phi_{\psi}^2 + N_3 / \left(\phi_{\psi}^{*}\right)^2}.
\end{equation}
\\
Therefore, we sampled each $\psi_j$, $j = 1, 2, \dots, N_2$ from its full conditional:
\begin{equation}
    \psi_j \mid \bm{\psi}_{-j}, \gamma, \bm{z}, \bm{Y}, \bm{\lambda}, \bm{\Lambda}, \bm{\Psi}, \mathcal{T}; u_1, u_2, a_o, b_o, \mu_{\lambda}, \phi_{\lambda}, \phi_{\lambda}^{*}, \mu_{\psi}, \phi_{\psi}, \phi_{\psi}^{*} \sim \mathcal{N}\left(\mu^{(\psi)}_{N_3, j}, \left(\tau_{N_3, j}^{(\psi)}\right)^2\right).
\end{equation}
\\
$\triangleright \ \textbf{Sampling } \bm{\Lambda}:$
\\
\\
Due to the presence of the logistic term, we cannot directly sample each $\lambda_{j, k}$ from its full conditional, whereas we can jointly sample $\lambda_{j, k}$ as well as the corresponding Pólya-Gamma variables.

Specifically, for each $(j, k) \in \left\{(j, k): j \in \{1, 2, \dots, N_2\}, k \in \{1, 2, \dots, N_3\}\right\}$, assuming that all variables except for $\lambda_{j, k}$ are given, we applied Pólya-Gamma Augmentation \citep{polson2013bayesian, donner2018efficient} as follows:
\begin{equation}
    \begin{aligned}
        & \prod_{i = 1}^{N_1} p(\mathcal{T}_{i, j, k} \mid y_{i, k}, \lambda_{j, k}, \psi_{j, k}) \\
        = \ & \prod_{i = 1}^{N_1} \left\{\left(\sigma(\lambda_{j, k})\right)^{y_{i, k}} \left(\sigma(\psi_{j, k})\right)^{1 - y_{i, k}}\right\}^{\mathcal{T}_{i, j, k}} \times \left\{\left(\sigma(-\lambda_{j, k})\right)^{y_{i, k}} \left(\sigma(-\psi_{j, k})\right)^{1 - y_{i, k}}\right\}^{1 - \mathcal{T}_{i, j, k}} \\
        = \ & \left\{\sigma(\lambda_{j, k})\right\}^{\delta_{j, k}^{(\lambda)}} \times \left\{\sigma(-\lambda_{j, k})\right\}^{\bar{\delta}_{j, k}^{(\lambda)}} \\
        = \ & \left(\int_{0}^{\infty} \exp(h(w, \lambda_{j, k})) p_{\text{PG}}(w \mid 1, 0) dw \right)^{\delta_{j, k}^{(\lambda)}} \times \left(\int_{0}^{\infty} \exp(h(w, -\lambda_{j, k})) p_{\text{PG}}(w \mid 1, 0) dw \right)^{\bar{\delta}_{j, k}^{(\lambda)}} \\
        = \ & \Bigg\{\underbrace{\int_{0}^{\infty} \dots \int_{0}^{\infty}}_{\delta_{j, k}^{(\lambda)}} \underbrace{\int_{0}^{\infty} \dots \int_{0}^{\infty}}_{\bar{\delta}_{j, k}^{(\lambda)}} \left[\prod_{d = 1}^{\delta_{j, k}^{(\lambda)}} \exp\left(h(w_{j, k}^{(\lambda, d)}, \lambda_{j, k})\right) p_{\text{PG}}\left(w_{j, k}^{(\lambda, d)} \mid 1, 0\right)\right] \\
        & \ \ \ \times \left[\prod_{\bar{d} = 1}^{\bar{\delta}_{j, k}^{(\lambda)}} \exp\left(h(\bar{w}_{j, k}^{(\lambda, \bar{d})}, -\lambda_{j, k})\right) p_{\text{PG}}\left(\bar{w}_{j, k}^{(\lambda, \bar{d})} \mid 1, 0\right)\right] dw_{j,k}^{(\lambda, 1)} \dots, dw_{j,k}^{\left(\lambda, \delta_{j, k}^{(\lambda)}\right)} d\bar{w}_{j,k}^{(\lambda, 1)} \dots, d\bar{w}_{j,k}^{\left(\lambda, \bar{\delta}_{j, k}^{(\lambda)}\right)}\Bigg\},
    \end{aligned}
    \label{eq:lambda_polya}
\end{equation}
where $\delta_{j, k}^{(\lambda)} = \sum_{i = 1}^{N_1} \left(y_{i, k} \mathcal{T}_{i, j, k}\right)$, $\bar{\delta}_{j, k}^{(\lambda)} = \sum_{i = 1}^{N_1} \left(y_{i, k} (1 - \mathcal{T}_{i, j, k})\right)$, $h(w, z) = \frac{z}{2} - \frac{z^2}{2} w - \ln{2}$,\\
$p_{\text{PG}}(w \mid b, 0)$ is the Pólya-Gamma density of random variable $w \in \mathbb{R}^{+}$ \citep{donner2018efficient}
and the tilted Pólya-Gamma densities are defined as:
\begin{equation}
    p_{\text{PG}}(w \mid b, c) \propto \exp\left(-\frac{c^2}{2} w\right) p_{\text{PG}}(w \mid b, 0),
\end{equation}
which will be very useful for our derivation.

From Equation \eqref{eq:lambda_polya}, if we regard $\lambda_{j, k}$ as well as its corresponding Pólya-Gamma variables $W_{j, k}^{(\lambda)}$ as a block, we can easily obtain their joint full conditional as follows:
\begin{equation}
    \begin{aligned}
        & p\left(\lambda_{j, k}, W_{j, k}^{(\lambda)} \mid \gamma, \bm{z}, \bm{Y}, \bm{\lambda}, \bm{\Lambda}_{-(j, k)}, \bm{\psi}, \bm{\Psi}, \mathcal{T}; u_1, u_2, a_o, b_o, \mu_{\lambda}, \phi_{\lambda}, \phi^{*}_{\lambda}, \mu_{\psi}, \phi_{\psi}, \phi^{*}_{\psi}\right) \\
        \propto \ & \left\{\prod_{d = 1}^{\delta_{j, k}^{(\lambda)}} \exp\left(h(w_{j, k}^{(\lambda, d)}, \lambda_{j, k})\right) p_{\text{PG}}\left(w_{j, k}^{(\lambda, d)} \mid 1, 0\right)\right\} \times \left\{\prod_{\bar{d} = 1}^{\delta_{j, k}^{(\lambda)}} \exp\left(h(\bar{w}_{j, k}^{(\lambda, \bar{d})}, -\lambda_{j, k})\right) p_{\text{PG}}\left(\bar{w}_{j, k}^{(\lambda, \bar{d})} \mid 1, 0\right)\right\} \\
        & \ \times \exp\left(-\frac{(\lambda_{j, k} - \lambda_j)^2}{2 \left(\phi_{\lambda}^{*}\right)^2}\right).
    \end{aligned}
\end{equation}

Applying a partially collapsed Gibbs sampling strategy mentioned in \cite{van2008partially, park2009partially}, our sampling scheme for $\lambda_{j, k}$ could be divided into two steps. The first step is to sample Pólya-Gamma variables $W_{j, k}^{(\lambda)}$ from:
\begin{equation}
    \begin{aligned}
        & p\left(W_{j, k}^{(\lambda)} \mid \lambda_{j, k}, \gamma, \bm{z}, \bm{Y}, \bm{\lambda}, \bm{\Lambda}_{-(j, k)}, \bm{\psi}, \bm{\Psi}, \mathcal{T}; u_1, u_2, a_o, b_o, \mu_{\lambda}, \phi_{\lambda}, \phi^{*}_{\lambda}, \mu_{\psi}, \phi_{\psi}, \phi^{*}_{\psi}\right) \\
        \propto \ & \left\{\prod_{d = 1}^{\delta_{j, k}^{(\lambda)}} \exp\left(h(w_{j, k}^{(\lambda, d)}, \lambda_{j, k})\right) p_{\text{PG}}\left(w_{j, k}^{(\lambda, d)} \mid 1, 0\right)\right\} \times \left\{\prod_{\bar{d} = 1}^{\delta_{j, k}^{(\lambda)}} \exp\left(h(\bar{w}_{j, k}^{(\lambda, \bar{d})}, -\lambda_{j, k})\right) p_{\text{PG}}\left(\bar{w}_{j, k}^{(\lambda, \bar{d})} \mid 1, 0\right)\right\} \\
        \propto \ & \left\{\prod_{d = 1}^{\delta_{j, k}^{(\lambda)}} p_{\text{PG}}\left(w_{j, k}^{(\lambda, d)} \mid 1, \lambda_{j, k}\right)\right\} \times \left\{\prod_{\bar{d} = 1}^{\delta_{j, k}^{(\lambda)}} p_{\text{PG}}\left(\bar{w}_{j, k}^{(\lambda, \bar{d})} \mid 1, \lambda_{j, k}\right)\right\}.
    \end{aligned}
\end{equation}

And the second step is to sample $\lambda_{j ,k}$, representing annotator $j$'s sound identification expertise for species $k$ when birds are present in recordings, from:
\begin{equation}
    \begin{aligned}
        & p\left(\lambda_{j, k} \mid W_{j, k}^{(\lambda)}, \gamma, \bm{z}, \bm{Y}, \bm{\lambda}, \bm{\Lambda}_{-(j, k)}, \bm{\psi}, \bm{\Psi}, \mathcal{T}; u_1, u_2, a_o, b_o, \mu_{\lambda}, \phi_{\lambda}, \phi^{*}_{\lambda}, \mu_{\psi}, \phi_{\psi}, \phi^{*}_{\psi}\right) \\
        \propto \ & \exp\left(-\frac{(\lambda_{j, k} - \lambda_j)^2}{2\left(\phi_{\lambda}^{*}\right)^2}\right) \times \left\{\prod_{d = 1}^{\delta_{j, k}^{(\lambda)}} \exp\left(\frac{\lambda_{j, k}}{2} - \frac{\lambda_{j, k}^2}{2} w_{j, k}^{(\lambda, d)}\right)\right\} \times \left\{\prod_{\bar{d} = 1}^{\bar{\delta}_{j, k}^{(\lambda)}} \exp\left(\frac{-\lambda_{j, k}}{2} - \frac{\lambda_{j, k}^2}{2} \bar{w}_{j, k}^{(\lambda, \bar{d})}\right)\right\} \\
        \propto \ & \exp\left\{-\frac{\left(\lambda_{j, k} - \mu_{j, k}^{(\lambda)}\right)^2}{2\left(\tau_{j, k}^{(\lambda)}\right)^2}\right\},
    \end{aligned}
\end{equation}
where
\begin{equation}
    \mu_{j, k}^{(\lambda)} = \frac{\lambda_j / \left(\phi_{\lambda}^{*}\right)^2 + \left(\delta_{j, k}^{(\lambda)} - \bar{\delta}_{j, k}^{(\lambda)}\right)/ 2}{1  / \left(\phi_{\lambda}^{*}\right)^2 + \sum_{d = 1}^{\delta_{j, k}^{(\lambda)}} w_{j, k}^{(\lambda, d)} + \sum_{\bar{d} = 1}^{\bar{\delta}_{j, k}^{(\lambda)}} \bar{w}_{j, k}^{(\lambda, \bar{d})}}, \ \left(\tau_{j, k}^{(\lambda)}\right)^2 = \frac{1}{1  / \left(\phi_{\lambda}^{*}\right)^2 + \sum_{d = 1}^{\delta_{j, k}^{(\lambda)}} w_{j, k}^{(\lambda, d)} + \sum_{\bar{d} = 1}^{\bar{\delta}_{j, k}^{(\lambda)}} \bar{w}_{j, k}^{(\lambda, \bar{d})}}.
\end{equation}
\\
$\triangleright \ \textbf{Sampling } \bm{\Psi}:$
\\
\\
Similar to $\bm{\Lambda}$, for each $(j, k)$, assuming all variables except for $\psi_{j, k}$ are given, we have:
\begin{equation}
    \begin{aligned}
        & \prod_{i = 1}^{N_1} p(\mathcal{T}_{i, j, k} \mid y_{i, k}, \lambda_{j, k}, \psi_{j, k}) \\
        = \ & \prod_{i = 1}^{N_1} \left\{\left(\sigma(\lambda_{j, k})\right)^{y_{i, k}} \left(\sigma(\psi_{j, k})\right)^{1 - y_{i, k}}\right\}^{\mathcal{T}_{i, j, k}} \times \left\{\left(\sigma(-\lambda_{j, k})\right)^{y_{i, k}} \left(\sigma(-\psi_{j, k})\right)^{1 - y_{i, k}}\right\}^{1 - \mathcal{T}_{i, j, k}} \\
        = \ & \left\{\sigma(\psi_{j, k})\right\}^{\delta_{j, k}^{(\psi)}} \times \left\{\sigma(-\psi_{j, k})\right\}^{\bar{\delta}_{j, k}^{(\psi)}} \\
        = \ & \left(\int_{0}^{\infty} \exp(h(w, \psi_{j, k})) p_{\text{PG}}(w \mid 1, 0) dw \right)^{\delta_{j, k}^{(\psi)}} \times \left(\int_{0}^{\infty} \exp(h(w, -\psi_{j, k})) p_{\text{PG}}(w \mid 1, 0) dw \right)^{\bar{\delta}_{j, k}^{(\psi)}} \\
        = \ & \Bigg\{\underbrace{\int_{0}^{\infty} \dots \int_{0}^{\infty}}_{\delta_{j, k}^{(\psi)}} \underbrace{\int_{0}^{\infty} \dots \int_{0}^{\infty}}_{\bar{\delta}_{j, k}^{(\psi)}} \left[\prod_{d = 1}^{\delta_{j, k}^{(\psi)}} \exp\left(h(w_{j, k}^{(\psi, d)}, \psi_{j, k})\right) p_{\text{PG}}\left(w_{j, k}^{(\psi, d)} \mid 1, 0\right)\right] \\
        & \ \ \ \times \left[\prod_{\bar{d} = 1}^{\bar{\delta}_{j, k}^{(\psi)}} \exp\left(h(\bar{w}_{j, k}^{(\psi, \bar{d})}, -\psi_{j, k})\right) p_{\text{PG}}\left(\bar{w}_{j, k}^{(\psi, \bar{d})} \mid 1, 0\right)\right] dw_{j,k}^{(\psi, 1)} \dots, dw_{j,k}^{\left(\psi, \delta_{j, k}^{(\psi)}\right)} d\bar{w}_{j,k}^{(\psi, 1)} \dots, d\bar{w}_{j,k}^{\left(\psi, \bar{\delta}_{j, k}^{(\psi)}\right)}\Bigg\},
    \end{aligned}
    \label{eq:psi_polya}
\end{equation}
where $\delta_{j, k}^{(\psi)} = \sum_{i = 1}^{N_1} ((1 - y_{i, k }) \mathcal{T}_{i, j, k})$ and $\bar{\delta}_{j, k}^{(\psi)} = \sum_{i = 1}^{N_1} ((1 - y_{i, k }) (1 - \mathcal{T}_{i, j, k}))$.

From Equation \eqref{eq:psi_polya}, we can obtain the joint full conditional of $\psi_{j, k}$ as well as its corresponding Pólya-Gamma variables $W_{j, k}^{(\psi)}$ as follows:
\begin{equation}
    \begin{aligned}
        & p\left(\psi_{j, k}, W_{j, k}^{(\psi)} \mid \gamma, \bm{z}, \bm{Y}, \bm{\lambda}, \bm{\Lambda}, \bm{\psi}, \bm{\Psi}_{-(j, k)}, \mathcal{T}; u_1, u_2, a_o, b_o, \mu_{\lambda}, \phi_{\lambda}, \phi^{*}_{\lambda}, \mu_{\psi}, \phi_{\psi}, \phi^{*}_{\psi}\right) \\
        \propto \ & \left\{\prod_{d = 1}^{\delta_{j, k}^{(\psi)}} \exp\left(h(w_{j, k}^{(\psi, d)}, \psi_{j, k})\right) p_{\text{PG}}\left(w_{j, k}^{(\psi, d)} \mid 1, 0\right)\right\} \times \left\{\prod_{\bar{d} = 1}^{\delta_{j, k}^{(\psi)}} \exp\left(h(\bar{w}_{j, k}^{(\psi, \bar{d})}, -\psi_{j, k})\right) p_{\text{PG}}\left(\bar{w}_{j, k}^{(\psi, \bar{d})} \mid 1, 0\right)\right\} \\
        & \ \times \exp\left(-\frac{(\psi_{j, k} - \psi_j)^2}{2 \left(\phi_{\psi}^{*}\right)^2}\right).
    \end{aligned}
\end{equation}

Again, by the partially collapsed Gibbs sampling strategy proposed in \cite{van2008partially, park2009partially}, the first step is to sample Pólya-Gamma $W_{j, k}^{(\psi)}$ from:
\begin{equation}
    \begin{aligned}
        & p\left(W_{j, k}^{(\psi)} \mid \psi_{j, k}, \gamma, \bm{z}, \bm{Y}, \bm{\lambda}, \bm{\Lambda}, \bm{\psi}, \bm{\Psi}_{-(j, k)}, \mathcal{T}; u_1, u_2, a_o, b_o, \mu_{\lambda}, \phi_{\lambda}, \phi^{*}_{\lambda}, \mu_{\psi}, \phi_{\psi}, \phi^{*}_{\psi}\right) \\
        \propto \ & \left\{\prod_{d = 1}^{\delta_{j, k}^{(\psi)}} \exp\left(h(w_{j, k}^{(\psi, d)}, \psi_{j, k})\right) p_{\text{PG}}\left(w_{j, k}^{(\psi, d)} \mid 1, 0\right)\right\} \times \left\{\prod_{\bar{d} = 1}^{\delta_{j, k}^{(\psi)}} \exp\left(h(\bar{w}_{j, k}^{(\psi, \bar{d})}, -\psi_{j, k})\right) p_{\text{PG}}\left(\bar{w}_{j, k}^{(\psi, \bar{d})} \mid 1, 0\right)\right\} \\
        \propto \ & \left\{\prod_{d = 1}^{\delta_{j, k}^{(\psi)}} p_{\text{PG}}\left(w_{j, k}^{(\psi, d)} \mid 1, \psi_{j, k} \right)\right\} \times \left\{\prod_{\bar{d} = 1}^{\delta_{j, k}^{(\psi)}} p_{\text{PG}}\left(\bar{w}_{j, k}^{(\psi, \bar{d})} \mid 1, \psi_{j, k}\right)\right\}.
    \end{aligned}
\end{equation}

The second step is to sample $\psi_{j, k}$, representing annotator $j$'s sound identification expertise for species $k$ when birds are not present in recordings, from:
\begin{equation}
    \begin{aligned}
        & p\left(\psi_{j, k} \mid W_{j, k}^{(\psi)}, \gamma, \bm{z}, \bm{Y}, \bm{\lambda}, \bm{\Lambda}, \bm{\psi}, \bm{\Psi}_{-(j, k)}, \mathcal{T}; u_1, u_2, a_o, b_o, \mu_{\lambda}, \phi_{\lambda}, \phi^{*}_{\lambda}, \mu_{\psi}, \phi_{\psi}, \phi^{*}_{\psi}\right) \\
        \propto \ & \exp\left(-\frac{(\psi_{j, k} - \psi_j)^2}{2\left(\phi_{\psi}^{*}\right)^2}\right) \times \left\{\prod_{d = 1}^{\delta_{j, k}^{(\psi)}} \exp\left(\frac{\psi_{j, k}}{2} - \frac{\psi_{j, k}^2}{2} w_{j, k}^{(\psi, d)}\right)\right\} \times \left\{\prod_{\bar{d} = 1}^{\bar{\delta}_{j, k}^{(\psi)}} \exp\left(\frac{-\psi_{j, k}}{2} - \frac{\psi_{j, k}^2}{2} \bar{w}_{j, k}^{(\psi, \bar{d})}\right)\right\} \\
        \propto \ & \exp\left\{-\frac{\left(\psi_{j, k} - \mu_{j, k}^{(\psi)}\right)^2}{2\left(\tau_{j, k}^{(\psi)}\right)^2}\right\},
    \end{aligned}
\end{equation}
where
\begin{equation}
    \mu_{j, k}^{(\psi)} = \frac{\psi_j / \left(\phi_{\psi}^{*}\right)^2 + \left(\delta_{j, k}^{(\psi)} - \bar{\delta}_{j, k}^{(\psi)}\right)/ 2}{1  / \left(\phi_{\psi}^{*}\right)^2 + \sum_{d = 1}^{\delta_{j, k}^{(\psi)}} w_{j, k}^{(\psi, d)} + \sum_{\bar{d} = 1}^{\bar{\delta}_{j, k}^{(\psi, \bar{d})}} \bar{w}_{j, k}^{(\psi, \bar{d})}}, \ \left(\tau_{j, k}^{(\psi)}\right)^2 = \frac{1}{1  / \left(\phi_{\psi}^{*}\right)^2 + \sum_{d = 1}^{\delta_{j, k}^{(\psi)}} w_{j, k}^{(\psi, d)} + \sum_{\bar{d} = 1}^{\bar{\delta}_{j, k}^{(\psi)}} \bar{w}_{j, k}^{(\psi, \bar{d})}}.
\end{equation}

%%%%% Section A2 %%%%%
\section{Additional results for simulation studies} \label{sec:_additional_simulation}
Here we provide additional results for simulation studies when different priors are adopted. Just like we've shown in Section \ref{sec:simulation}, all MSEs are in units of $10^{-3}$ and \#Anns represents the number of annotations for each recording. Additionally, in our experiments, we found that when the number of annotations per recording is relatively low, our models' performances in terms of annotation aggregation and estimation of annotators' identification expertise would be less than satisfactory if we reduced the sample size of the occurrence probabilities' prior.

\subsection{Species identification} \label{sec:_additional_simulation_agg}
{
\tiny
\setlength{\tabcolsep}{1.2pt}
\renewcommand{\arraystretch}{1.0}
\begin{longtable}{c|c|c|ccccc|ccccc|ccccc|ccccc}
\caption{AUCs of different methods under four scenarios with different priors for annotators' TPRs and occurrence probabilities.} \\
\label{table:agg_auc}
\multirow{2}{*}{Scen.} & \multirow{2}{*}{$\frac{a_p}{a_p + b_p}$} & Method & \multicolumn{5}{c|}{\textbf{Base}} & \multicolumn{5}{c|}{\textbf{Base-Hierarchical}} & \multicolumn{5}{c|}{\textbf{DP-BMM}} & \multicolumn{5}{c}{\textbf{DP-BMM-Hierarchical}} \\
\cline{3-23}
& & \#Anns & 0.8 & 1.6 & 2.4 & 3.2 & 4.0 & 0.8 & 1.6 & 2.4 & 3.2 & 4.0 & 0.8 & 1.6 & 2.4 & 3.2 & 4.0 & 0.8 & 1.6 & 2.4 & 3.2 & 4.0 \\
\hline
% Scenario 1
%% 0.01
\multirow{8}{*}{1} & \multirow{4}{*}{0.01} & 0.75 & 0.862 & 0.937 & 0.976 & 0.993 & 0.997 & 0.864 & 0.939 & 0.976 & 0.993 & 0.997 & 0.863 & 0.937 & 0.976 & 0.993 & 0.997 & 0.866 & 0.939 & 0.976 & 0.993 & 0.997  \\
 & & 0.78 & 0.863 & 0.938 & 0.976 & 0.993 & 0.997 & 0.866 & 0.940 & 0.976 & 0.993 & 0.996 & 0.863 & 0.938 & 0.976 & 0.993 & 0.997 & 0.865 & 0.939 & 0.976 & 0.993 & 0.997 \\
 & & 0.84 & 0.865 & 0.939 & 0.977 & 0.993 & 0.997 & 0.866 & 0.941 & 0.976 & 0.993 & 0.997 & 0.864 & 0.939 & 0.977 & 0.993 & 0.997 & 0.866 & 0.940 & 0.977 & 0.993 & 0.996 \\
 & & 0.87 & 0.866 & 0.940 & 0.977 & 0.993 & 0.997 & 0.867 & 0.941 & 0.977 & 0.993 & 0.996 & 0.866 & 0.940 & 0.977 & 0.993 & 0.997 & 0.867 & 0.941 & 0.977 & 0.993 & 0.996 \\
\cline{2-23}
%% 0.03
 & \multirow{4}{*}{0.03} & 0.75 & 0.862 & 0.936 & 0.976 & 0.993 & 0.997 & 0.865 & 0.939 & 0.976 & 0.993 & 0.997 & 0.862 & 0.936 & 0.976 & 0.993 & 0.997 & 0.865 & 0.939 & 0.976 & 0.993 & 0.996 \\
 & & 0.78 & 0.863 & 0.936 & 0.976 & 0.993 & 0.997 & 0.864 & 0.938 & 0.976 & 0.993 & 0.997 & 0.862 & 0.937 & 0.976 & 0.993 & 0.997 & 0.865 & 0.939 & 0.976 & 0.993 & 0.997 \\
 & & 0.84 & 0.864 & 0.938 & 0.976 & 0.993 & 0.997 & 0.864 & 0.940 & 0.976 & 0.993 & 0.997 & 0.863 & 0.938 & 0.977 & 0.993 & 0.997 & 0.864 & 0.940 & 0.977 & 0.993 & 0.996 \\
 & & 0.87 & 0.865 & 0.938 & 0.977 & 0.993 & 0.997 & 0.868 & 0.952 & 0.978 & 0.993 & 0.998 & 0.864 & 0.938 & 0.977 & 0.993 & 0.997 & 0.864 & 0.940 & 0.977 & 0.993 & 0.997 \\
\hline
% Scenario 2
%% 0.01
\multirow{8}{*}{2} & \multirow{4}{*}{0.01} & 0.75 & 0.872 & 0.952 & 0.977 & 0.993 & 0.997 & 0.870 & 0.953 & 0.977 & 0.994 & 0.997 & 0.873 & 0.957 & 0.981 & 0.996 & 0.999 & 0.870 & 0.955 & 0.981 & 0.994 & 0.999 \\
 & & 0.78 & 0.873 & 0.952 & 0.977 & 0.993 & 0.997 & 0.871 & 0.953 & 0.977 & 0.993 & 0.997 & 0.874 & 0.958 & 0.981 & 0.995 & 0.999 & 0.871 & 0.957 & 0.981 & 0.996 & 0.999 \\
 & & 0.84 & 0.875 & 0.953 & 0.978 & 0.994 & 0.998 & 0.871 & 0.953 & 0.977 & 0.994 & 0.998 & 0.875 & 0.959 & 0.981 & 0.995 & 0.999 & 0.872 & 0.958 & 0.981 & 0.995 & 0.999 \\
 & & 0.87 & 0.866 & 0.940 & 0.977 & 0.993 & 0.997 & 0.871 & 0.954 & 0.977 & 0.993 & 0.997 & 0.876 & 0.960 & 0.981 & 0.996 & 0.999 & 0.874 & 0.958 & 0.981 & 0.996 & 0.999 \\
\cline{2-23}
%% 0.03
 & \multirow{4}{*}{0.03} & 0.75 & 0.872 & 0.951 & 0.977 & 0.993 & 0.997 & 0.869 & 0.952 & 0.977 & 0.993 & 0.997 & 0.873 & 0.954 & 0.980 & 0.995 & 0.998 & 0.872 & 0.953 & 0.979 & 0.994 & 0.998 \\
 & & 0.78 & 0.872 & 0.952 & 0.977 & 0.993 & 0.997 & 0.871 & 0.953 & 0.977 & 0.993 & 0.997 & 0.873 & 0.954 & 0.980 & 0.995 & 0.998 & 0.872 & 0.953 & 0.978 & 0.994 & 0.998 \\
 & & 0.84 & 0.874 & 0.953 & 0.978 & 0.994 & 0.998 & 0.869 & 0.953 & 0.978 & 0.993 & 0.998 & 0.875 & 0.954 & 0.980 & 0.995 & 0.998 & 0.871 & 0.954 & 0.980 & 0.994 & 0.997 \\
 & & 0.87 & 0.875 & 0.953 & 0.978 & 0.994 & 0.998 & 0.868 & 0.952 & 0.978 & 0.993 & 0.998 & 0.876 & 0.954 & 0.980 & 0.995 & 0.998 & 0.869 & 0.954 & 0.980 & 0.994 & 0.998 \\
\hline
% Scenario 3
%% 0.01
\multirow{8}{*}{3} & \multirow{4}{*}{0.01} & 0.75 & 0.826 & 0.895 & 0.949 & 0.975 & 0.984 & 0.832 & 0.902 & 0.957 & 0.983 & 0.990 & 0.827 & 0.896 & 0.949 & 0.975 & 0.984 & 0.833 & 0.902 & 0.955 & 0.980 & 0.989 \\
 & & 0.78 & 0.827 & 0.897 & 0.950 & 0.976 & 0.984 & 0.833 & 0.902 & 0.957 & 0.981 & 0.991 & 0.827 & 0.897 & 0.950 & 0.975 & 0.984 & 0.831 & 0.902 & 0.958 & 0.981 & 0.988 \\
 & & 0.84 & 0.830 & 0.899 & 0.951 & 0.976 & 0.984 & 0.836 & 0.902 & 0.958 & 0.983 & 0.989 & 0.829 & 0.899 & 0.951 & 0.976 & 0.984 & 0.837 & 0.904 & 0.957 & 0.983 & 0.988 \\
 & & 0.87 & 0.831 & 0.900 & 0.952 & 0.976 & 0.984 & 0.836 & 0.904 & 0.957 & 0.982 & 0.987 & 0.830 & 0.900 & 0.952 & 0.976 & 0.984 & 0.838 & 0.904 & 0.955 & 0.983 & 0.986 \\
\cline{2-23}
%% 0.03
 & \multirow{4}{*}{0.03} & 0.75 & 0.826 & 0.893 & 0.947 & 0.975 & 0.984 & 0.828 & 0.905 & 0.957 & 0.982 & 0.989 & 0.825 & 0.893 & 0.947 & 0.975 & 0.984 & 0.835 & 0.905 & 0.957 & 0.984 & 0.988 \\
 & & 0.78 & 0.825 & 0.894 & 0.949 & 0.975 & 0.984 & 0.830 & 0.906 & 0.957 & 0.982 & 0.987 & 0.826 & 0.894 & 0.948 & 0.976 & 0.984 & 0.834 & 0.908 & 0.958 & 0.980 & 0.987 \\
 & & 0.84 & 0.828 & 0.896 & 0.950 & 0.976 & 0.984 & 0.829 & 0.907 & 0.958 & 0.983 & 0.989 & 0.828 & 0.896 & 0.950 & 0.976 & 0.984 & 0.830 & 0.908 & 0.957 & 0.984 & 0.988 \\
 & & 0.87 & 0.829 & 0.897 & 0.951 & 0.976 & 0.984 & 0.829 & 0.908 & 0.959 & 0.984 & 0.990 & 0.828 & 0.897 & 0.952 & 0.976 & 0.984 & 0.833 & 0.908 & 0.958 & 0.984 & 0.987 \\
\hline
% Scenario 4
%% 0.01
\multirow{8}{*}{4} & \multirow{4}{*}{0.01} & 0.75 & 0.800 & 0.913 & 0.956 & 0.972 & 0.982 & 0.842 & 0.929 & 0.963 & 0.976 & 0.985 & 0.801 & 0.913 & 0.960 & 0.975 & 0.986 & 0.843 & 0.931 & 0.966 & 0.977 & 0.988 \\
 & & 0.78 & 0.802 & 0.916 & 0.957 & 0.973 & 0.982 & 0.840 & 0.929 & 0.963 & 0.977 & 0.978 & 0.803 & 0.916 & 0.963 & 0.975 & 0.986 & 0.843 & 0.932 & 0.967 & 0.979 & 0.987 \\
 & & 0.84 & 0.807 & 0.920 & 0.959 & 0.973 & 0.983 & 0.842 & 0.930 & 0.963 & 0.979 & 0.987 & 0.809 & 0.922 & 0.964 & 0.976 & 0.986 & 0.846 & 0.935 & 0.968 & 0.980 & 0.982 \\
 & & 0.87 & 0.815 & 0.922 & 0.959 & 0.973 & 0.983 & 0.842 & 0.931 & 0.964 & 0.977 & 0.983 & 0.814 & 0.925 & 0.964 & 0.978 & 0.986 & 0.843 & 0.934 & 0.968 & 0.980 & 0.986 \\
\cline{2-23}
%% 0.03
 & \multirow{4}{*}{0.03} & 0.75 & 0.795 & 0.905 & 0.954 & 0.972 & 0.983 & 0.831 & 0.931 & 0.962 & 0.975 & 0.982 & 0.797 & 0.905 & 0.954 & 0.973 & 0.985 & 0.837 & 0.930 & 0.963 & 0.980 & 0.988 \\
 & & 0.78 & 0.799 & 0.907 & 0.955 & 0.973 & 0.983 & 0.837 & 0.930 & 0.963 & 0.978 & 0.985 & 0.797 & 0.907 & 0.957 & 0.973 & 0.985 & 0.838 & 0.931 & 0.962 & 0.978 & 0.986 \\
 & & 0.84 & 0.799 & 0.912 & 0.958 & 0.974 & 0.983 & 0.839 & 0.929 & 0.963 & 0.977 & 0.986 & 0.799 & 0.912 & 0.960 & 0.976 & 0.985 & 0.835 & 0.930 & 0.968 & 0.978 & 0.987 \\
 & & 0.87 & 0.803 & 0.916 & 0.959 & 0.974 & 0.983 & 0.839 & 0.931 & 0.964 & 0.977 & 0.986 & 0.802 & 0.916 & 0.960 & 0.976 & 0.985 & 0.833 & 0.932 & 0.968 & 0.978 & 0.984 \\
\hline
\end{longtable}
}

\subsection{Assessment of sound identification expertise}
Here we provide our models' performances in terms of inferring annotators' TPRs in Tables \ref{table:ability_coverage} and \ref{table:ability_mse}. For the sake of completeness, we also provide the coverages as well as the MSEs of annotators' FPRs of our models in Tables \ref{table:fpr_coverage} and \ref{table:fpr_mse}. Though our models could achieve relatively high coverages and low MSEs of annotators' TPRs, they didn't do well in inferring annotators' FPRs.

\label{sec:_additional_simulation_ability}
% tpr
%% - coverage
{
\tiny
\setlength{\tabcolsep}{2.0pt}
\renewcommand{\arraystretch}{1.0}
\begin{longtable}{c|c|c|ccccc|ccccc|ccccc|ccccc}
\caption{The coverages for annotators' TPRs of different methods under four scenarios with different priors for TPRs and occurrence probabilities.} \\
\label{table:ability_coverage}
\multirow{2}{*}{Scen.} & \multirow{2}{*}{$\frac{a_p}{a_p + b_p}$} & Method & \multicolumn{5}{c|}{\textbf{Base}} & \multicolumn{5}{c|}{\textbf{Base-Hierarchical}} & \multicolumn{5}{c|}{\textbf{DP-BMM}} & \multicolumn{5}{c}{\textbf{DP-BMM-Hierarchical}} \\
\cline{3-23}
& & \#Anns & 0.8 & 1.6 & 2.4 & 3.2 & 4.0 & 0.8 & 1.6 & 2.4 & 3.2 & 4.0 & 0.8 & 1.6 & 2.4 & 3.2 & 4.0 & 0.8 & 1.6 & 2.4 & 3.2 & 4.0 \\
\hline
% Scenario 1
%% 0.01
\multirow{8}{*}{1} & \multirow{4}{*}{0.01} & 0.75 & 0.50 & 0.45 & 0.60 & 0.90 & 0.95 & 0.80 & 0.75 & 0.80 & 0.95 & 0.85 & 0.45 & 0.45 & 0.65 & 0.95 & 0.95 & 0.80 & 0.75 & 0.75 & 0.80 & 0.80 \\
 & & 0.78 & 0.55 & 0.55 & 0.75 & 1.00 & 0.95 & 0.90 & 0.90 & 0.85 & 0.90 & 0.90 & 0.60 & 0.60 & 0.75 & 0.95 & 0.95 & 0.85 & 0.75 & 0.75 & 0.95 & 0.90 \\
 & & 0.84 & 0.80 & 0.95 & 0.90 & 1.00 & 0.95 & 1.00 & 1.00 & 0.95 & 1.00 & 0.95 & 0.80 & 1.00 & 0.90 & 1.00 & 0.95 &  0.95 & 0.95 & 0.95 & 1.00 & 0.95 \\
 & & 0.87 & 0.90 & 1.00 & 0.90 & 1.00 & 0.95 & 1.00 & 0.95 & 0.95 & 0.75 & 1.00 & 0.95 & 1.00 & 0.90 & 1.00 & 0.95 & 0.90 & 0.95 & 0.95 & 1.00 & 0.95 \\
\cline{2-23}
%% 0.03
 & \multirow{4}{*}{0.03} & 0.75 & 0.35 & 0.35 & 0.45 & 0.75 & 0.95 & 0.75 & 0.70 & 0.80 & 0.75 & 0.85 & 0.40 & 0.35 & 0.45 & 0.85 & 0.95 & 0.70 & 0.60 & 0.75 & 0.85 & 0.55 \\
 & & 0.78 & 0.45 & 0.45 & 0.55 & 0.90 & 0.95 & 0.80 & 0.60 & 0.85 & 0.95 & 0.85 & 0.45 & 0.40 & 0.55 & 0.90 & 0.95 & 0.75 & 0.75 & 0.60 & 0.90 & 0.95 \\
 & & 0.84 & 0.50 & 0.65 & 0.85 & 1.00 & 0.95 & 0.90 & 0.95 & 0.95 & 0.90 & 0.95 & 0.50 & 0.65 & 0.80 & 1.00 & 0.95 & 0.95 & 0.90 & 0.90 & 1.00 & 0.90 \\
 & & 0.87 & 0.70 & 0.80 & 0.90 & 1.00 & 0.95 & 0.95 & 1.00 & 0.95 & 1.00 & 0.95 & 0.70 & 0.85 & 0.90 & 1.00 & 0.95 & 0.95 & 1.00 & 0.95 & 1.00 & 0.90 \\
\hline
% Scenario 2
%% 0.01
\multirow{8}{*}{2} & \multirow{4}{*}{0.01} & 0.75 & 0.40 & 0.60 & 0.75 & 0.90 & 0.90 & 0.70 & 0.85 & 0.85 & 0.90 & 0.85 & 0.40 & 0.70 & 0.85 & 0.90 & 0.90 & 0.80 & 0.80 & 0.85 & 0.85 & 0.90 \\
 & & 0.78 & 0.60 & 0.80 & 0.85 & 0.90 & 0.95 & 0.85 & 0.65 & 0.80 & 0.90 & 0.95 & 0.60 & 0.95 & 0.85 & 0.95 & 0.90 & 0.80 & 0.90 & 0.80 & 0.90 & 1.00 \\
 & & 0.84 & 0.70 & 1.00 & 0.85 & 1.00 & 1.00 & 0.95 & 0.90 & 0.90 & 0.90 & 0.95 & 0.70 & 1.00 & 0.90 & 1.00 & 1.00 & 0.95 & 0.95 & 0.90 & 1.00 & 1.00 \\
 & & 0.87 & 0.85 & 1.00 & 0.90 & 1.00 & 1.00 & 0.95 & 0.80 & 0.95 & 1.00 & 0.90 & 0.85 & 0.95 & 0.90 & 1.00 & 1.00 & 1.00 & 0.85 & 0.95 & 0.95 & 0.85 \\
\cline{2-23}
%% 0.03
 & \multirow{4}{*}{0.03} & 0.75 & 0.20 & 0.45 & 0.70 & 0.90 & 0.75 & 0.60 & 0.55 & 0.80 & 0.80 & 0.85 & 0.20 & 0.45 & 0.65 & 0.90 & 0.75 & 0.65 & 0.65 & 0.80 & 0.85 & 0.75 \\
 & & 0.78 & 0.40 & 0.45 & 0.75 & 0.90 & 0.90 & 0.55 & 0.70 & 0.85 & 0.85 & 0.85 & 0.35 & 0.45 & 0.70 & 0.90 & 0.85 & 0.80 & 0.90 & 0.85 & 0.80 & 0.90 \\
 & & 0.84 & 0.60 & 0.90 & 0.85 & 0.95 & 1.00 & 0.90 & 0.95 & 0.90 & 0.90 & 1.00 & 0.40 & 0.80 & 0.85 & 0.90 & 1.00 & 0.90 & 0.90 & 0.90 & 0.90 & 1.00 \\
 & & 0.87 & 0.65 & 0.95 & 0.90 & 1.00 & 1.00 & 0.95 & 0.75 & 1.00 & 1.00 & 0.95 & 0.65 & 0.95 & 0.90 & 1.00 & 1.00 & 0.95 & 0.95 & 1.00 & 1.00 & 1.00 \\
\hline
% Scenario 3
%% 0.01
\multirow{8}{*}{3} & \multirow{4}{*}{0.01} & 0.75 & 0.20 & 0.15 & 0.20 & 0.25 & 0.40 & 0.80 & 0.85 & 0.85 & 0.85 & 0.85 & 0.20 & 0.20 & 0.20 & 0.25 & 0.40 & 0.75 & 0.85 & 0.85 & 0.85 & 0.85 \\
 & & 0.78 & 0.30 & 0.20 & 0.30 & 0.40 & 0.40 & 0.80 & 0.85 & 0.90 & 0.90 & 0.90 & 0.30 & 0.20 & 0.30 & 0.35 & 0.40 & 0.80 & 0.90 & 0.85 & 0.90 & 0.85 \\
 & & 0.84 & 0.45 & 0.45 & 0.60 & 0.55 & 0.55 & 1.00 & 1.00 & 0.95 & 0.95 & 1.00 & 0.35 & 0.40 & 0.60 & 0.55 & 0.55 & 0.95 & 1.00 & 0.95 & 0.95 & 1.00 \\
 & & 0.87 & 0.55 & 0.65 & 0.70 & 0.65 & 0.60 & 1.00 & 1.00 & 1.00 & 1.00 & 1.00 & 0.50 & 0.65 & 0.65 & 0.65 & 0.60 & 1.00 & 1.00 & 1.00 & 1.00 & 1.00 \\
\cline{2-23}
%% 0.03
 & \multirow{4}{*}{0.03} & 0.75 & 0.15 & 0.05 & 0.00 & 0.10 & 0.35 & 0.75 & 0.80 & 0.80 & 0.85 & 0.85 & 0.15 & 0.05 & 0.00 & 0.10 & 0.35 & 0.75 & 0.80 & 0.80 & 0.85 & 0.85 \\
 & & 0.78 & 0.15 & 0.10 & 0.00 & 0.15 & 0.35 & 0.80 & 0.85 & 0.90 & 0.90 & 0.85 & 0.15 & 0.10 & 0.00 & 0.20 & 0.35 & 0.80 & 0.85 & 0.85 & 0.85 & 0.85 \\
 & & 0.84 & 0.25 & 0.15 & 0.15 & 0.40 & 0.50 & 1.00 & 1.00 & 0.95 & 0.95 & 0.95 & 0.25 & 0.15 & 0.20 & 0.40 & 0.45 & 1.00 & 1.00 & 0.95 & 0.95 & 0.95 \\
 & & 0.87 & 0.30 & 0.20 & 0.35 & 0.50 & 0.55 & 1.00 & 1.00 & 1.00 & 1.00 & 1.00 & 0.30 & 0.20 & 0.40 & 0.50 & 0.45 & 1.00 & 1.00 & 1.00 & 1.00 & 1.00 \\
\hline
% Scenario 4
%% 0.01
\multirow{8}{*}{4} & \multirow{4}{*}{0.01} & 0.75 & 0.25 & 0.15 & 0.45 & 0.40 & 0.55 & 0.80 & 0.80 & 0.85 & 0.85 & 0.85 & 0.25 & 0.15 & 0.45 & 0.40 & 0.55 & 0.85 & 0.80 & 0.85 & 0.85 & 0.85 \\
 & & 0.78 & 0.30 & 0.30 & 0.45 & 0.50 & 0.60 & 0.85 & 0.85 & 0.90 & 0.90 & 0.85 & 0.25 & 0.30 & 0.55 & 0.60 & 0.60 & 0.85 & 0.85 & 0.85 & 0.90 & 0.85 \\
 & & 0.84 & 0.45 & 0.45 & 0.60 & 0.55 & 0.55 & 1.00 & 1.00 & 0.95 & 0.95 & 1.00 & 0.45 & 0.70 & 0.70 & 0.85 & 0.70 & 0.95 & 1.00 & 0.95 & 0.90 & 0.90 \\
 & & 0.87 & 0.70 & 0.70 & 0.65 & 0.85 & 0.70 & 1.00 & 1.00 & 0.95 & 1.00 & 1.00 & 0.65 & 0.80 & 0.70 & 0.85 & 0.75 & 1.00 & 0.95 & 1.00 & 1.00 & 1.00 \\
\cline{2-23}
%% 0.03
 & \multirow{4}{*}{0.03} & 0.75 & 0.15 & 0.00 & 0.05 & 0.10 & 0.45 & 0.80 & 0.75 & 0.75 & 0.80 & 0.85 & 0.10 & 0.00 & 0.15 & 0.15 & 0.45 & 0.80 & 0.75 & 0.75 & 0.80 & 0.85 \\
 & & 0.78 & 0.10 & 0.00 & 0.20 & 0.20 & 0.55 & 0.80 & 0.75 & 0.85 & 0.90 & 0.85 & 0.20 & 0.00 & 0.20 & 0.15 & 0.50 & 0.80 & 0.85 & 0.85 & 0.85 & 0.85 \\
 & & 0.84 & 0.25 & 0.10 & 0.45 & 0.50 & 0.60 & 1.00 & 0.95 & 0.95 & 0.90 & 0.90 & 0.25 & 0.10 & 0.45 & 0.40 & 0.60 & 1.00 & 0.90 & 0.90 & 0.90 & 0.85 \\
 & & 0.87 & 0.30 & 0.30 & 0.50 & 0.75 & 0.65 & 1.00 & 1.00 & 1.00 & 0.95 & 1.00 & 0.25 & 0.25 & 0.50 & 0.60 & 0.65 & 1.00 & 1.00 & 1.00 & 0.95 & 0.95 \\
\hline
\end{longtable}
}

%% - MSE
{
\tiny
\setlength{\tabcolsep}{2.0pt}
\renewcommand{\arraystretch}{1.0}
\begin{longtable}{c|c|c|ccccc|ccccc|ccccc|ccccc}
\caption{The MSEs for annotators' TPRs of different methods under four scenarios with different priors for TPRs and occurrence probabilities.} \\
\label{table:ability_mse}
\multirow{2}{*}{Scen.} & \multirow{2}{*}{$\frac{a_p}{a_p + b_p}$} & Method & \multicolumn{5}{c|}{\textbf{Base}} & \multicolumn{5}{c|}{\textbf{Base-Hierarchical}} & \multicolumn{5}{c|}{\textbf{DP-BMM}} & \multicolumn{5}{c}{\textbf{DP-BMM-Hierarchical}} \\
\cline{3-23}
& & \#Anns & 0.8 & 1.6 & 2.4 & 3.2 & 4.0 & 0.8 & 1.6 & 2.4 & 3.2 & 4.0 & 0.8 & 1.6 & 2.4 & 3.2 & 4.0 & 0.8 & 1.6 & 2.4 & 3.2 & 4.0 \\
\hline
% Scenario 1
%% 0.01
\multirow{8}{*}{1} & \multirow{4}{*}{0.01} & 0.75 & 29.6 & 18.9 & 10.1 & 2.18 & 2.79 & 10.3 & 7.40 & 5.82 & 1.98 & 2.51 & 30.1 & 19.3 & 10.5 & 2.19 & 2.81 & 9.03 & 7.58 & 7.33 & 2.72 & 2.62 \\
 & & 0.78 & 21.8 & 13.8 & 7.60 & 1.61 & 2.20 & 7.44 & 5.35 & 4.59 & 1.77 & 2.38 & 21.3 & 12.6 & 7.60 & 1.67 & 2.19 & 8.08 & 6.69 & 4.19 & 1.52 & 1.61 \\
 & & 0.84 & 10.2 & 5.92 & 4.49 & 1.35 & 1.49 & 5.17 & 3.07 & 3.28 & 1.58 & 2.47 & 9.99 & 5.83 & 4.43 & 1.30 & 1.50 & 5.67 & 3.67 & 3.39 & 1.64 & 1.70 \\
 & & 0.87 & 7.35 & 4.23 & 3.49 & 1.39 & 1.28 & 6.75 & 3.04 & 3.32 & 2.62 & 2.11 & 7.15 & 4.02 & 3.56 & 1.41 & 1.31 & 5.42 & 3.47 & 3.80 & 1.74 & 2.08 \\
\cline{2-23}
%% 0.03
 & \multirow{4}{*}{0.03} & 0.75 & 55.3 & 37.8 & 16.4 & 2.92 & 3.58 & 16.0 & 11.9 & 8.72 & 2.35 & 3.69 & 52.4 & 32.2 & 17.3 & 3.01 & 3.57 & 15.7 & 14.7 & 8.83 & 2.43 & 4.66 \\
 & & 0.78 & 40.9 & 26.7 & 12.9 & 2.04 & 2.81 & 13.2 & 15.0 & 5.78 & 1.90 & 4.02 & 43.5 & 26.4 & 12.6 & 2.20 & 2.80 & 12.9 & 10.1 & 10.2 & 1.36 & 2.57 \\
 & & 0.84 & 24.9 & 13.2 & 7.04 & 1.29 & 1.88 & 6.40 & 5.23 & 3.23 & 1.47 & 1.74 & 23.9 & 13.2 & 7.04 & 1.35 & 1.86 & 5.83 & 4.45 & 3.44 & 1.43 & 1.97 \\
 & & 0.87 & 16.1 & 9.58 & 5.13 & 1.23 & 1.53 & 5.38 & 4.23 & 3.36 & 1.51 & 2.49 & 15.5 & 8.67 & 4.80 & 1.30 & 1.62 & 5.55 & 4.42 & 3.43 & 1.48 & 2.32 \\
\hline
% Scenario 2
%% 0.01
\multirow{8}{*}{2} & \multirow{4}{*}{0.01} & 0.75 & 45.0 & 12.9 & 7.56 & 3.23 & 3.16 & 15.3 & 6.10 & 4.66 & 3.21 & 4.17 & 45.1 & 10.6 & 6.02 & 2.95 & 3.14 & 12.5 & 6.11 & 4.27 & 3.90 & 2.53 \\
 & & 0.78 & 36.2 & 9.09 & 5.49 & 2.48 & 2.57 & 9.51 & 8.80 & 4.19 & 2.85 & 2.01 & 33.1 & 6.63 & 4.46 & 2.29 & 2.67 & 7.46 & 4.99 & 3.60 & 3.14 & 2.29 \\
 & & 0.84 & 17.3 & 3.82 & 3.51 & 1.89 & 1.67 & 4.87 & 4.38 & 3.55 & 2.61 & 1.53 & 17.7 & 3.28 & 3.24 & 1.88 & 1.80 & 4.42 & 3.56 & 3.92 & 2.49 & 1.81 \\
 & & 0.87 & 10.6 & 3.42 & 3.45 & 2.05 & 1.64 & 5.07 & 5.61 & 4.28 & 2.73 & 1.81 & 11.8 & 3.12 & 3.49 & 2.06 & 1.73 & 5.47 & 5.69 & 5.54 & 3.70 & 1.81 \\
\cline{2-23}
%% 0.03
 & \multirow{4}{*}{0.03} & 0.75 & 77.1 & 28.6 & 12.6 & 5.30 & 4.09 & 32.6 & 13.6 & 7.74 & 6.37 & 3.81 & 75.3 & 27.5 & 12.6 & 4.94 & 4.37 & 31.2 & 9.95 & 9.42 & 4.31 & 4.23 \\
 & & 0.78 & 64.8 & 20.8 & 9.03 & 3.71 & 3.12 & 29.8 & 7.12 & 6.00 & 3.14 & 3.24 & 64.8 & 20.5 & 9.38 & 3.73 & 3.24 & 17.2 & 5.21 & 6.08 & 3.71 & 2.40 \\
 & & 0.84 & 42.4 & 8.64 & 4.18 & 2.14 & 1.98 & 9.57 & 6.02 & 3.23 & 2.75 & 1.44 & 43.4 & 10.7 & 4.41 & 2.28 & 2.30 & 9.82 & 5.64 & 3.13 & 2.38 & 1.53 \\
 & & 0.87 & 31.8 & 5.78 & 3.49 & 1.95 & 1.78 & 7.88 & 8.59 & 3.76 & 2.13 & 1.76 & 31.1 & 6.75 & 3.35 & 2.04 & 1.97 & 7.66 & 4.46 & 3.76 & 2.84 & 1.85 \\
\hline
% Scenario 3
%% 0.01
\multirow{8}{*}{3} & \multirow{4}{*}{0.01} & 0.75 & 93.5 & 79.0 & 43.7 & 23.2 & 14.9 & 10.6 & 8.76 & 6.36 & 6.48 & 3.84 & 94.1 & 71.8 & 42.3 & 23.7 & 14.9 & 10.9 & 8.51 & 7.00 & 5.85 & 3.82 \\
 & & 0.78 & 80.6 & 60.3 & 36.3 & 20.5 & 12.3 & 7.14 & 6.15 & 4.49 & 3.71 & 2.90 & 81.4 & 58.9 & 33.1 & 19.6 & 12.4 & 8.00 & 5.71 & 4.77 & 4.32 & 3.30 \\
 & & 0.84 & 56.7 & 34.2 & 20.0 & 13.2 & 8.69 & 4.86 & 3.23 & 3.50 & 1.91 & 2.45 & 64.9 & 33.9 & 20.1 & 12.8 & 8.9 & 5.36 & 3.32 & 3.09 & 1.82 & 2.55 \\
 & & 0.87 & 49.7 & 20.9 & 13.6 & 10.0 & 7.26 & 6.11 & 4.31 & 4.47 & 1.79 & 3.12 & 49.8 & 22.3 & 13.6 & 10.3 & 6.94 & 6.41 & 3.82 & 4.06 & 1.97 & 2.44 \\
\cline{2-23}
%% 0.03
 & \multirow{4}{*}{0.03} & 0.75 & 117 & 134 & 83.5 & 36.5 & 19.9 & 13.4 & 13.6 & 11.2 & 9.44 & 4.89 & 117 & 129 & 82.8 & 36.3 & 20.1 & 14.5 & 12.7 & 10.1 & 9.32 & 5.47 \\
 & & 0.78 & 107 & 114 & 68.0 & 30.3 & 16.6 & 9.81 & 8.27 & 5.69 & 5.44 & 3.94 & 109 & 114 & 69.2 & 30.2 & 16.9 & 8.37 & 7.82 & 6.00 & 6.17 & 4.51 \\
 & & 0.84 & 93.1 & 79.2 & 42.3 & 20.5 & 11.8 & 5.28 & 2.88 & 3.14 & 2.27 & 2.53 & 90.3 & 82.2 & 43.0 & 20.1 & 11.9 & 5.35 & 2.96 & 2.73 & 2.89 & 2.56 \\
 & & 0.87 & 80.3 & 64.8 & 32.7 & 17.4 & 9.36 & 5.64 & 3.09 & 3.24 & 2.10 & 2.71 & 81.3 & 65.2 & 29.5 & 16.6 & 9.54 & 5.79 & 2.80 & 3.48 & 2.00 & 2.71 \\
\hline
% Scenario 4
%% 0.01
\multirow{8}{*}{4} & \multirow{4}{*}{0.01} & 0.75 & 84.7 & 71.1 & 37.8 & 23.9 & 15.5 & 9.12 & 11.9 & 6.95 & 8.64 & 6.62 & 86.6 & 70.4 & 32.6 & 19.9 & 14.2 & 7.79 & 11.6 & 7.03 & 9.50 & 7.22 \\
 & & 0.78 & 72.9 & 52.0 & 30.5 & 18.9 & 13.3 & 5.62 & 9.55 & 4.72 & 5.91 & 5.54 & 75.1 & 50.0 & 21.2 & 16.1 & 12.9 & 5.04 & 7.67 & 4.58 & 5.76 & 5.90 \\
 & & 0.84 & 51.9 & 24.9 & 15.9 & 10.9 & 10.1 & 3.59 & 5.80 & 3.07 & 3.81 & 4.28 & 51.8 & 20.1 & 12.3 & 10.1 & 9.74 & 3.41 & 5.39 & 3.19 & 4.00 & 4.06 \\
 & & 0.87 & 36.8 & 13.7 & 13.0 & 7.84 & 8.62 & 5.52 & 6.69 & 4.22 & 4.10 & 4.36 & 37.1 & 11.1 & 9.15 & 7.44 & 8.46 & 4.98 & 7.00 & 3.51 & 4.03 & 4.57 \\
\cline{2-23}
%% 0.03
 & \multirow{4}{*}{0.03} & 0.75 & 112 & 144 & 83.1 & 43.2 & 20.3 & 11.4 & 16.6 & 11.2 & 13.3 & 9.13 & 113 & 140 & 80.0 & 41.3 & 21.5 & 12.4 & 17.7 & 10.3 & 12.5 & 8.60 \\
 & & 0.78 & 103 & 127 & 64.3 & 33.3 & 18.0 & 8.70 & 12.3 & 7.13 & 7.17 & 7.32 & 104 & 128 & 61.4 & 33.9 & 18.3 & 7.48 & 10.5 & 8.29 & 8.47 & 6.28 \\
 & & 0.84 & 87.3 & 89.8 & 35.4 & 20.3 & 12.6 & 3.60 & 5.36 & 4.40 & 4.37 & 4.23 & 85.3 & 91.7 & 34.6 & 21.0 & 12.7 & 3.53 & 5.89 & 3.78 & 4.19 & 4.35 \\
 & & 0.87 & 72.8 & 57.7 & 27.4 & 15.1 & 10.7 & 4.40 & 4.79 & 3.39 & 3.53 & 4.18 & 73.6 & 61.2 & 26.6 & 15.6 & 11.1 & 4.37 & 5.50 & 3.53 & 3.90 & 4.28 \\
\hline
\end{longtable}
}

% ---------------------------------------------------
% fpr
%% - coverage
{
\tiny
\setlength{\tabcolsep}{2.0pt}
\renewcommand{\arraystretch}{1.0}
\begin{longtable}{c|c|c|ccccc|ccccc|ccccc|ccccc}
\caption{The coverages for annotators' FPRs of different methods under four scenarios with different priors for TPRs and occurrence probabilities.} \\
\label{table:fpr_coverage}
\multirow{2}{*}{Scen.} & \multirow{2}{*}{$\frac{a_p}{a_p + b_p}$} & Method & \multicolumn{5}{c|}{\textbf{Base}} & \multicolumn{5}{c|}{\textbf{Base-Hierarchical}} & \multicolumn{5}{c|}{\textbf{DP-BMM}} & \multicolumn{5}{c}{\textbf{DP-BMM-Hierarchical}} \\
\cline{3-23}
& & \#Anns & 0.8 & 1.6 & 2.4 & 3.2 & 4.0 & 0.8 & 1.6 & 2.4 & 3.2 & 4.0 & 0.8 & 1.6 & 2.4 & 3.2 & 4.0 & 0.8 & 1.6 & 2.4 & 3.2 & 4.0 \\
\hline
% Scenario 1
%% 0.01
\multirow{15}{*}{1} & \multirow{5}{*}{0.01} & 0.75 & 0.10 & 0.15 & 0.25 & 0.25 & 0.40 & 0.05 & 0.40 & 0.80 & 0.90 & 0.85 & 0.10 & 0.15 & 0.20 & 0.30 & 0.35 & 0.10 & 0.45 & 0.75 & 0.80 & 0.80 \\
 & & 0.78 & 0.10 & 0.15 & 0.25 & 0.30 & 0.35 & 0.05 & 0.45 & 0.85 & 0.80 & 0.80 & 0.10 & 0.10 & 0.25 & 0.30 & 0.35 & 0.10 & 0.45 & 0.80 & 0.75 & 0.90 \\
 & & 0.81 & 0.10 & 0.10 & 0.25 & 0.30 & 0.40 & 0.10 & 0.50 & 0.75 & 0.90 & 0.85 & 0.10 & 0.10 & 0.25 & 0.30 & 0.40 & 0.10 & 0.40 & 0.80 & 0.85 & 0.75 \\
 & & 0.84 & 0.10 & 0.15 & 0.30 & 0.30 & 0.40 & 0.10 & 0.50 & 0.80 & 0.70 & 0.75 & 0.10 & 0.10 & 0.25 & 0.30 & 0.40 & 0.10 & 0.50 & 0.80 & 0.90 & 0.90 \\
 & & 0.87 & 0.10 & 0.10 & 0.30 & 0.30 & 0.40 & 0.05 & 0.55 & 0.85 & 0.80 & 0.85 & 0.10 & 0.10 & 0.25 & 0.30 & 0.40 & 0.05 & 0.40 & 0.85 & 0.75 & 0.70 \\
\cline{2-23}
%% 0.02
 & \multirow{5}{*}{0.02} & 0.75 & 0.10 & 0.15 & 0.20 & 0.25 & 0.40 & 0.10 & 0.40 & 0.70 & 0.65 & 0.85 & 0.10 & 0.15 & 0.25 & 0.25 & 0.40 & 0.05 & 0.45 & 0.75 & 0.85 & 0.85 \\
 & & 0.78 & 0.10 & 0.15 & 0.25 & 0.25 & 0.35 & 0.10 & 0.45 & 0.85 & 0.75 & 0.80 & 0.10 & 0.15 & 0.20 & 0.25 & 0.35 & 0.10 & 0.45 & 0.75 & 0.70 & 0.85 \\
 & & 0.81 & 0.10 & 0.15 & 0.25 & 0.25 & 0.40 & 0.10 & 0.45 & 0.75 & 0.80 & 0.70 & 0.10 & 0.10 & 0.25 & 0.25 & 0.40 & 0.10 & 0.40 & 0.85 & 0.75 & 0.90 \\
 & & 0.84 & 0.10 & 0.15 & 0.25 & 0.25 & 0.35 & 0.10 & 0.40 & 0.85 & 0.75 & 0.85 & 0.10 & 0.15 & 0.20 & 0.25 & 0.35 & 0.10 & 0.45 & 0.80 & 0.80 & 0.80 \\
 & & 0.87 & 0.10 & 0.10 & 0.25 & 0.30 & 0.35 & 0.10 & 0.40 & 0.75 & 0.80 & 0.80 & 0.10 & 0.10 & 0.25 & 0.30  & 0.40 & 0.10 & 0.45 & 0.85 & 0.80 & 0.90 \\
\cline{2-23}
%% 0.03
 & \multirow{5}{*}{0.03} & 0.75 & 0.10 & 0.15 & 0.20 &  0.25 & 0.35 & 0.10 & 0.35 & 0.75 & 0.70 & 0.85 & 0.10 & 0.15 & 0.20 & 0.25 & 0.35 & 0.10 & 0.30 & 0.75 & 0.90 & 0.80 \\
 & & 0.78 & 0.10 & 0.15 & 0.20 & 0.30 & 0.40 & 0.05 & 0.40 & 0.70 & 0.85 & 0.85 & 0.10 & 0.15 & 0.20 & 0.25 & 0.40 & 0.10 & 0.35 & 0.70 & 0.75 & 0.90 \\
 & & 0.81 & 0.10 & 0.15 & 0.20 & 0.25 & 0.35 & 0.10 & 0.35 & 0.80 & 0.80 & 0.90 & 0.10 & 0.15 & 0.20 & 0.25 & 0.35 & 0.05 & 0.45 & 0.80 & 0.85 & 0.90 \\
 & & 0.84 & 0.10 & 0.15 & 0.20 & 0.25 & 0.40 & 0.05 & 0.40 & 0.65 & 0.85 & 0.90 & 0.10 & 0.15 & 0.20 & 0.25 & 0.35 & 0.10 & 0.45 & 0.70 & 0.70 & 0.90 \\
 & & 0.87 & 0.10 & 0.15 & 0.25 & 0.30 & 0.40 & 0.05 & 0.35 & 0.80 & 0.75 & 0.80 & 0.10 & 0.15 & 0.20 & 0.25 & 0.40 & 0.10 & 0.30 & 0.75 & 0.75 & 0.90 \\
\hline
% Scenario 2
%% 0.01
\multirow{15}{*}{2} & \multirow{5}{*}{0.01} & 0.75 & 0.15 & 0.15 & 0.35 & 0.25 & 0.30 & 0.10 & 0.35 & 0.60 & 0.70 & 0.75 & 0.15 & 0.15 & 0.35 & 0.25 & 0.30 & 0.10 & 0.55 & 0.55 & 0.60 & 0.80 \\
 & & 0.78 & 0.15 & 0.15 & 0.35 & 0.30 & 0.35 & 0.10 & 0.35 & 0.55 & 0.45 & 0.80 & 0.15 & 0.15 & 0.35 & 0.25 & 0.30 & 0.05 & 0.50 & 0.60 & 0.65 & 0.75 \\
 & & 0.81 & 0.15 & 0.15 & 0.35 & 0.30 & 0.35 & 0.10 & 0.45 & 0.65 & 0.60 & 0.85 & 0.15 & 0.15 & 0.35 & 0.25 & 0.30 & 0.10 & 0.35 & 0.55 & 0.55 & 0.85 \\
 & & 0.84 & 0.15 & 0.15 & 0.35 & 0.30 & 0.30 & 0.10 & 0.45 & 0.60 & 0.65 & 0.75 & 0.15 & 0.15 & 0.35 & 0.25 & 0.35 & 0.05 & 0.45 & 0.60 & 0.60 & 0.85 \\
 & & 0.87 & 0.15 & 0.15 & 0.35 & 0.35 & 0.35 & 0.10 & 0.45 & 0.60 & 0.75 & 0.75 & 0.15 & 0.20 & 0.35 & 0.25 & 0.30 & 0.10 & 0.50 & 0.50 & 0.55 & 0.80 \\
\cline{2-23}
%% 0.02
 & \multirow{5}{*}{0.02} & 0.75 & 0.15 & 0.20 & 0.35 & 0.20 & 0.30 & 0.10 & 0.40 & 0.50 & 0.55 & 0.75 & 0.15 & 0.10 & 0.30 & 0.20 & 0.30 & 0.10 & 0.35 & 0.50 & 0.55 & 0.85 \\
 & & 0.78 & 0.15 & 0.15 & 0.35 & 0.25 & 0.35 & 0.10 & 0.35 & 0.50 & 0.75 & 0.75 & 0.15 & 0.15 & 0.35 & 0.20 & 0.30 & 0.10 & 0.35 & 0.55 & 0.70 & 0.80 \\
 & & 0.81 & 0.15 & 0.15 & 0.35 & 0.25 & 0.30 & 0.10 & 0.40 & 0.55 & 0.55 & 0.75 & 0.15 & 0.15 & 0.35 & 0.20 & 0.30 & 0.10 & 0.35 & 0.45 & 0.60 & 0.80 \\
 & & 0.84 & 0.15 & 0.15 & 0.35 & 0.25 & 0.30 & 0.10 & 0.45 & 0.55 & 0.60 & 0.75 & 0.15 & 0.15 & 0.35 & 0.30 & 0.35 & 0.10 & 0.45 & 0.55 & 0.45 & 0.80 \\
 & & 0.87 & 0.15 & 0.15 & 0.35 & 0.25 & 0.30 & 0.10 & 0.45 & 0.60 & 0.60 & 0.80 & 0.15 & 0.15 & 0.35 & 0.30 & 0.30 & 0.10 & 0.40 & 0.50 & 0.65 & 0.80 \\
\cline{2-23}
%% 0.03
 & \multirow{5}{*}{0.03} & 0.75 & 0.15 & 0.10 & 0.25 & 0.20 & 0.30 & 0.05 & 0.30 & 0.45 & 0.60 & 0.85 & 0.15 & 0.10 & 0.25 & 0.20 & 0.30 & 0.10 & 0.30 & 0.50 & 0.60 & 0.65 \\
 & & 0.78 & 0.15 & 0.10 & 0.30 & 0.15 & 0.30 & 0.10 & 0.30 & 0.60 & 0.55 & 0.85 & 0.15 & 0.15 & 0.30 & 0.20 & 0.30 & 0.10 & 0.25 & 0.50 & 0.45 & 0.75 \\
 & & 0.81 & 0.15 & 0.05 & 0.35 & 0.20 & 0.35 & 0.10 & 0.30 & 0.55 & 0.55 & 0.75 & 0.15 & 0.05 & 0.30 & 0.20 & 0.35 & 0.10 & 0.35 & 0.55 & 0.55 & 0.80 \\
 & & 0.84 & 0.15 & 0.10 & 0.35 & 0.25 & 0.35 & 0.10 & 0.40 & 0.40 & 0.55 & 0.80 & 0.15 & 0.10 & 0.30 & 0.25 & 0.30 & 0.10 & 0.35 & 0.45 & 0.65 & 0.80 \\
 & & 0.87 & 0.15 & 0.15 & 0.35 & 0.25 & 0.35 & 0.10 & 0.30 & 0.50 & 0.70 & 0.85 & 0.15 & 0.10 & 0.35 & 0.20 & 0.35 & 0.10 & 0.35 & 0.55 & 0.60 & 0.85 \\
\hline
% Scenario 3
%% 0.01
\multirow{15}{*}{3} & \multirow{5}{*}{0.01} & 0.75 & 0.10 & 0.25 & 0.70 & 0.55 & 0.25 & 0.10 & 0.00 & 0.20 & 0.15 & 0.20 & 0.10 & 0.25 & 0.70 & 0.55 & 0.25 & 0.10 & 0.05 & 0.15 & 0.05 & 0.20 \\
 & & 0.78 & 0.10 & 0.35 & 0.70 & 0.55 & 0.25 & 0.10 & 0.00 & 0.15 & 0.15 & 0.20 & 0.10 & 0.30 & 0.65 & 0.55 & 0.25 & 0.10 & 0.05 & 0.15 & 0.15 & 0.15 \\
 & & 0.81 & 0.10 & 0.25 & 0.70 & 0.55 & 0.25 & 0.10 & 0.10 & 0.15 & 0.15 & 0.20 & 0.10 & 0.35 & 0.65 & 0.55 & 0.25 & 0.10 & 0.05 & 0.20 & 0.15 & 0.20 \\
 & & 0.84 & 0.10 & 0.30 & 0.65 & 0.50 & 0.25 & 0.10 & 0.05 & 0.15 & 0.15 & 0.15 & 0.10 & 0.25 & 0.60 & 0.55 & 0.25 & 0.10 & 0.00 & 0.15 & 0.15 & 0.15 \\
 & & 0.87 & 0.10 & 0.30 & 0.55 & 0.50 & 0.20 & 0.10 & 0.05 & 0.15 & 0.10 & 0.20 & 0.10 & 0.30 & 0.55 & 0.50 & 0.20 & 0.10 & 0.05 & 0.15 & 0.15 & 0.15 \\
\cline{2-23}
%% 0.02
 & \multirow{5}{*}{0.02} & 0.75 & 0.10 & 0.35 & 0.60 & 0.55 & 0.30 & 0.10 & 0.00 & 0.20 & 0.10 & 0.20 & 0.10 & 0.35 & 0.70 & 0.60 & 0.25 & 0.10 & 0.10 & 0.20 & 0.05 & 0.15 \\
 & & 0.78 & 0.10 & 0.35 & 0.75 & 0.60 & 0.25 & 0.10 & 0.05 & 0.20 & 0.10 & 0.20 & 0.10 & 0.35 & 0.65 & 0.55 & 0.25 & 0.10 & 0.05 & 0.15 & 0.10 & 0.15 \\
 & & 0.81 & 0.10 & 0.35 & 0.70 & 0.55 & 0.25 & 0.10 & 0.05 & 0.15 & 0.10 & 0.15 & 0.10 & 0.30 & 0.70 & 0.55 & 0.25 & 0.10 & 0.05 & 0.15 & 0.05 & 0.15 \\
 & & 0.84 & 0.10 & 0.35 & 0.70 & 0.55 & 0.25 & 0.10 & 0.05 & 0.15 & 0.20 & 0.15 & 0.10 & 0.30 & 0.70 & 0.55 & 0.25 & 0.10 & 0.05 & 0.20 & 0.10 & 0.15 \\
 & & 0.87 & 0.10 & 0.30 & 0.60 & 0.55 & 0.25 & 0.10 & 0.05 & 0.15 & 0.15 & 0.15 & 0.10 & 0.30 & 0.65 & 0.55 & 0.20 & 0.10 & 0.05 & 0.20 & 0.15 & 0.15 \\
\cline{2-23}
%% 0.03
 & \multirow{5}{*}{0.03} & 0.75 & 0.10 & 0.30 & 0.50 & 0.60 & 0.35 & 0.10 & 0.05 & 0.20 & 0.05 & 0.15 & 0.10 & 0.30 & 0.50 & 0.70 & 0.25 & 0.10 & 0.05 & 0.15 & 0.05 & 0.15 \\
 & & 0.78 & 0.10 & 0.30 & 0.50 & 0.60 & 0.25 & 0.10 & 0.05 & 0.15 & 0.10 & 0.20 & 0.10 & 0.30 & 0.55 & 0.60 & 0.30 & 0.10 & 0.05 & 0.15 & 0.05 & 0.15 \\
 & & 0.81 & 0.10 & 0.25 & 0.55 & 0.55 & 0.25 & 0.10 & 0.05 & 0.15 & 0.10 & 0.15 & 0.10 & 0.30 & 0.60 & 0.55 & 0.30 & 0.10 & 0.05 & 0.20 & 0.15 & 0.20 \\
 & & 0.84 & 0.10 & 0.35 & 0.65 & 0.55 & 0.25 & 0.10 & 0.05 & 0.15 & 0.15 & 0.15 & 0.10 & 0.35 & 0.65 & 0.60 & 0.20 & 0.10 & 0.05 & 0.15 & 0.05 & 0.20 \\
 & & 0.87 & 0.10 & 0.30 & 0.70 & 0.55 & 0.25 & 0.10 & 0.05 & 0.15 & 0.15 & 0.15 & 0.10 & 0.35 & 0.70 & 0.55 & 0.20 & 0.10 & 0.05 & 0.20 & 0.15 & 0.15 \\
\hline
% Scenario 4
%% 0.01
\multirow{15}{*}{4} & \multirow{5}{*}{0.01} & 0.75 & 0.15 & 0.40 & 0.70 & 0.50 & 0.35 & 0.10 & 0.10 & 0.15 & 0.20 & 0.20 & 0.15 & 0.40 & 0.75 & 0.50 & 0.35 & 0.10 & 0.15 & 0.15 & 0.20 & 0.20 \\
 & & 0.78 & 0.20 & 0.40 & 0.75 & 0.45 & 0.30 & 0.10 & 0.10 & 0.15 & 0.20 & 0.15 & 0.20 & 0.40 & 0.80 & 0.45 & 0.30 & 0.10 & 0.10 & 0.10 & 0.20 & 0.15 \\
 & & 0.81 & 0.20 & 0.50 & 0.75 & 0.45 & 0.30 & 0.10 & 0.10 & 0.15 & 0.20 & 0.15 & 0.20 & 0.40 & 0.80 & 0.45 & 0.30 & 0.15 & 0.15 & 0.15 & 0.20 & 0.15 \\
 & & 0.84 & 0.20 & 0.45 & 0.75 & 0.45 & 0.30 & 0.15 & 0.10 & 0.15 & 0.20 & 0.15 & 0.15 & 0.60 & 0.80 & 0.45 & 0.30 & 0.10 & 0.20 & 0.15 & 0.20 & 0.20 \\
 & & 0.87 & 0.20 & 0.50 & 0.75 & 0.45 & 0.30 & 0.10 & 0.10 & 0.15 & 0.20 & 0.15 & 0.20 & 0.55 & 0.80 & 0.45 & 0.30 & 0.10 & 0.15 & 0.15 & 0.15 & 0.15 \\
\cline{2-23}
%% 0.02
 & \multirow{5}{*}{0.02} & 0.75 & 0.15 & 0.30 & 0.70 & 0.55 & 0.40 & 0.10 & 0.10 & 0.10 & 0.20 & 0.15 & 0.10 & 0.30 & 0.70 & 0.55 & 0.40 & 0.10 & 0.15 & 0.15 & 0.20 & 0.15 \\
 & & 0.78 & 0.15 & 0.35 & 0.75 & 0.55 & 0.40 & 0.10 & 0.10 & 0.10 & 0.20 & 0.15 & 0.15 & 0.30 & 0.75 & 0.55 & 0.35 & 0.10 & 0.15 & 0.15 & 0.20 & 0.20 \\
 & & 0.81 & 0.15 & 0.40 & 0.75 & 0.50 & 0.30 & 0.10 & 0.10 & 0.15 & 0.20 & 0.20 & 0.15 & 0.40 & 0.75 & 0.55 & 0.35 & 0.10 & 0.15 & 0.10 & 0.20 & 0.20 \\
 & & 0.84 & 0.15 & 0.40 & 0.75 & 0.50 & 0.30 & 0.10 & 0.10 & 0.15 & 0.20 & 0.15 & 0.15 & 0.40 & 0.75 & 0.50 & 0.35 & 0.10 & 0.15 & 0.10 & 0.20 & 0.15 \\
 & & 0.87 & 0.20 & 0.45 & 0.80 & 0.45 & 0.30 & 0.10 & 0.10 & 0.15 & 0.20 & 0.15 & 0.20 & 0.50 & 0.75 & 0.50 & 0.35 & 0.10 & 0.15 & 0.20 & 0.20 & 0.20 \\
\cline{2-23}
%% 0.03
 & \multirow{5}{*}{0.03} & 0.75 & 0.15 & 0.30 & 0.65 & 0.60 & 0.40 & 0.10 & 0.15 & 0.15 & 0.20 & 0.15 & 0.15 & 0.25 & 0.65 & 0.60 & 0.40 & 0.10 & 0.15 & 0.15 & 0.20 & 0.15 \\
 & & 0.78 & 0.15 & 0.30 & 0.65 & 0.55 & 0.35 & 0.10 & 0.15 & 0.15 & 0.20 & 0.15 & 0.15 & 0.30 & 0.65 & 0.60 & 0.40 & 0.15 & 0.10 & 0.15 & 0.20 & 0.15 \\
 & & 0.81 & 0.15 & 0.30 & 0.70 & 0.55 & 0.40 & 0.10 & 0.10 & 0.15 & 0.20 & 0.15 & 0.15 & 0.30 & 0.70 & 0.55 & 0.40 & 0.10 & 0.10 & 0.15 & 0.20 & 0.15 \\
 & & 0.84 & 0.15 & 0.30 & 0.75 & 0.55 & 0.40 & 0.10 & 0.15 & 0.15 & 0.20 & 0.15 & 0.15 & 0.30 & 0.70 & 0.50 & 0.40 & 0.10 & 0.10 & 0.15 & 0.20 & 0.20 \\
 & & 0.87 & 0.15 & 0.40 & 0.70 & 0.50 & 0.30 & 0.15 & 0.10 & 0.15 & 0.20 & 0.15 & 0.15 & 0.40 & 0.70 & 0.50 & 0.35 & 0.10 & 0.10 & 0.15 & 0.20 & 0.20 \\
\hline
\end{longtable}
}

%% - MSE
{
\tiny
\setlength{\tabcolsep}{0.9pt}
\renewcommand{\arraystretch}{1.0}
\begin{longtable}{c|c|c|ccccc|ccccc|ccccc|ccccc}
\caption{The MSEs for annotators' FPRs of different methods under four scenarios with different priors for TPRs and occurrence probabilities.} \\
\label{table:fpr_mse}
\multirow{2}{*}{Scen.} & \multirow{2}{*}{$\frac{a_p}{a_p + b_p}$} & Method & \multicolumn{5}{c|}{\textbf{Base}} & \multicolumn{5}{c|}{\textbf{Base-Hierarchical}} & \multicolumn{5}{c|}{\textbf{DP-BMM}} & \multicolumn{5}{c}{\textbf{DP-BMM-Hierarchical}} \\
\cline{3-23}`
& & \#Anns & 0.8 & 1.6 & 2.4 & 3.2 & 4.0 & 0.8 & 1.6 & 2.4 & 3.2 & 4.0 & 0.8 & 1.6 & 2.4 & 3.2 & 4.0 & 0.8 & 1.6 & 2.4 & 3.2 & 4.0 \\
\hline
% Scenario 1
%% 0.01
\multirow{15}{*}{1} & \multirow{5}{*}{0.01} & 0.75 & 1.29 & 0.523 & 0.236 & 0.134 & 0.0940 & 0.692 & 0.132 & 0.0357 & 0.0174 & 0.0136 & 1.29 & 0.526 & 0.238 & 0.135 & 0.0946 & 0.651 & 0.140 & 0.0340 & 0.0217 & 0.0143 \\
 & & 0.78 & 1.24 & 0.508 & 0.228 & 0.133 & 0.0922 & 0.727 & 0.108 & 0.0258 & 0.0188 & 0.0159 & 1.23 & 0.499 & 0.229 & 0.133 & 0.0926 & 0.699 & 0.121 & 0.0280 & 0.0248 & 0.0103 \\
 & & 0.81 & 1.20 & 0.490 & 0.224 & 0.133 & 0.0918 & 0.627 & 0.106 & 0.0242 & 0.0188 & 0.0160 & 1.20 & 0.488 & 0.225 & 0.133 & 0.0916 & 0.660 & 0.141 & 0.0333 & 0.0202 & 0.0175 \\
 & & 0.84 & 1.15 & 0.472 & 0.219 & 0.131 & 0.0917 & 0.656 & 0.106 & 0.0275 & 0.0185 & 0.0139 & 1.14 & 0.470 & 0.219 & 0.130 & 0.0911 & 0.668 & 0.134 & 0.0246 & 0.0186 & 0.0145 \\
 & & 0.87 & 1.12 & 0.457 & 0.211 & 0.130 & 0.0896 & 0.566 & 0.102 & 0.0220 & 0.0148 & 0.0154 & 1.12 & 0.458 & 0.213 & 0.129 & 0.0890 & 0.634 & 0.100 & 0.0222 & 0.0175 & 0.0152 \\
\cline{2-23}
%% 0.02
 & \multirow{5}{*}{0.02} & 0.75 & 1.41 & 0.567 & 0.252 & 0.140 & 0.0957 & 0.800 & 0.144 & 0.0380 & 0.0209 & 0.0192 & 1.41 & 0.570 & 0.253 & 0.139 & 0.0965 & 0.907 & 0.133 & 0.0413 & 0.0229 & 0.0199 \\
 & & 0.78 & 1.36 & 0.550 & 0.244 & 0.137 & 0.0950 & 0.885 & 0.135 & 0.0369 & 0.0165 & 0.0171 & 1.38 & 0.551 & 0.245 & 0.138 & 0.0955 & 0.874 & 0.145 & 0.0399 & 0.0200 & 0.0165 \\
 & & 0.81 & 1.30 & 0.531 & 0.238 & 0.136 & 0.0933 & 0.780 & 0.133 & 0.0361 & 0.0258 & 0.0119 & 1.32 & 0.520 & 0.237 & 0.135 & 0.0940 & 0.827 & 0.124 & 0.0362 & 0.0185 & 0.0245 \\
 & & 0.84 & 1.27 & 0.508 & 0.229 & 0.133 & 0.0927 & 0.764 & 0.132 & 0.0277 & 0.0221 & 0.0246 & 1.27 & 0.495 & 0.231 &  0.135 & 0.0927 & 0.728 & 0.112 & 0.0342 & 0.0221 & 0.0158 \\
 & & 0.87 & 1.18 & 0.492 & 0.225 & 0.133 & 0.0918 & 0.779 & 0.132 & 0.0292 & 0.0155 & 0.0165 & 1.20 & 0.488 & 0.223 & 0.132 & 0.0911 & 0.744 & 0.142 & 0.0278 & 0.0167 & 0.0119 \\
\cline{2-23}
%% 0.03
 & \multirow{5}{*}{0.03} & 0.75 & 1.54 & 0.632 & 0.268 & 0.144 & 0.0987 & 0.935 & 0.171 & 0.0431 & 0.0235 & 0.0160 & 1.51 & 0.605 & 0.273 & 0.145 & 0.0991 & 0.948 & 0.187 & 0.0501 & 0.0225 & 0.0233 \\
 & & 0.78 & 1.45 & 0.593 & 0.261 & 0.140 & 0.0966 & 0.944 & 0.173 & 0.0438 & 0.0250 & 0.0230 & 1.47 & 0.591 & 0.259 & 0.141 & 0.0968 & 0.970 & 0.179 & 0.0458 & 0.0228 & 0.0143 \\
 & & 0.81 & 1.42 & 0.569 & 0.251 & 0.141 & 0.0962 & 0.926 & 0.161 & 0.0465 & 0.0218 & 0.0185 & 1.41 & 0.568 & 0.252 & 0.138 & 0.0963 & 0.878 & 0.163 & 0.0434 & 0.0202 & 0.0228 \\
 & & 0.84 & 1.36 & 0.545 & 0.244 & 0.136 & 0.0948 & 0.794 & 0.157 & 0.0396 & 0.0232 & 0.0157 & 1.35 & 0.542 & 0.244 & 0.137 & 0.0949 & 0.914 & 0.131 & 0.0305 & 0.0148 & 0.0123 \\
 & & 0.87 & 1.30 & 0.529 & 0.235 & 0.135 & 0.0941 & 0.825 & 0.128 & 0.0342 & 0.0216 & 0.0174 & 1.29 & 0.523 & 0.235 &  0.136 & 0.0932 & 0.771 & 0.171 & 0.0254 & 0.0210 & 0.0141 \\
\hline
% Scenario 2
%% 0.01
\multirow{15}{*}{2} & \multirow{5}{*}{0.01} & 0.75 & 1.29 & 0.481 & 0.285 & 0.177 & 0.0990 & 0.816 & 0.106 & 0.0617 & 0.0405 & 0.0219 & 1.29 & 0.467 & 0.278 & 0.177 & 0.0997 & 0.846 & 0.111 & 0.0720 & 0.0843 & 0.0176 \\
 & & 0.78 & 1.25 & 0.468 & 0.279 & 0.175 & 0.0986 & 0.754 & 0.100 & 0.0670 & 0.0379 & 0.0216 & 1.22 & 0.452 & 0.274 & 0.175 & 0.0999 & 0.833 & 0.0881 & 0.0714 & 0.0491 & 0.0194 \\
 & & 0.81 & 1.20 & 0.458 & 0.274 & 0.172 & 0.0973 & 0.711 & 0.103 & 0.0729 & 0.0351 & 0.0139 & 1.18 & 0.440 & 0.269 & 0.172 & 0.0984 & 0.833 & 0.122 & 0.0665 & 0.0401 & 0.0305 \\
 & & 0.84 & 1.14 & 0.441 & 0.268 & 0.170 & 0.0956 & 0.771 & 0.104 & 0.0588 & 0.0327 & 0.0176 & 1.14 & 0.433 & 0.267 & 0.172 & 0.0974 & 0.781 & 0.128 & 0.0793 & 0.0515 & 0.0261 \\
 & & 0.87 & 1.07 & 0.432 & 0.263 & 0.170 & 0.0951 & 0.749 & 0.0761 & 0.0764 & 0.0326 & 0.0175 & 1.09 & 0.423 & 0.263 & 0.170 & 0.0958 & 0.753 & 0.100 & 0.0708 & 0.0421 & 0.0135 \\
\cline{2-23}
%% 0.02
 & \multirow{5}{*}{0.02} & 0.75 & 1.43 & 0.518 & 0.297 & 0.182 & 0.101 & 1.08 & 0.125 & 0.0692 & 0.0415 & 0.0202 & 1.42 & 0.514 & 0.302 & 0.185 & 0.103 & 1.01 & 0.155 & 0.0969 & 0.0494 & 0.0289 \\
 & & 0.78 & 1.39 & 0.503 & 0.292 & 0.180 & 0.100 & 0.992 & 0.135 & 0.0937 & 0.0430 & 0.0168 & 1.39 & 0.497 & 0.294 & 0.181 & 0.102 & 0.902 & 0.150 & 0.116 & 0.0384 & 0.0250 \\
 & & 0.81 & 1.31 & 0.489 & 0.284 & 0.178 & 0.0995 & 0.848 & 0.140 & 0.0733 & 0.0380 & 0.0199 & 1.33 & 0.488 & 0.286 & 0.180 & 0.101 & 0.989 & 0.122 & 0.0817 & 0.0480 & 0.0207 \\
 & & 0.84 & 1.30 & 0.475 & 0.278 & 0.176 & 0.0979 & 0.889 & 0.119 & 0.0804 & 0.0395 & 0.0159 & 1.30 & 0.476 & 0.281 & 0.177 & 0.0995 & 0.929 & 0.128 & 0.0859 & 0.0377 & 0.0303 \\
 & & 0.87 & 1.21 & 0.460 & 0.274 & 0.173 & 0.0966 & 0.897 & 0.0939 & 0.0787 & 0.0417 & 0.0213 & 1.19 & 0.464 & 0.274 & 0.176 & 0.0995 & 0.968 & 0.122 & 0.0788 & 0.0368 & 0.0148 \\
\cline{2-23}
%% 0.03
 & \multirow{5}{*}{0.03} & 0.75 & 1.57 & 0.570 & 0.310 & 0.191 & 0.104 & 1.15 & 0.158 & 0.103 & 0.0410 & 0.0202 & 1.55 & 0.572 & 0.316 & 0.190 & 0.107 & 1.25 & 0.172 & 0.112 & 0.0674 & 0.0422 \\
 & & 0.78 & 1.52 & 0.548 & 0.303 & 0.186 & 0.102 & 1.15 & 0.145 & 0.0915 & 0.0455 & 0.0177 & 1.51 & 0.554 & 0.309 & 0.187 & 0.104 & 1.15 & 0.152 & 0.105 & 0.0431 & 0.0251 \\
 & & 0.81 & 1.45 & 0.520 & 0.296 & 0.183 & 0.102 & 1.06 & 0.153 & 0.0968 & 0.0514 & 0.0194 & 1.45 & 0.544 & 0.300 & 0.184 & 0.104 & 1.12 & 0.125 & 0.0843 & 0.0469 & 0.0301 \\
 & & 0.84 & 1.40 & 0.503 & 0.289 & 0.180 & 0.0999 & 1.03 & 0.132 & 0.0770 & 0.0387 & 0.0164 & 1.41 & 0.523 & 0.292 & 0.181 & 0.102 & 1.08 & 0.151 & 0.0909 & 0.0549 & 0.0255 \\
 & & 0.87 & 1.34 & 0.487 & 0.283 & 0.178 & 0.0994 & 1.04 & 0.137 & 0.0888 & 0.0428 & 0.0183 & 1.35 & 0.501 & 0.286 & 0.178 & 0.101 & 1.05 & 0.154 & 0.0958 & 0.0358 & 0.0300 \\
\hline
% Scenario 3
%% 0.01
\multirow{15}{*}{3} & \multirow{5}{*}{0.01} & 0.75 & 1.37 & 0.399 & 0.0748 & 0.0668 & 0.145 & 1.68 & 1.31 & 1.05 & 1.05 & 0.838 & 1.36 & 0.376 & 0.0747 & 0.0655 & 0.146 & 1.70 & 1.29 & 1.06 & 1.08 & 0.843 \\
 & & 0.78 & 1.29 & 0.355 & 0.0733 & 0.0684 & 0.148 & 1.71 & 1.30 & 1.04 & 1.02 & 0.822 & 1.29 & 0.347 & 0.0727 & 0.0679 & 0.148 & 1.73 & 1.28 & 1.01 & 1.06 & 0.890 \\
 & & 0.81 & 1.27 & 0.319 & 0.0707 & 0.0690 & 0.151 & 1.73 & 1.29 & 1.05 & 1.02 & 0.830 & 1.23 & 0.319 & 0.0725 & 0.0702 & 0.150 & 1.75 & 1.29 & 1.05 & 1.03 & 0.816 \\
 & & 0.84 & 1.13 & 0.293 & 0.0715 & 0.0711 & 0.153 & 1.71 & 1.29 & 1.01 & 1.04 & 0.867 & 1.21 & 0.291 & 0.0710 & 0.0718 & 0.152 & 1.76 & 1.29 & 1.04 & 1.04 & 0.828 \\
 & & 0.87 & 1.10 & 0.258 & 0.0724 & 0.0734 & 0.156 & 1.67 & 1.27 & 1.01 & 1.05 & 0.855 & 1.09 & 0.263 & 0.0704 & 0.0731 & 0.155 & 1.69 & 1.26 & 1.03 & 1.04 & 0.869 \\
\cline{2-23}
%% 0.02
 & \multirow{5}{*}{0.02} & 0.75 & 1.52 & 0.491 & 0.0852 & 0.0628 & 0.142 & 1.79 & 1.35 & 1.08 & 1.10 & 0.871 & 1.51 & 0.477 & 0.0815 & 0.0629 & 0.143 & 1.77 & 1.32 & 1.08 & 1.09 & 0.909 \\
 & & 0.78 & 1.45 & 0.451 & 0.0789 & 0.0654 & 0.145 & 1.77 & 1.37 & 1.12 & 1.04 & 0.898 & 1.44 & 0.460 & 0.0801 & 0.0652 & 0.144 & 1.77 & 1.35 & 1.05 & 1.10 & 0.882 \\
 & & 0.81 & 1.38 & 0.414 & 0.0757 & 0.0664 & 0.147 & 1.78 & 1.33 & 1.04 & 1.03 & 0.862 & 1.39 & 0.406 & 0.0762 & 0.0665 & 0.147 & 1.77 & 1.34 & 1.08 & 1.11 & 0.856 \\
 & & 0.84 & 1.34 & 0.377 & 0.0727 & 0.0678 & 0.149 & 1.77 & 1.33 & 1.02 & 1.07 & 0.857 & 1.35 & 0.372 & 0.0730 & 0.0681 & 0.150 & 1.81 & 1.32 & 1.08 & 1.05 & 0.856 \\
 & & 0.87 & 1.25 & 0.346 & 0.0716 & 0.0695 & 0.152 & 1.80 & 1.32 & 1.03 & 1.02 & 0.888 & 1.24 & 0.322 & 0.0716 & 0.0696 & 0.152 & 1.78 & 1.35 & 1.05 & 1.01 & 0.831 \\
\cline{2-23}
%% 0.03
 & \multirow{5}{*}{0.03} & 0.75 & 1.61 & 0.636 & 0.101 & 0.0604 & 0.138 & 1.86 & 1.41 & 1.17 & 1.12 & 0.887 & 1.61 & 0.619 & 0.101 & 0.0595 & 0.137 & 1.83 & 1.40 & 1.12 & 1.12 & 0.864 \\
 & & 0.78 & 1.56 & 0.565 & 0.0912 & 0.0629 & 0.141 & 1.86 & 1.40 & 1.12 & 1.04 & 0.912 & 1.57 & 0.568 & 0.0919 & 0.0628 & 0.140 & 1.80 & 1.40 & 1.10 & 1.12 & 0.926 \\
 & & 0.81 & 1.50 & 0.521 & 0.0861 & 0.0637 & 0.143 & 1.82 & 1.36 & 1.09 & 1.07 & 0.848 & 1.49 & 0.527 & 0.0844 & 0.0630 & 0.142 & 1.85 & 1.41 & 1.08 & 1.09 & 0.861 \\
 & & 0.84 & 1.48 & 0.454 & 0.0789 & 0.0656 & 0.146 & 1.81 & 1.41 & 1.06 & 1.05 & 0.854 & 1.46 & 0.467 & 0.0803 & 0.0653 & 0.144 & 1.84 & 1.39 & 1.07 & 1.06 & 0.870 \\
 & & 0.87 & 1.40 & 0.413 & 0.0761 & 0.0661 & 0.148 & 1.80 & 1.38 & 1.06 & 1.09 & 0.861 & 1.41 & 0.415 & 0.0741 & 0.0663 & 0.147 & 1.82 & 1.37 & 1.06 & 1.05 & 0.874 \\
\hline
% Scenario 4
%% 0.01
\multirow{15}{*}{4} & \multirow{5}{*}{0.01} & 0.75 & 1.36 & 0.252 & 0.0916 & 0.0907 & 0.149 & 1.69 & 1.29 & 1.14 & 1.01 & 0.932 & 1.36 & 0.248 & 0.0890 & 0.0942 & 0.149 & 1.71 & 1.35 & 1.09 & 1.02 & 0.94 \\
 & & 0.78 & 1.29 & 0.215 & 0.0890 & 0.0942 & 0.150 & 1.73 & 1.32 & 1.12 & 0.934 & 0.971 & 1.30 & 0.208 & 0.0875 & 0.0972 & 0.149 & 1.70 & 1.31 & 1.13 & 0.955 & 0.943 \\
 & & 0.81 & 1.23 & 0.175 & 0.0874 & 0.0971 & 0.153 & 1.71 &  1.30 & 1.12 & 0.993 & 0.895 & 1.22 & 0.184 & 0.0866 & 0.0978 & 0.150 & 1.70 & 1.28 & 1.13 & 0.967 & 0.960 \\
 & & 0.84 & 1.16 & 0.162 & 0.0871 & 0.0995 & 0.153 & 1.73 & 1.31 & 1.10 & 0.904 & 0.906 & 1.15 & 0.151 & 0.0868 & 0.0992 & 0.152 & 1.69 & 1.29 & 1.10 & 0.922 & 1.00 \\
 & & 0.87 & 1.03 & 0.140 & 0.0862 & 0.103 & 0.156 & 1.69 &  1.31 & 1.06 & 0.954 & 0.943 & 1.04 & 0.130 & 0.0876 & 0.1032 & 0.155 & 1.67 & 1.33 & 1.11 & 0.956 & 0.956 \\
\cline{2-23}
%% 0.02
 & \multirow{5}{*}{0.02} & 0.75 & 1.53 & 0.388 & 0.0993 & 0.0859 & 0.145 & 1.76 & 1.37 & 1.17 & 1.02 & 0.955 & 1.51 & 0.358 & 0.0957 & 0.0854 & 0.143 & 1.83 & 1.36 & 1.17 & 1.02 & 1.04 \\
 & & 0.78 & 1.50 & 0.328 & 0.0949 & 0.0884 & 0.147 & 1.80 & 1.34 & 1.10 & 0.980 & 0.957 & 1.47 & 0.331 & 0.0918 & 0.0887 & 0.144 & 1.80 & 1.35 & 1.15 & 0.997 & 0.933 \\
 & & 0.81 & 1.44 & 0.270 & 0.0907 & 0.0904 & 0.148 & 1.77 &  1.33 & 1.20 & 0.945 & 0.938 & 1.41 & 0.266 & 0.0902 & 0.0917 & 0.146 & 1.77 & 1.38 & 1.15 & 1.02 & 0.911 \\
 & & 0.84 & 1.36 & 0.216 & 0.0890 & 0.0939 & 0.151 & 1.75 & 1.34 & 1.12 & 0.960 & 0.957 & 1.35 & 0.226 & 0.0882 & 0.0924 & 0.150 & 1.79 & 1.34 & 1.12 & 0.979 & 0.931 \\
 & & 0.87 & 1.25 & 0.184 & 0.0874 & 0.0985 & 0.153 & 1.77 &  1.35 & 1.10 & 0.970 & 0.948 & 1.24 & 0.168 & 0.0887 & 0.0962 & 0.150 & 1.77 & 1.33 & 1.16 & 1.00 & 0.933 \\
\cline{2-23}
%% 0.03
 & \multirow{5}{*}{0.03} & 0.75 & 1.63 & 0.512 & 0.118 & 0.0797 & 0.141 & 1.85 & 1.41 & 1.21 & 1.04 & 0.996 & 1.64 & 0.500 & 0.117 & 0.0808 & 0.138 & 1.83 & 1.44 & 1.16 & 1.08 & 0.950 \\
 & & 0.78 & 1.58 & 0.459 & 0.107 & 0.0829 & 0.142 & 1.85 &  1.42 & 1.21 & 1.04 & 0.981 & 1.59 & 0.464 & 0.105 & 0.0831 & 0.140 & 1.86 & 1.41 & 1.19 & 1.05 & 0.930 \\
 & & 0.81 & 1.55 & 0.388 & 0.100 & 0.0857 & 0.143 & 1.85 &  1.41 & 1.15 & 0.965 & 0.974 & 1.58 & 0.411 & 0.0988 & 0.0857 & 0.143 & 1.83 & 1.39 & 1.19 & 0.992 & 0.964 \\
 & & 0.84 & 1.51 & 0.355 & 0.0916 & 0.0885 & 0.148 & 1.82 &  1.38 & 1.20 & 0.998 & 0.918 & 1.49 & 0.356 & 0.0924 & 0.0866 & 0.145 & 1.83 & 1.36 & 1.18 & 0.972 & 0.945 \\
 & & 0.87 & 1.41 & 0.269 & 0.0916 & 0.0921 & 0.147 & 1.81 & 1.38 & 1.17 & 0.988 & 0.945 & 1.41 & 0.278 & 0.0898 & 0.0902 & 0.147 & 1.86 & 1.38 & 1.19 & 0.993 & 0.976 \\
\hline
\end{longtable}
}

%%%%% Section A3 %%%%%
\section{MCMC diagnostics for application to species annotation data} \label{sec:diagnostic_application}
In addition to computing the Gelman-Rubin statistics for model parameters, we assessed the convergence of the MCMC scheme by examining traceplots of model parameters across chains. As briefly discussed in Section \ref{sec:application}, our specific focus is on the following parameters:
\begin{itemize}
    \item \textbf{Base}: the occurrence probabilities of different species $\bm{o} = (o_1, o_2, \dots, o_{N_3})^T$, the TPRs of annotators' $\bm{\lambda} = (\lambda_1, \lambda_2, \dots, \lambda_{N_2})^T$ and the FPRs of annotators' $\bm{\psi} = (\psi_1, \psi_2, \dots, \psi_{N2})^T$;
    \item \textbf{Base-Hierarchical}: the occurrence probabilities of different species $\bm{o} = (o_1, o_2, \dots, o_{N_3})^T$, the annotator-specific parameters of our annotators $\bm{\lambda} = (\lambda_1, \lambda_2, \dots, \lambda_{N_2})^T$ as well as $\bm{\psi} = (\psi_1, \psi_2, \dots, \psi_{N_2})^T$, and the species-specific parameters of the annotators $\bm{\Lambda} = (\lambda_{j, k})$ as well as $\bm{\Psi} = (\psi_{j, k})$;
    \item \textbf{DP-BMM}: the concentration parameter $\gamma$, the TPRs of annotators' $\bm{\lambda} = (\lambda_1, \lambda_2, \dots, \lambda_{N_2})^T$ and the FPRs of annotators' $\bm{\psi} = (\psi_1, \psi_2, \dots, \psi_{N2})^T$;
    \item \textbf{DP-BMM-Hierarchical}: the concentration parameter $\gamma$, the annotator-specific parameters of our annotators $\bm{\lambda} = (\lambda_1, \lambda_2, \dots, \lambda_{N_2})^T$ as well as $\bm{\psi} = (\psi_1, \psi_2, \dots, \psi_{N_2})^T$, and the species-specific parameters of the annotators $\bm{\Lambda} = (\lambda_{j, k})$ as well as $\bm{\Psi} = (\psi_{j, k})$.
\end{itemize}
The traceplots for the parameters of interest in each model are presented separately in Section \ref{sec:app_diag_base}, Section \ref{sec:app_diag_base_hierarchy}, Section \ref{sec:app_diag_dpbmm}, and Section \ref{sec:app_diag_dpbmm_hierarchy}. 
To improve clarity and visualization, we display the traceplots for $1000$ iterations out of all the iterations. Additionally, for simplicity, we use the taxon ID in our dataset to reference the bird species. We also provide the bird species that are included in our analysis along with their taxonomic information in Appendix \ref{sec:bird_list}.

\subsection{Diagnostics for Base}
\label{sec:app_diag_base}
\begin{figure}
\centering
\includegraphics[scale = 0.60]{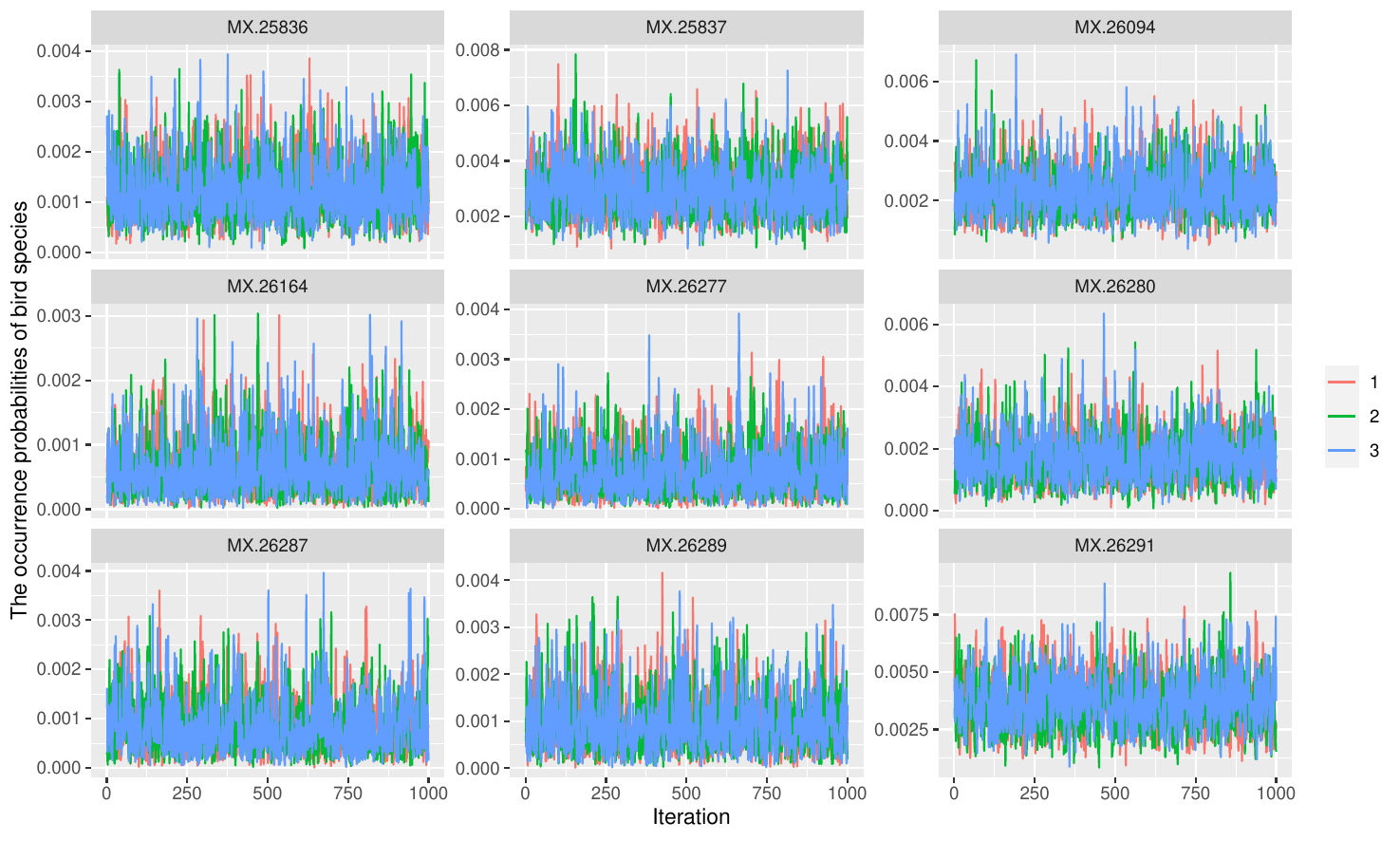}
\caption{Traceplots for the occurrence probabilities of nine randomly chosen birds in \textbf{Base}. Colors correspond to different MCMC chains.}
\label{fig:app_base_diag_p}
\end{figure}

\begin{figure}
\centering
\includegraphics[scale = 0.60]{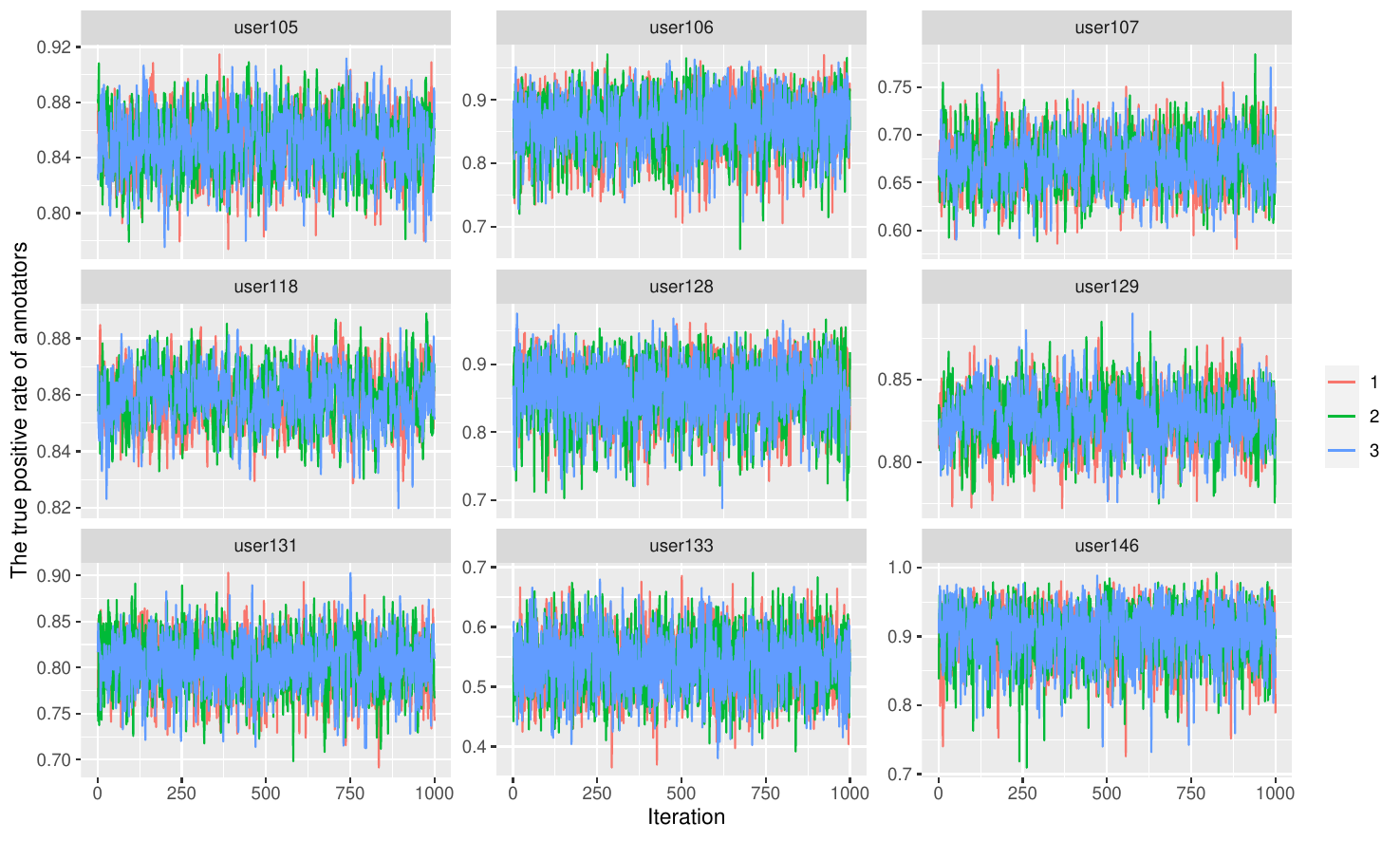}
\caption{Traceplots for the true positive rate of nine randomly chosen annotators in \textbf{Base}. Colors correspond to different MCMC chains.}
\label{fig:app_base_diag_lambda}
\end{figure}

\begin{figure}
\centering
\includegraphics[scale = 0.60]{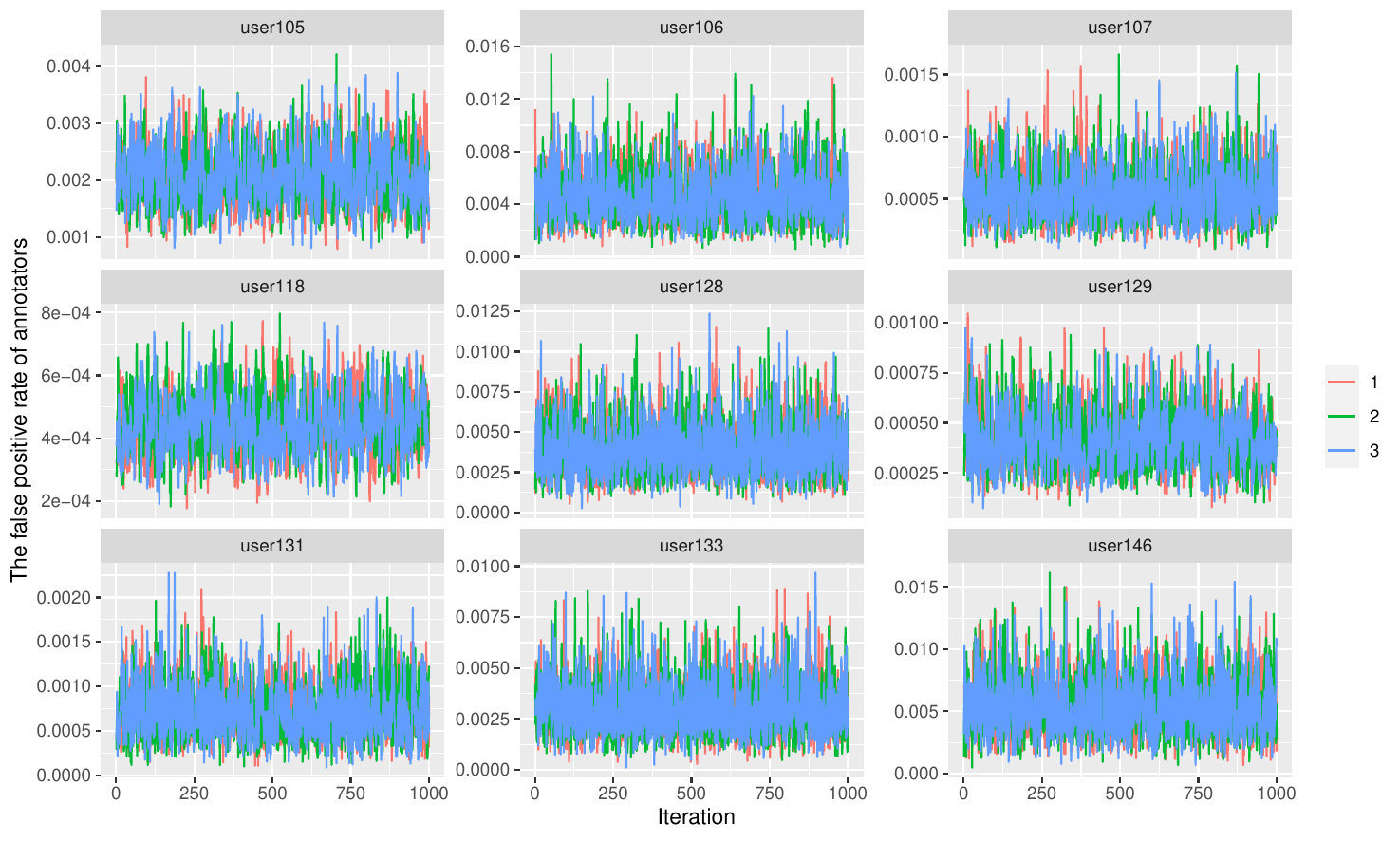}
\caption{Traceplots for the false positive rate of nine randomly chosen annotators in \textbf{Base}. Colors correspond to different MCMC chains.}
\label{fig:app_base_diag_psi}
\end{figure}

The traceplots for the occurrence probabilities of nine randomly selected bird species are presented in Figure \ref{fig:app_base_diag_p}. Additionally, the traceplots for the TPRs and the FPRs of nine randomly chosen annotators are shown in Figures \ref{fig:app_base_diag_lambda} and \ref{fig:app_base_diag_psi}, respectively. In general, there is a strong indication of favorable overall convergence and mixing, with the MCMC samples concentrating around similar values.

\subsection{Diagnostics for Base-Hierarchical}
\label{sec:app_diag_base_hierarchy}
\begin{figure}
\centering
\includegraphics[scale = 0.60]{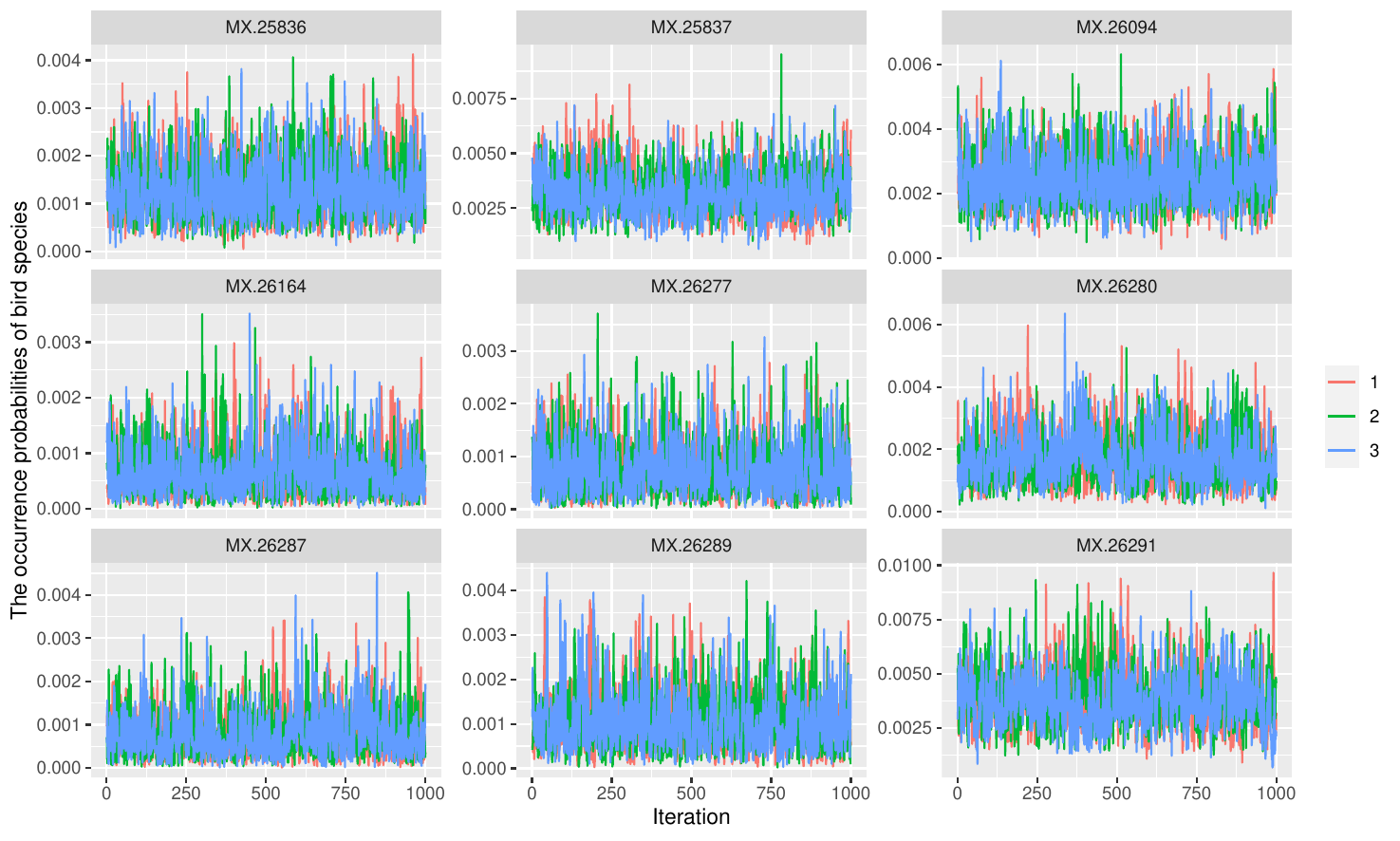}
\caption{Traceplots for the occurrence probabilities of nine randomly chosen birds in \textbf{Base-Hierarchical}. Colors correspond to different MCMC chains.}
\label{fig:app_base_hierarchy_diag_p}
\end{figure}

\begin{figure}
\centering
\includegraphics[scale = 0.60]{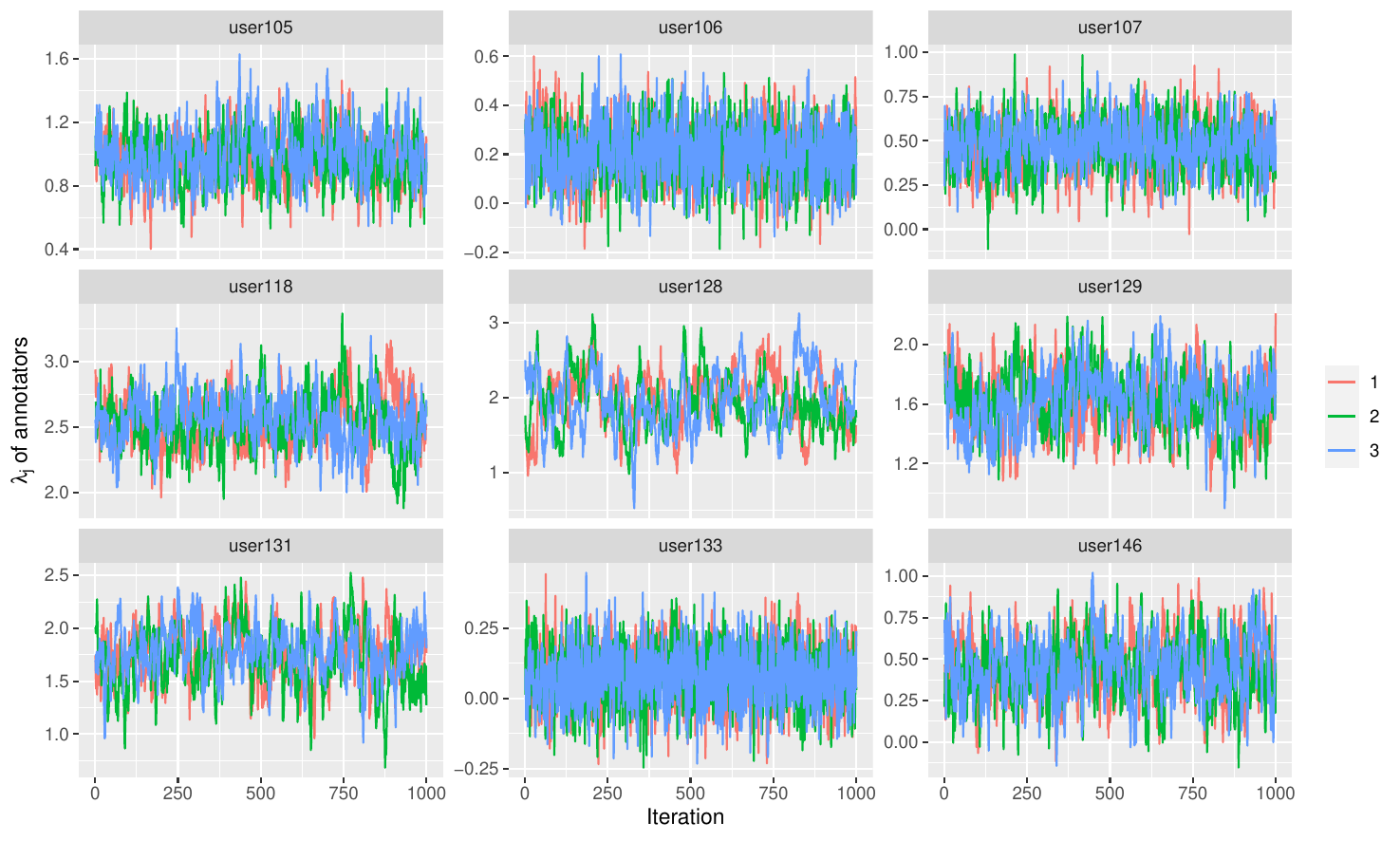}
\caption{Traceplots for the $\lambda_j$'s of nine randomly chosen annotators in \textbf{Base-Hierarchical}. Colors correspond to different MCMC chains.}
\label{fig:app_base_hierarchy_diag_lambda}
\end{figure}

\begin{figure}
\centering
\includegraphics[scale = 0.60]{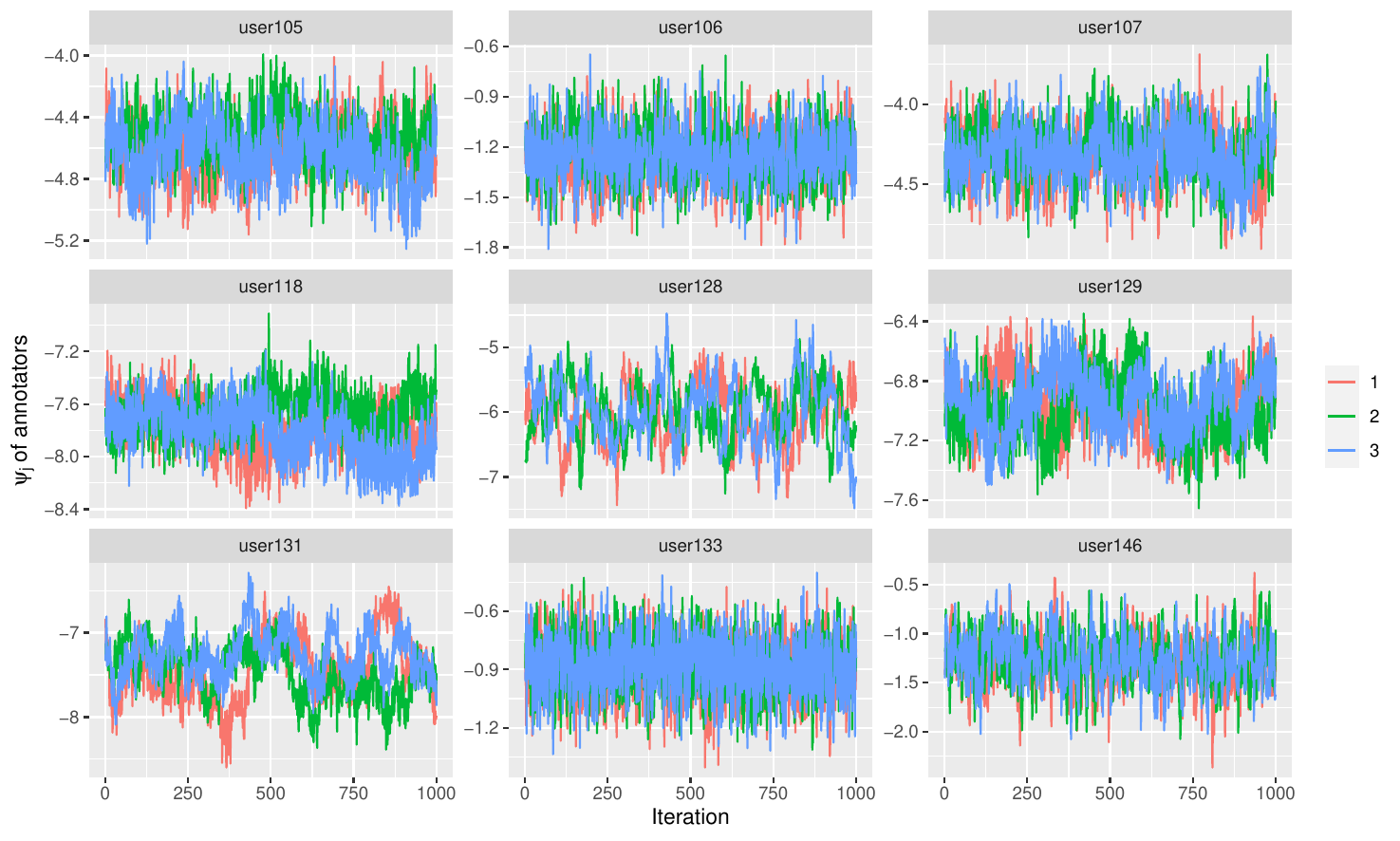}
\caption{Traceplots for the $\psi_j$'s of nine randomly chosen annotators in \textbf{Base-Hierarchical}. Colors correspond to different MCMC chains.}
\label{fig:app_base_hierarchy_diag_psi}
\end{figure}

\begin{figure}
\centering
\includegraphics[scale = 0.60]{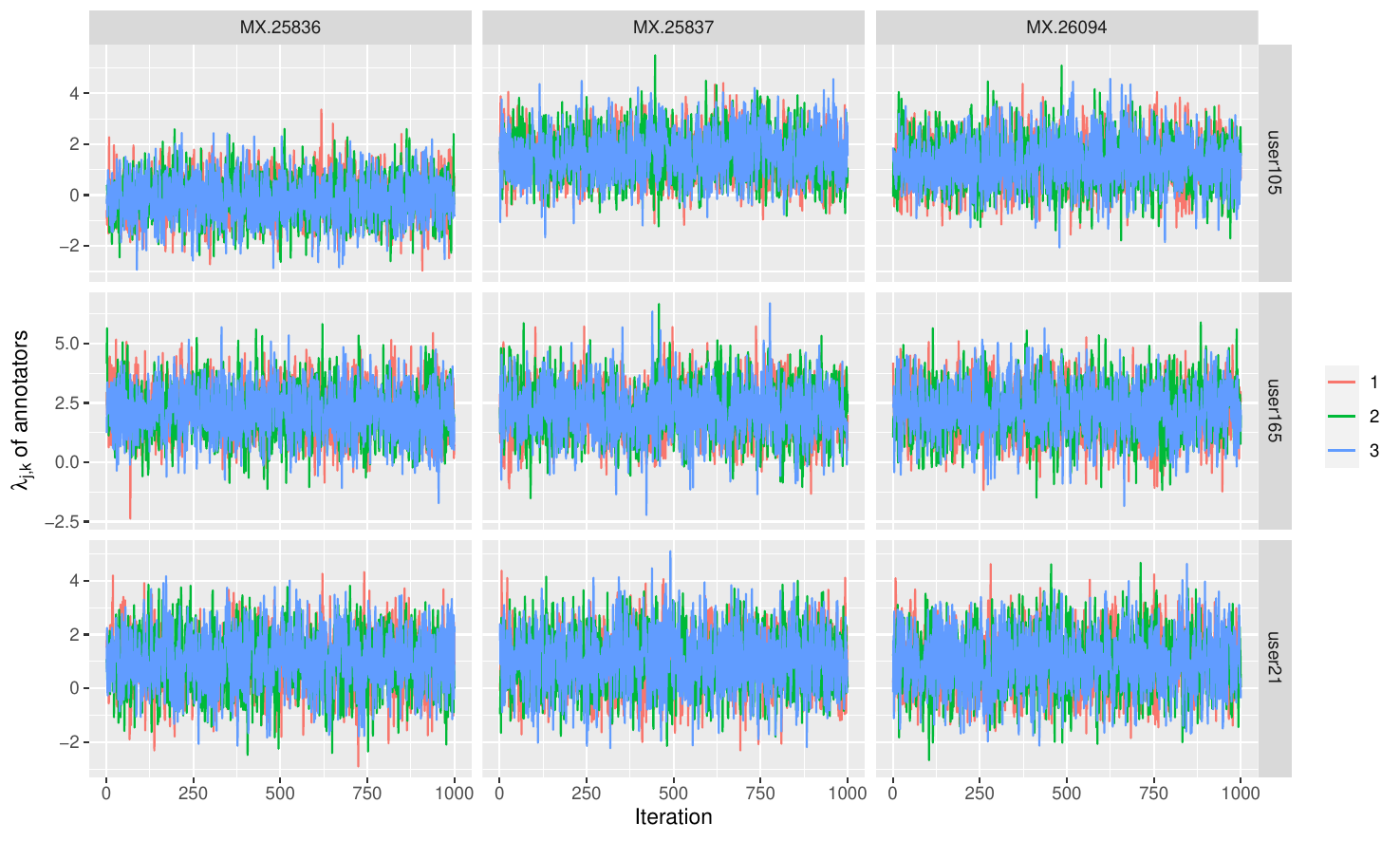}
\caption{Traceplots for the $\lambda_{j,k}$'s of three randomly chosen annotators on three randomly chosen bird species in \textbf{Base-Hierarchical}. Colors correspond to different MCMC chains.}
\label{fig:app_base_hierarchy_diag_lambda_mat}
\end{figure}

\begin{figure}
\centering
\includegraphics[scale = 0.60]{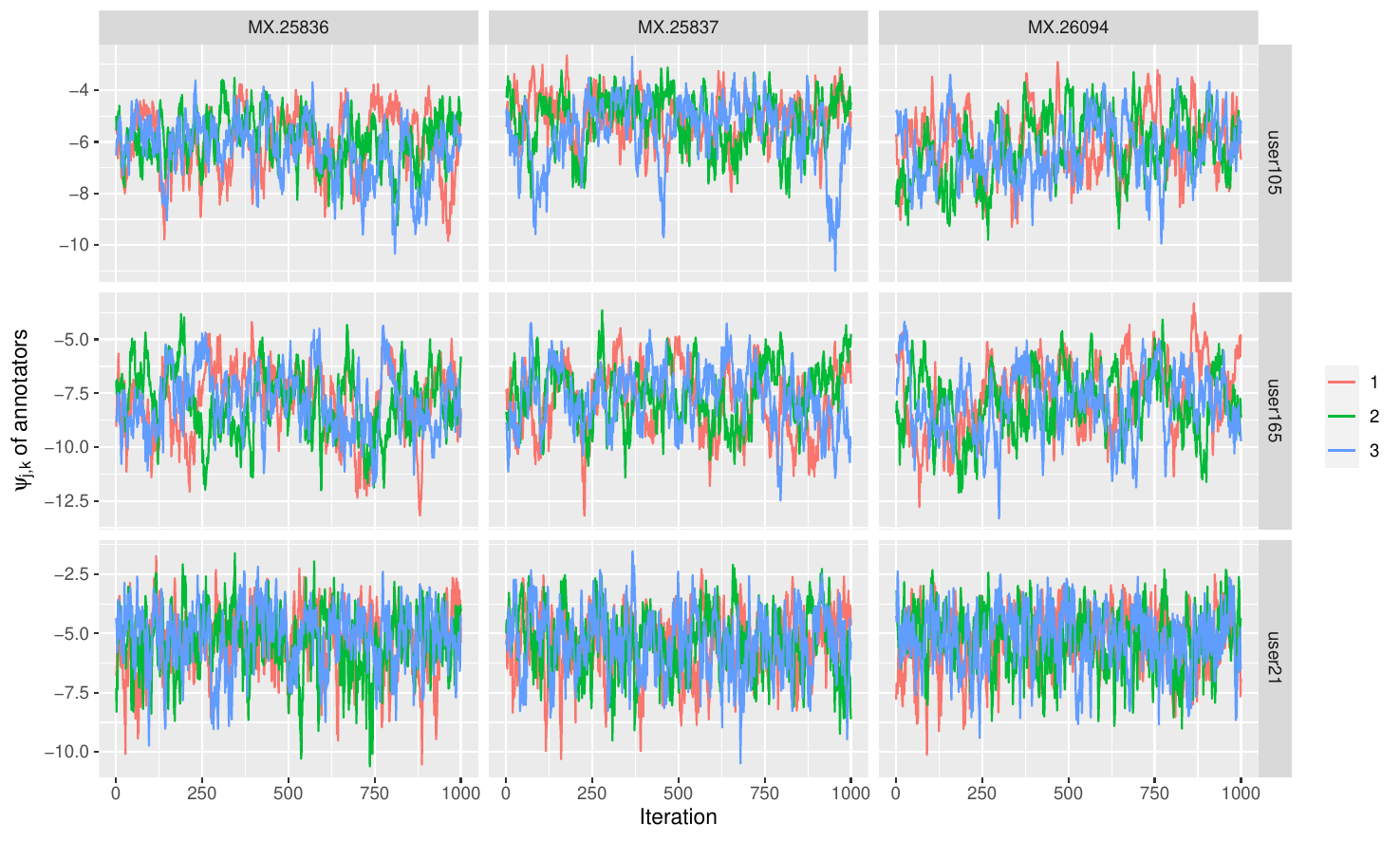}
\caption{Traceplots for the $\psi_{j,k}$'s of three randomly chosen annotators on three randomly chosen bird species in \textbf{Base-Hierarchical}. Colors correspond to different MCMC chains.}
\label{fig:app_base_hierarchy_diag_psi_mat}
\end{figure}

The traceplots for the occurrence probabilities of nine randomly chosen bird species are shown in Figure \ref{fig:app_base_hierarchy_diag_p}. As for the parameters related to annotators' bird song identification expertise, the traceplots for $\lambda_j$'s and $\psi_j$'s of nine randomly chosen annotators are shown in Figures \ref{fig:app_base_hierarchy_diag_lambda} and \ref{fig:app_base_hierarchy_diag_psi}, respectively. In addition, the traceplots for three randomly chosen annotators' $\lambda_{j, k}$ and $\psi_{j, k}$ on three randomly selected species are shown in Figures \ref{fig:app_base_hierarchy_diag_lambda_mat} and \ref{fig:app_base_hierarchy_diag_psi_mat}, respectively. In general, though some parameters exhibit a certain degree of autocorrelation, the Markov chains have converged sufficiently well and mixing seems to be good.

\subsection{Diagnostics for DP-BMM}
\label{sec:app_diag_dpbmm}
\begin{figure}
\centering
\includegraphics[scale = 0.60]{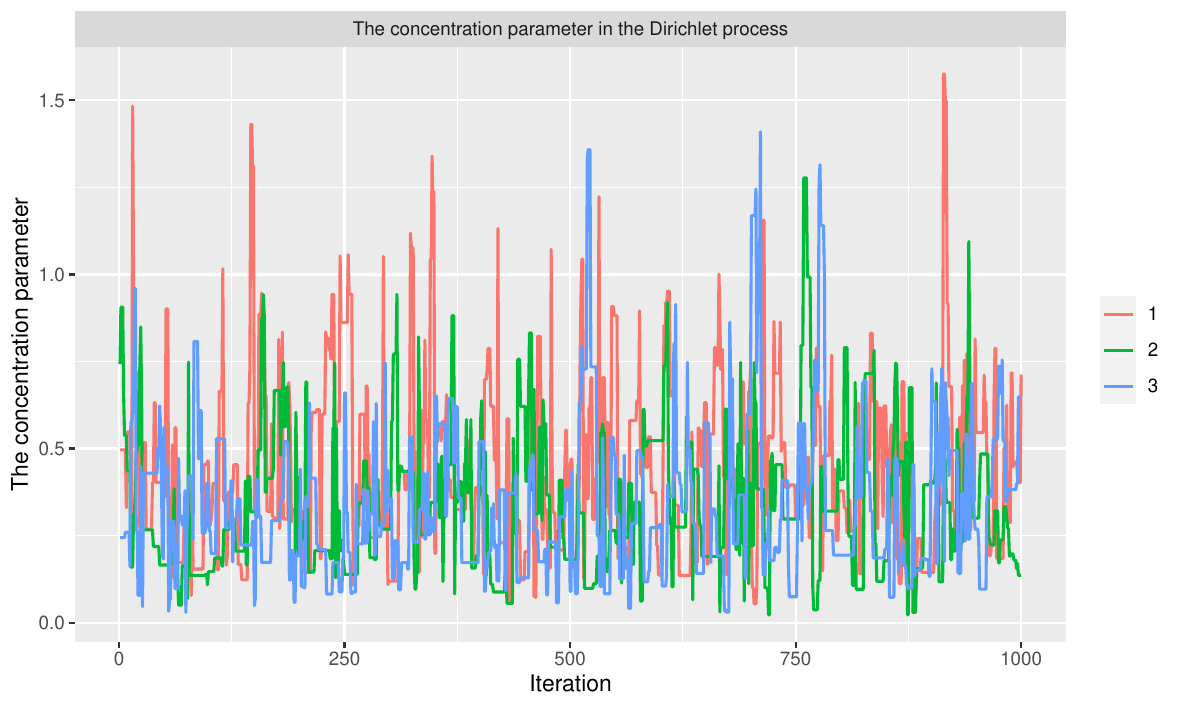}
\caption{Traceplots for the concentration parameter in the Dirichlet process in \textbf{DP-BMM}. Colors correspond to different MCMC chains.}
\label{fig:app_dpbmm_diag_gamma}
\end{figure}

\begin{figure}
\centering
\includegraphics[scale = 0.60]{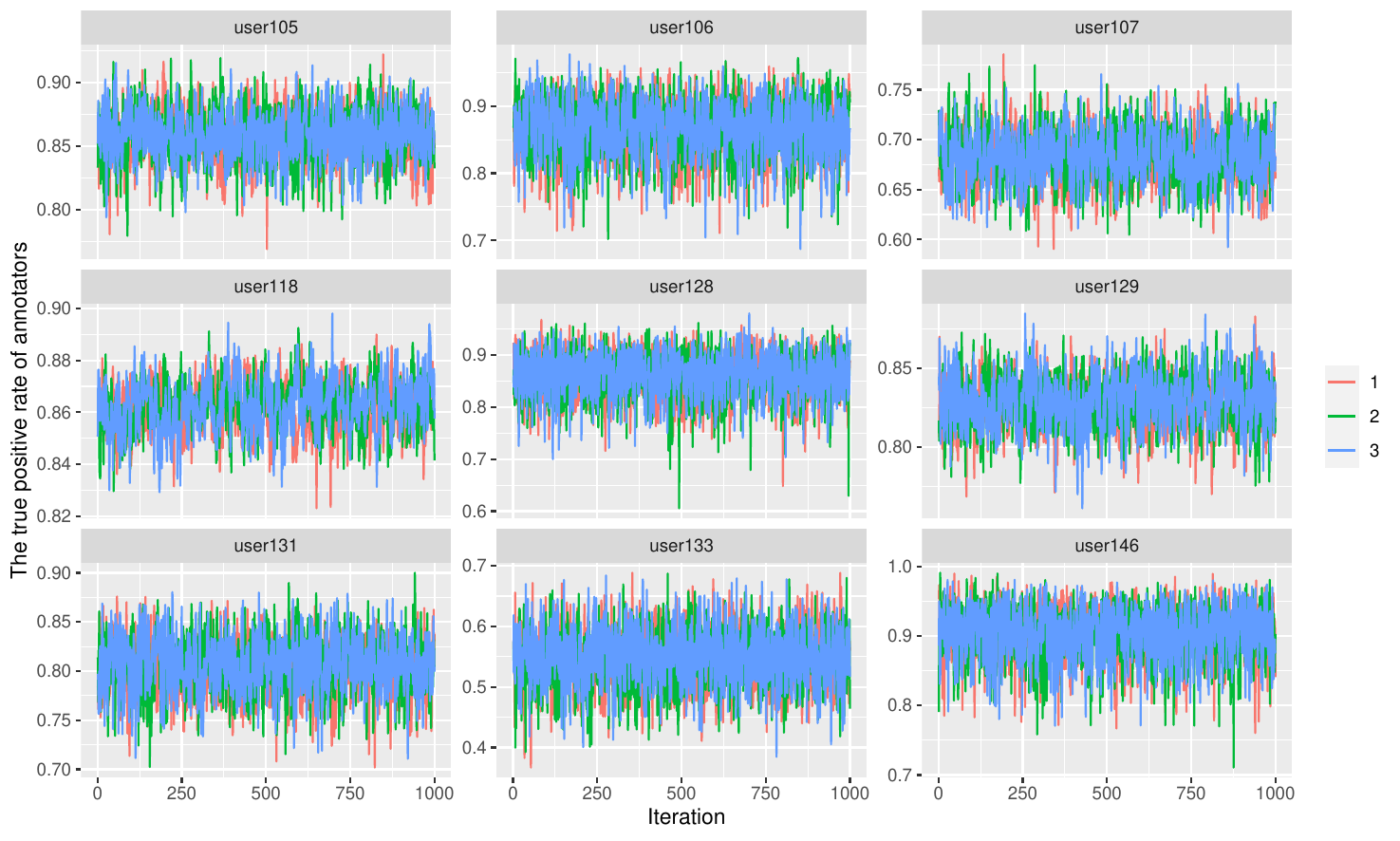}
\caption{Traceplots for the true positive rate of nine randomly chosen annotators in \textbf{DP-BMM}. Colors correspond to different MCMC chains.}
\label{fig:app_dpbmm_diag_lambda}
\end{figure}

\begin{figure}
\centering
\includegraphics[scale = 0.60]{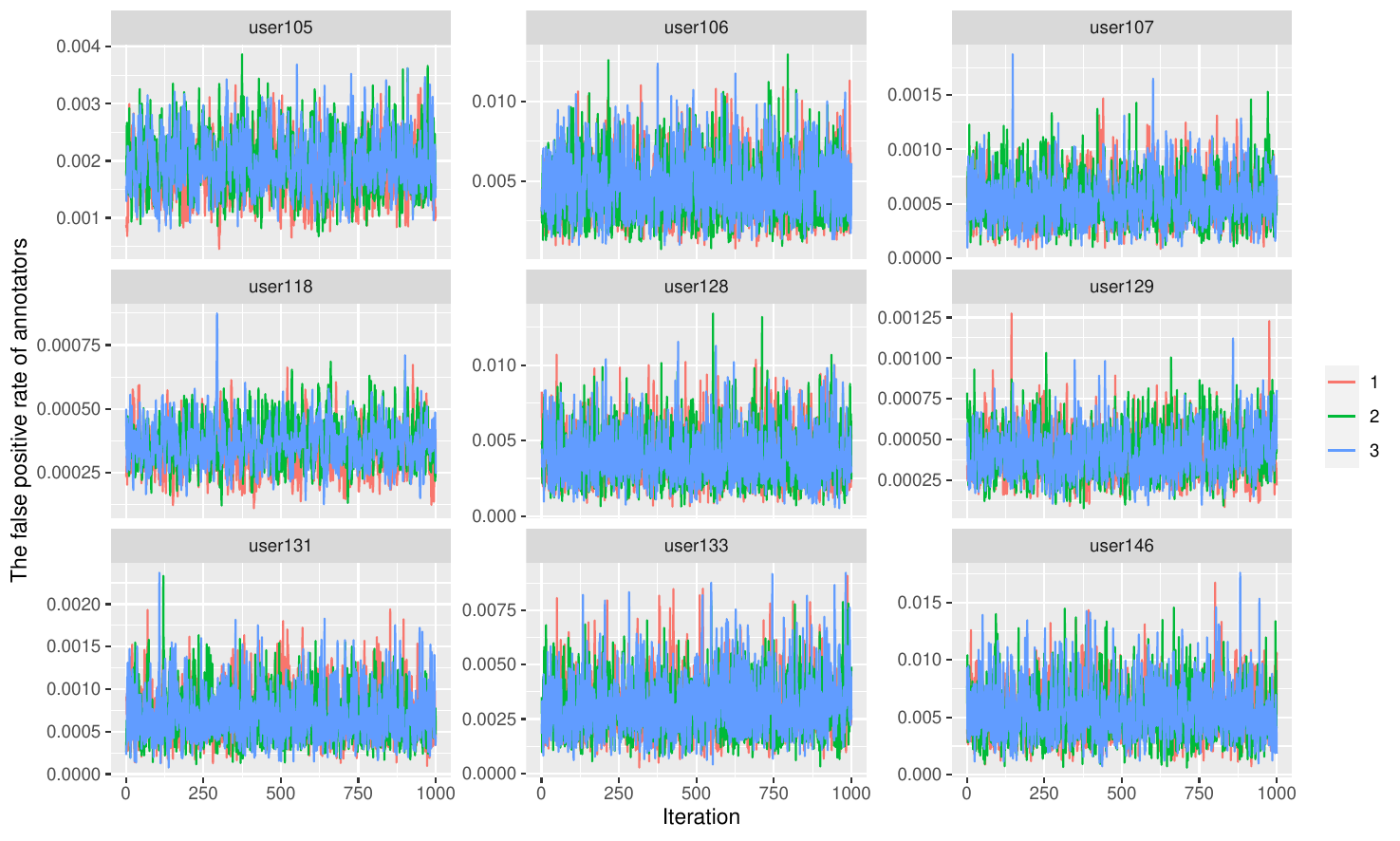}
\caption{Traceplots for the false positive rate of nine randomly chosen annotators in \textbf{DP-BMM}. Colors correspond to different MCMC chains.}
\label{fig:app_dpbmm_diag_psi}
\end{figure}

The traceplot for the concentration parameter is presented in Figure \ref{fig:app_dpbmm_diag_gamma}. Additionally, the traceplots for the TPRs and FPRs of nine randomly chosen annotators are shown in Figures \ref{fig:app_dpbmm_diag_lambda} and \ref{fig:app_dpbmm_diag_psi}, respectively. In general, there is a strong indication of favorable overall convergence and mixing, with the MCMC samples concentrating around similar values.

\subsection{Diagnostics for DP-BMM-Hierarchial}
\label{sec:app_diag_dpbmm_hierarchy}
\begin{figure}
\centering
\includegraphics[scale = 0.60]{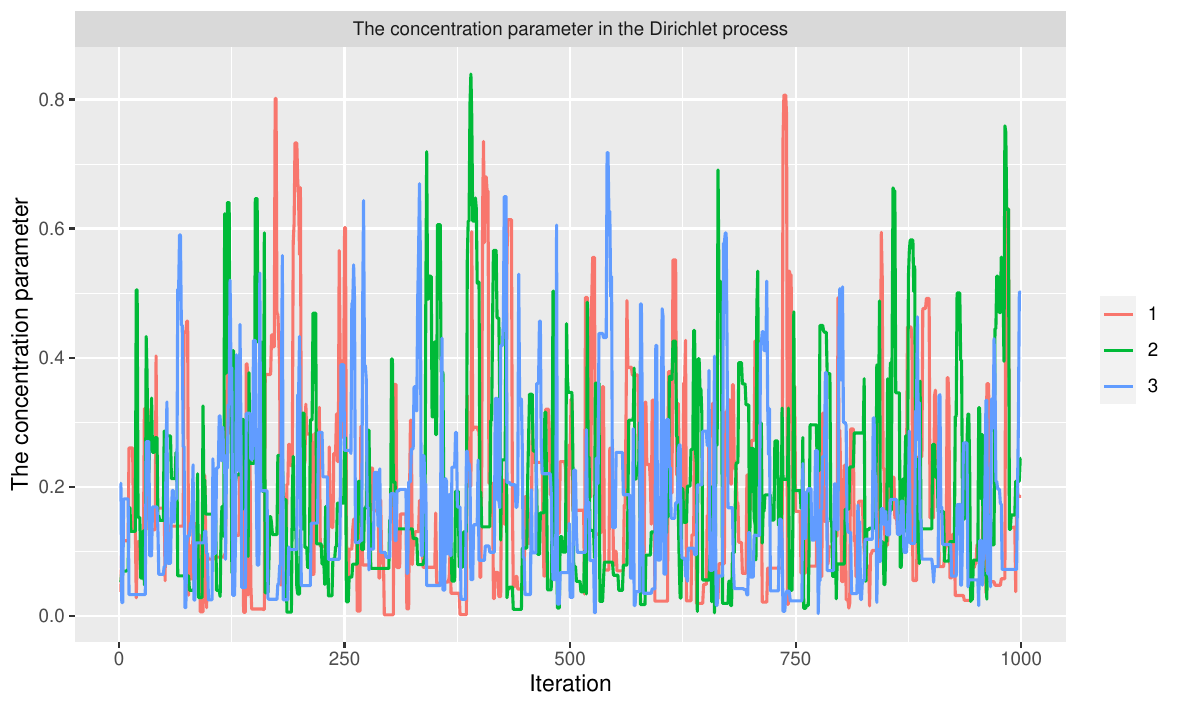}
\caption{Traceplots for the concentration parameter in the Dirichlet process in \textbf{DP-BMM-Hierarchical}. Colors correspond to different MCMC chains.}
\label{fig:app_dpbmm_hierarchy_diag_gamma}
\end{figure}

\begin{figure}
\centering
\includegraphics[scale = 0.60]{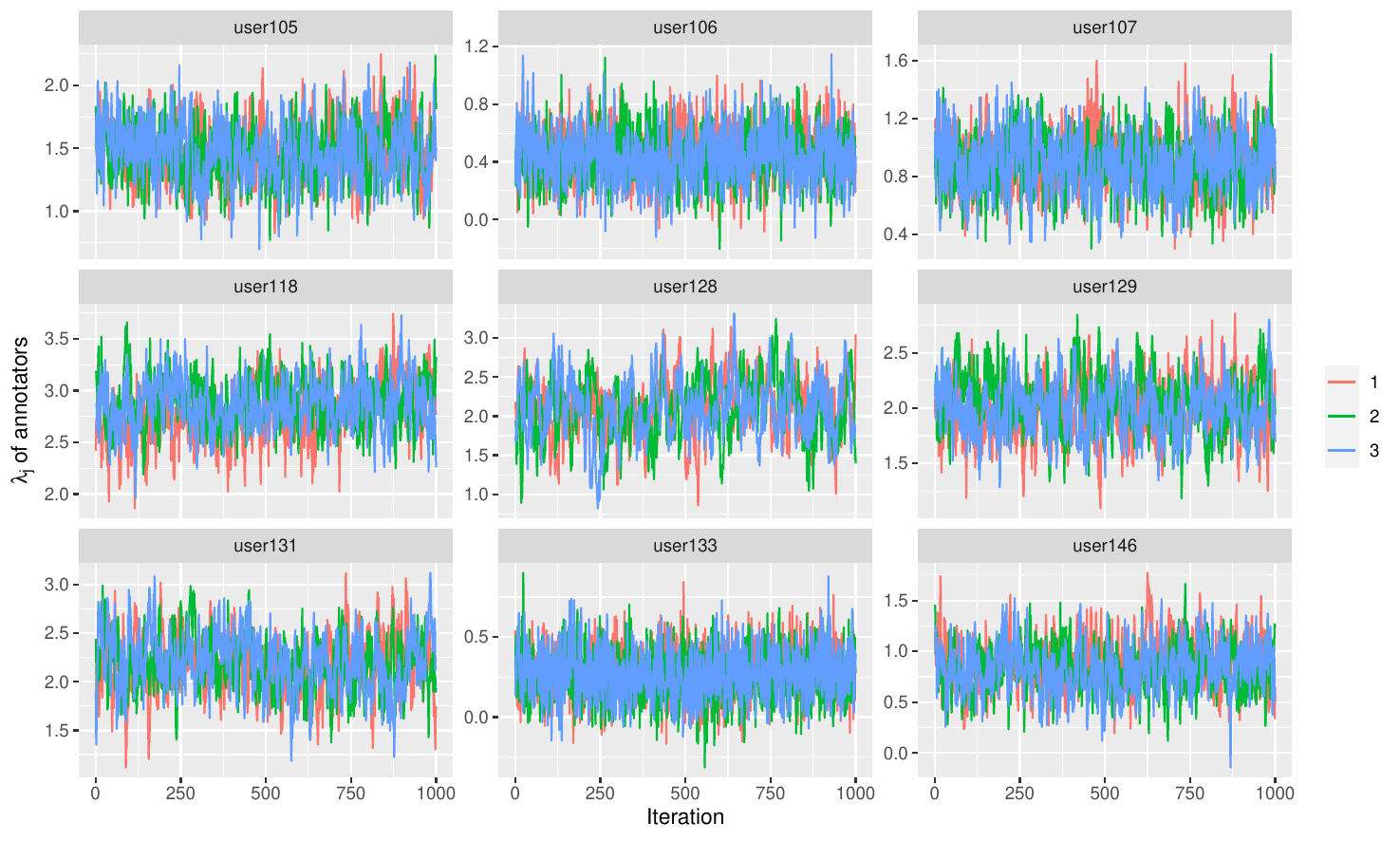}
\caption{Traceplots for the $\lambda_j$'s of nine randomly chosen annotators in \textbf{DP-BMM-Hierarchical}. Colors correspond to different MCMC chains.}
\label{fig:app_dpbmm_hierarchy_diag_lambda}
\end{figure}

\begin{figure}
\centering
\includegraphics[scale = 0.60]{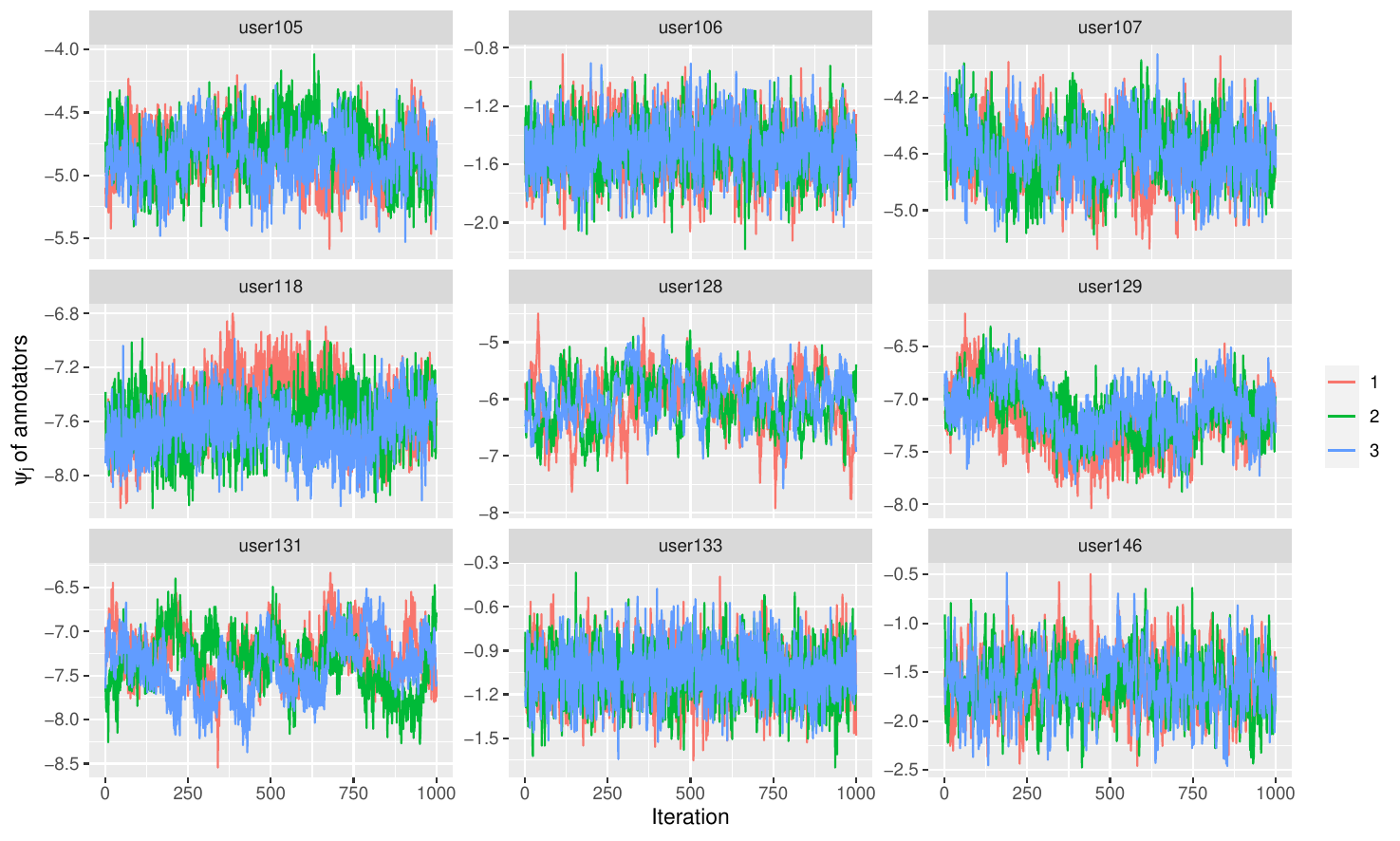}
\caption{Traceplots for the $\psi_j$'s of nine randomly chosen annotators in \textbf{DP-BMM-Hierarchical}. Colors correspond to different MCMC chains.}
\label{fig:app_dpbmm_hierarchy_diag_psi}
\end{figure}

\begin{figure}
\centering
\includegraphics[scale = 0.60]{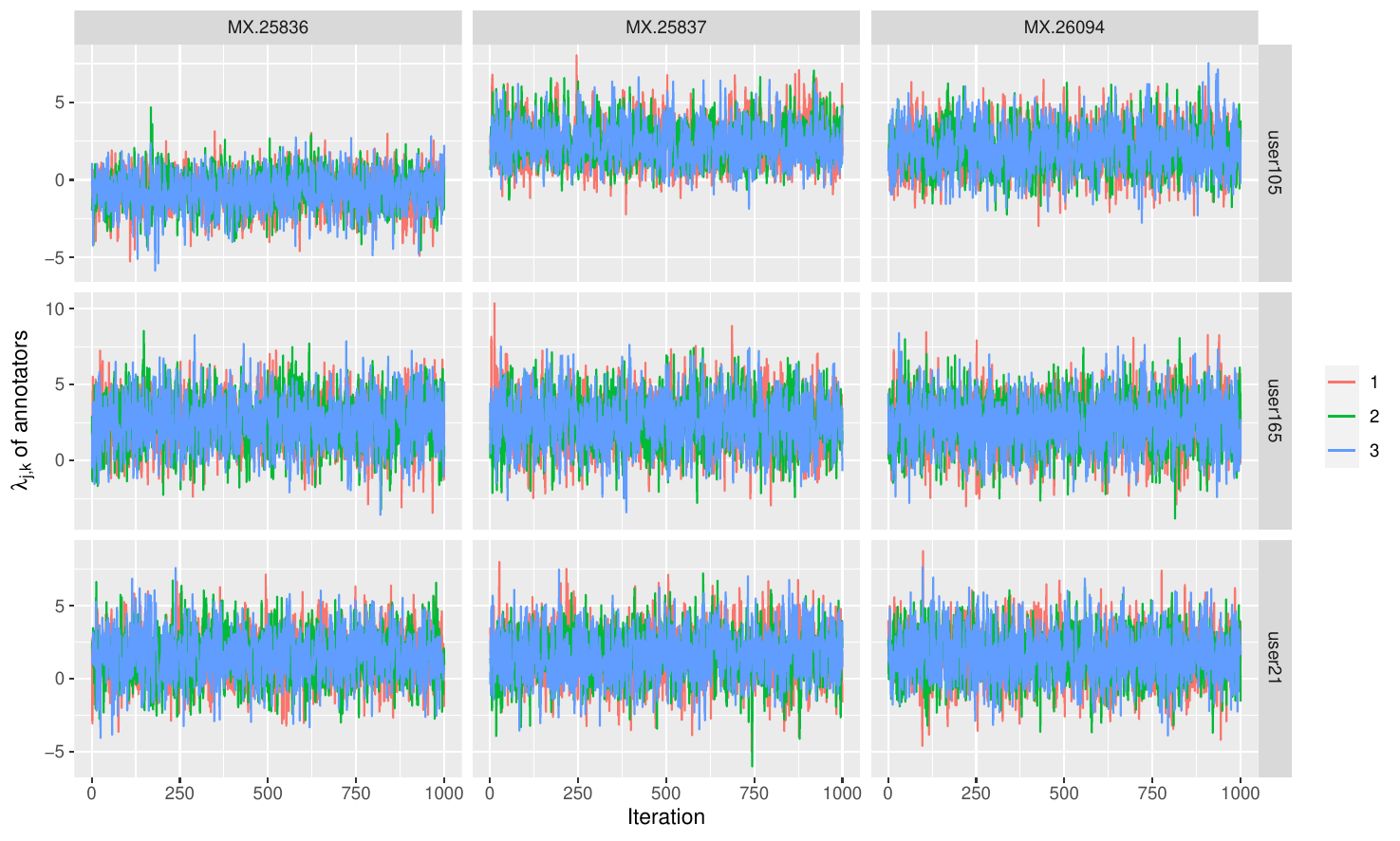}
\caption{Traceplots for the $\lambda_{j,k}$'s of three randomly chosen annotators on three randomly chosen bird species in \textbf{DP-BMM-Hierarchical}. Colors correspond to different MCMC chains.}
\label{fig:app_dpbmm_hierarchy_diag_lambda_mat}
\end{figure}

\begin{figure}
\centering
\includegraphics[scale = 0.60]{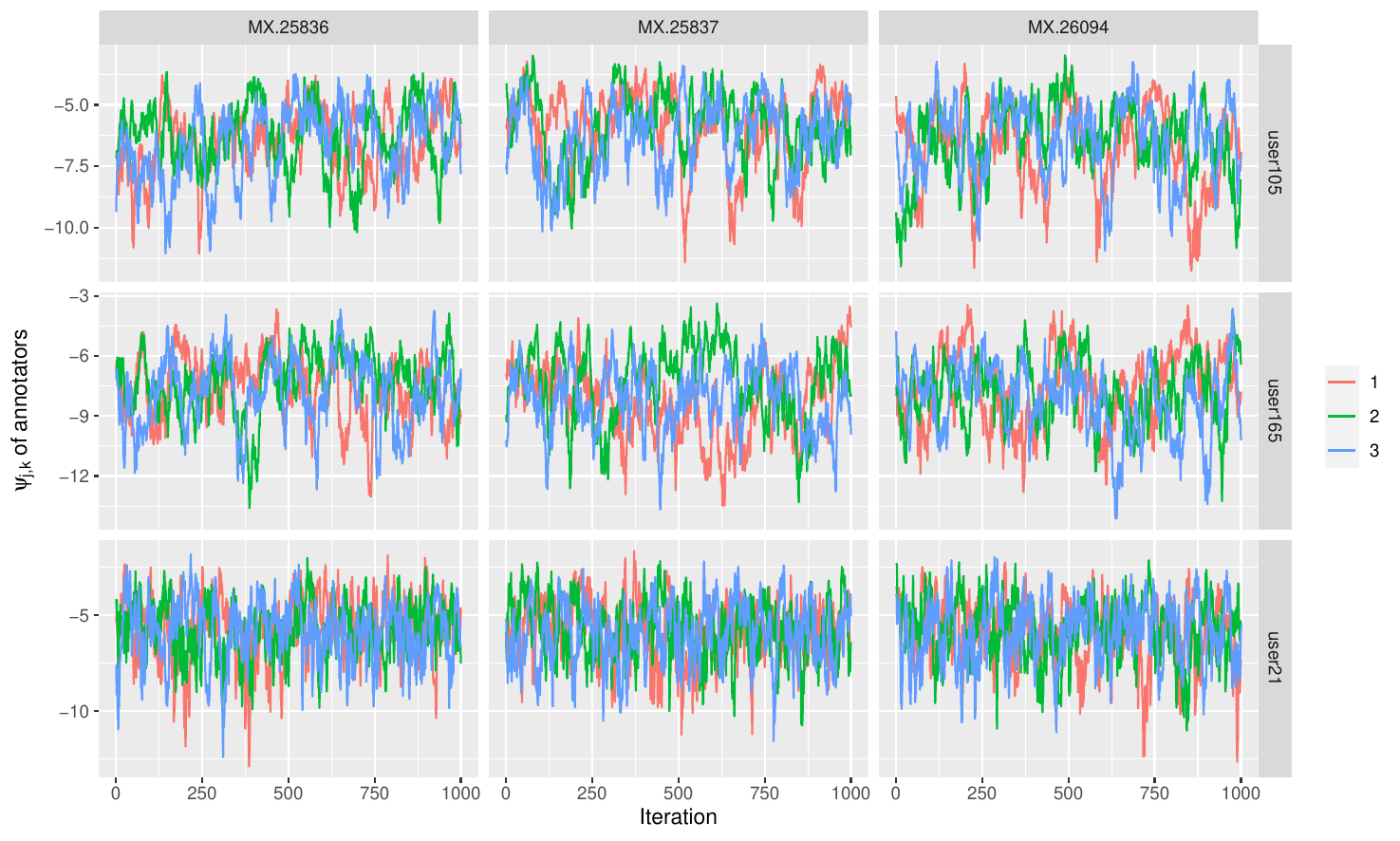}
\caption{Traceplots for the $\psi_{j,k}$'s of three randomly chosen annotators on three randomly chosen bird species in \textbf{DP-BMM-Hierarchical}. Colors correspond to different MCMC chains.}
\label{fig:app_dpbmm_hierarchy_diag_psi_mat}
\end{figure}

The traceplot for the concentration parameter is presented in Figure \ref{fig:app_dpbmm_hierarchy_diag_gamma}. As for the parameters related to annotators' bird song identification expertise, the traceplots for $\lambda_j$'s and $\psi_j$'s of nine randomly chosen annotators are shown in Figures \ref{fig:app_dpbmm_hierarchy_diag_lambda} and \ref{fig:app_dpbmm_hierarchy_diag_psi}, respectively. In addition, the traceplots for three randomly chosen annotators' $\lambda_{j, k}$ and $\psi_{j, k}$ on three randomly selected species are shown in Figures \ref{fig:app_dpbmm_hierarchy_diag_lambda_mat} and \ref{fig:app_dpbmm_hierarchy_diag_psi_mat}, respectively. In general, though some parameters exhibit a certain degree of autocorrelation, the Markov chains have converged sufficiently well and mixing seems to be good.

%%%%% Section A4 %%%%%
\section{Additional results for application to species annotation data} \label{suppsec:additional_app}
\begin{figure}
\centering
\includegraphics[scale = 0.55]{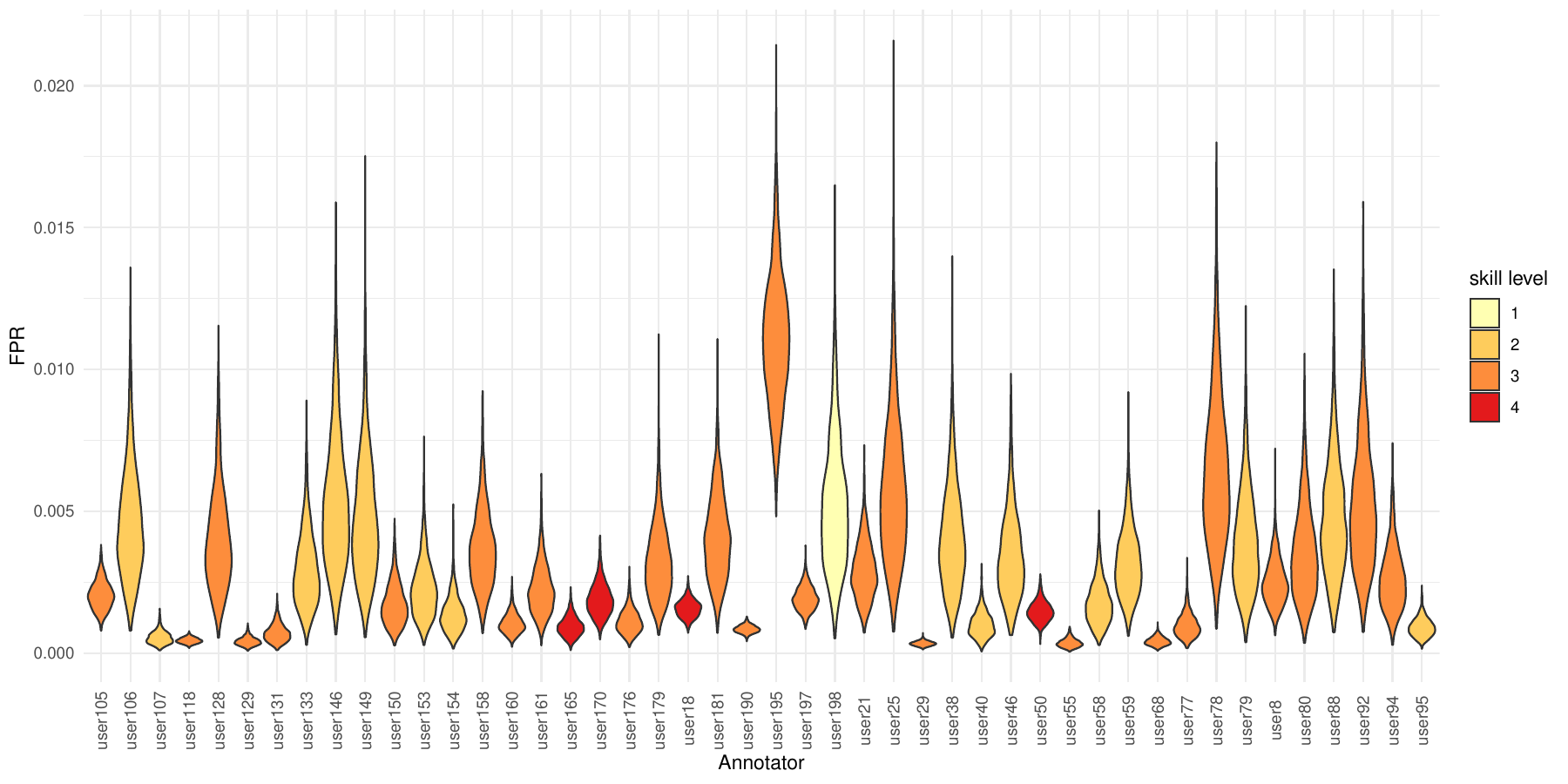}
\caption{The posterior distributions of annotators' TPRs in \textbf{Base}. Levels $1, 2, 3, 4$ represent the different levels of annotators' bird song identification expertise. Level $4$ corresponds to the highest skill level, followed by decreasing levels of 3, 2, and 1 in that order.}
\label{fig:app_additional_base_psi}
\end{figure}

\begin{figure}
\centering
\includegraphics[scale = 0.55]{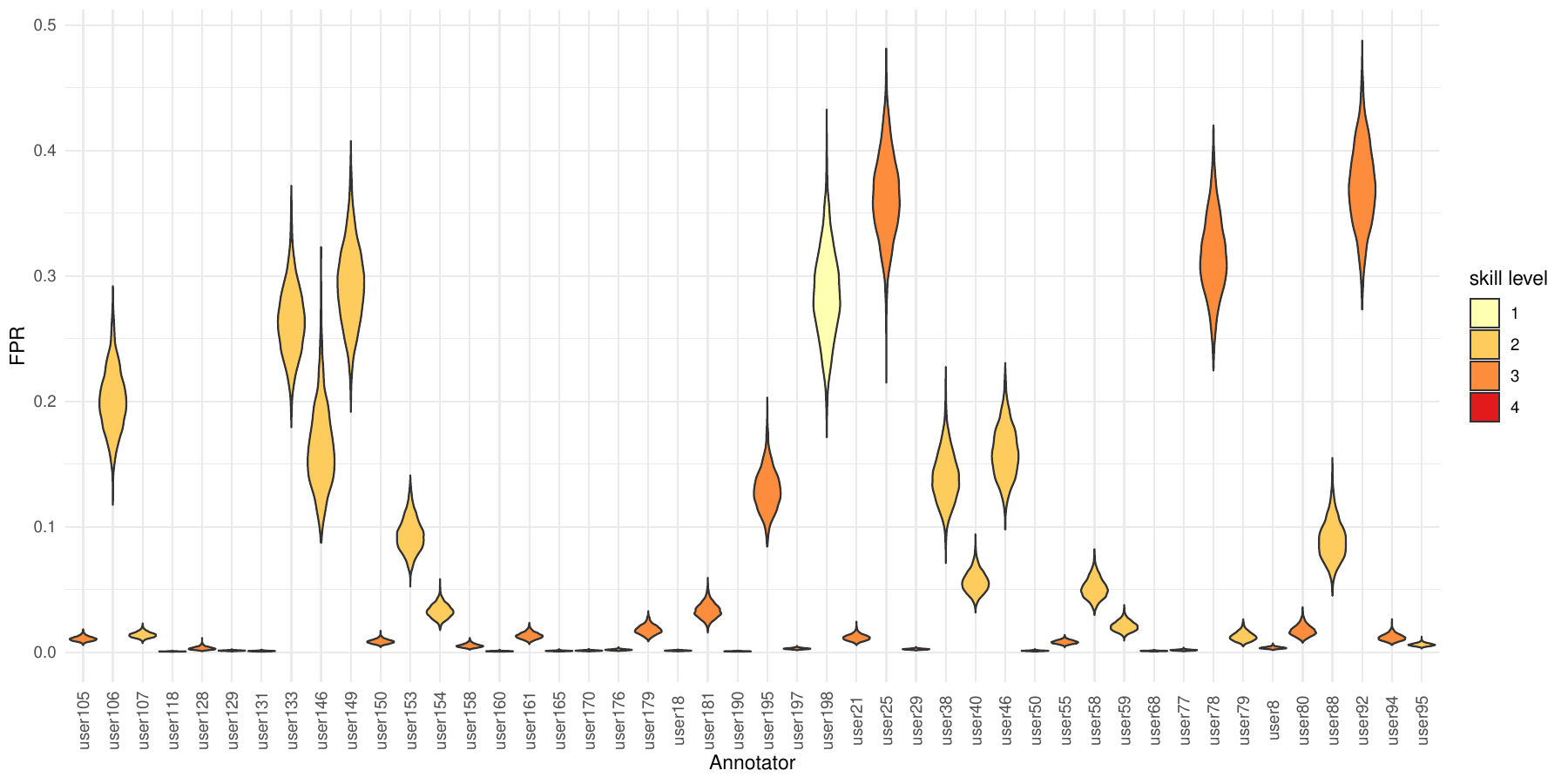}
\caption{The posterior distributions of annotators' TPRs in \textbf{Base-Hierarchical}. Levels $1, 2, 3, 4$ represent the different levels of annotators' bird song identification expertise. Level $4$ corresponds to the highest skill level, followed by decreasing levels of 3, 2, and 1 in that order.}
\label{fig:app_additional_base_hierarchy_psi}
\end{figure}

\begin{figure}
\centering
\includegraphics[scale = 0.55]{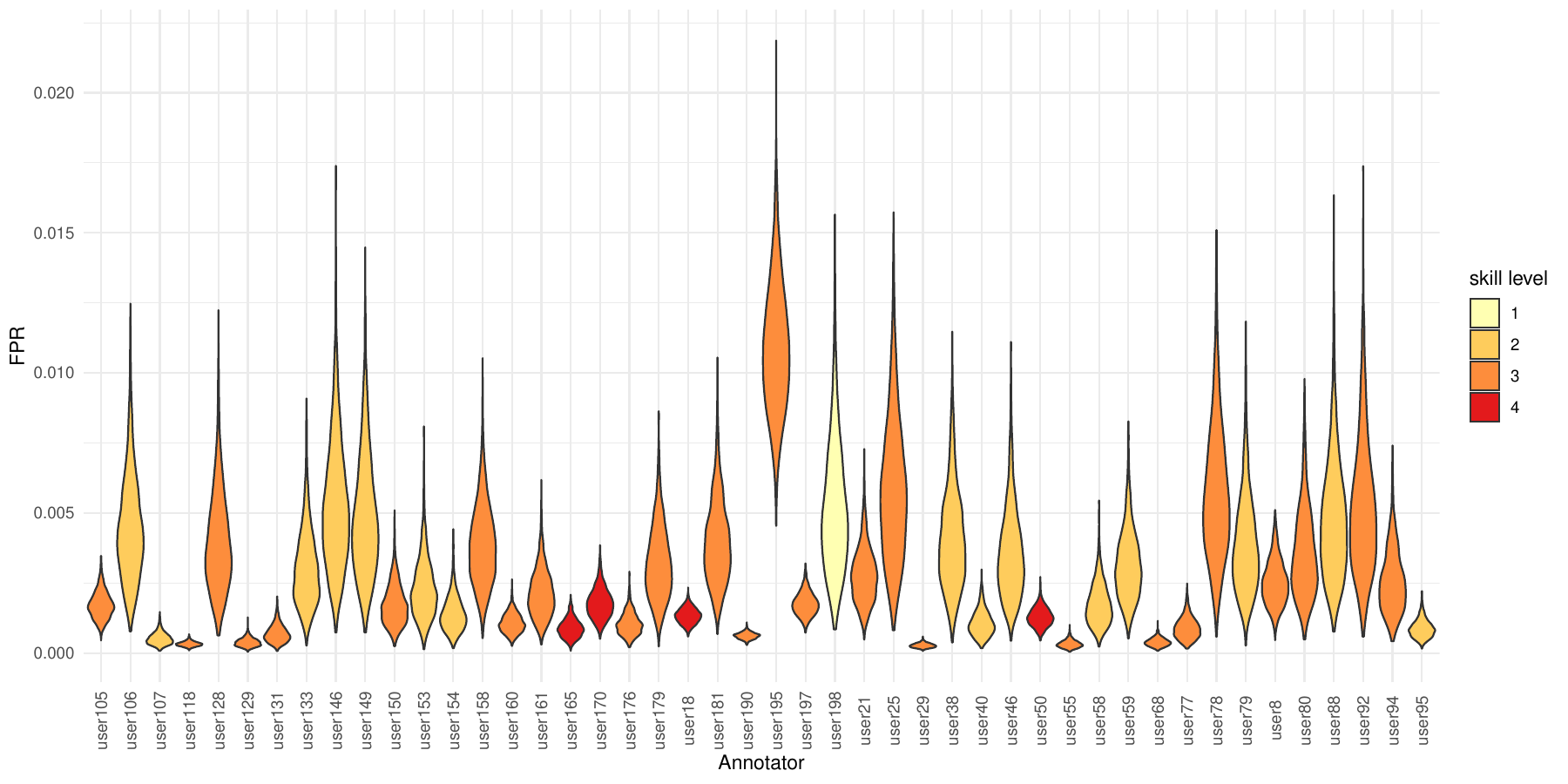}
\caption{The posterior distributions of annotators' TPRs in \textbf{DP-BMM}. Levels $1, 2, 3, 4$ represent the different levels of annotators' bird song identification expertise. Level $4$ corresponds to the highest skill level, followed by decreasing levels of 3, 2, and 1 in that order.}
\label{fig:app_additional_dpbmm_psi}
\end{figure}

\begin{figure}
\centering
\includegraphics[scale = 0.55]{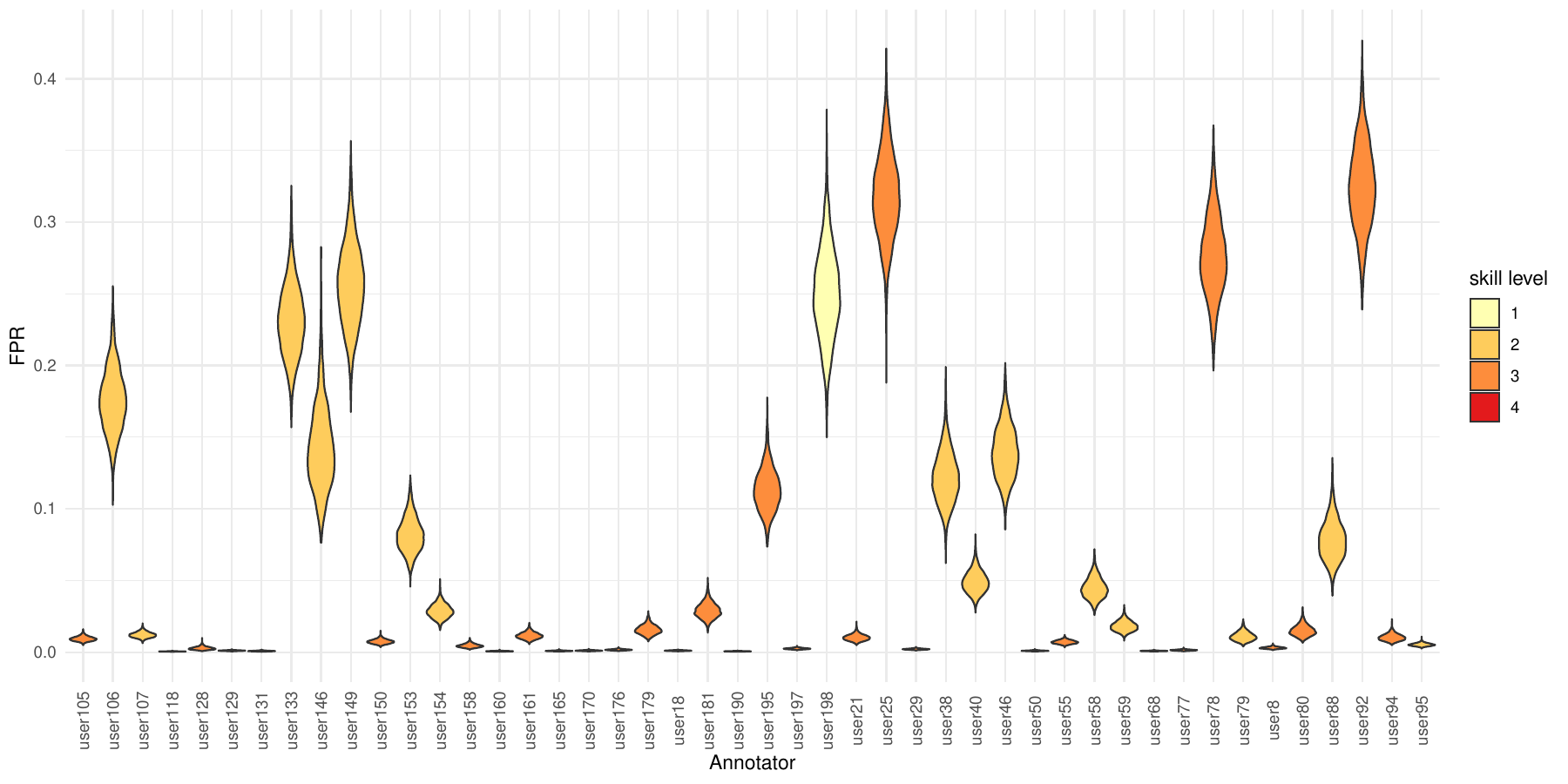}
\caption{The posterior distributions of annotators' TPRs in \textbf{DP-BMM-Hierarchical}. Levels $1, 2, 3, 4$ represent the different levels of annotators' bird song identification expertise. Level $4$ corresponds to the highest skill level, followed by decreasing levels of 3, 2, and 1 in that order.}
\label{fig:app_additional_dpbmm_hierarchy_psi}
\end{figure}

Similar to the discussions in Section \ref{sec:assess_skills}, because of the convergence of Markov chains and the resemblance of posterior distributions obtained from different Markov chains for each model, here we present the posterior distribution of each annotator's $\psi_j$ acquired from one chain for each model in Figures \ref{fig:app_additional_base_psi}, \ref{fig:app_additional_base_hierarchy_psi}, \ref{fig:app_additional_dpbmm_psi}, \ref{fig:app_additional_dpbmm_hierarchy_psi}.

%%%%% Section A5 %%%%%
\section{List of bird species included in our analysis} \label{sec:bird_list}
Table \ref{table:species_info} shows the list of $117$ bird species that are included in our analysis.
{\singlespacing
\footnotesize
\begin{longtable}{rlllc}
\caption{Bird species included in our data.} \\
\label{table:species_info}
\textbf{Taxon ID} & \textbf{Scientific Name} & \textbf{Vernacular Name} \\ 
  \hline
MX.25836&Gavia stellata&Red-throated Loon  \\
MX.25837&Gavia arctica&Black-throated Loon  \\
MX.26094&Ardea cinerea&Grey Heron  \\
MX.26164&Botaurus stellaris&Eurasian Bittern  \\
MX.26277&Cygnus olor&Mute Swan  \\
MX.26280&Cygnus cygnus&Whooper Swan  \\
MX.26287&Anser fabalis&Bean Goose  \\
MX.26289&Anser albifrons&Greater White-fronted Goose  \\
MX.26291&Anser anser&Greylag Goose  \\
MX.26298&Branta canadensis&Canada Goose  \\
MX.26299&Branta leucopsis&Barnacle Goose  \\
MX.26360&Anas penelope&Eurasian Wigeon  \\
MX.26373&Anas platyrhynchos&Mallard  \\
MX.26427&Clangula hyemalis&Long-tailed Duck  \\
MX.26429&Melanitta nigra&Common Scoter  \\
MX.26440&Mergus serrator&Red-breasted Merganser  \\
MX.26442&Mergus merganser&Common Merganser  \\
MX.26530&Haliaeetus albicilla&White-tailed Sea Eagle  \\
MX.26639&Accipiter nisus&Eurasian Sparrowhawk  \\
MX.26647&Accipiter gentilis&Northern Goshawk  \\
MX.26701&Buteo buteo&Common Buzzard  \\
MX.26796&Falco tinnunculus&Common Kestrel  \\
MX.26921&Lagopus lagopus&Willow Ptarmigan  \\
MX.26926&Tetrao tetrix&Black Grouse  \\
MX.26928&Tetrao urogallus&Western Capercaillie  \\
MX.26931&Tetrastes bonasia&Hazel Grouse  \\
MX.27152&Phasianus colchicus&Common Pheasant  \\
MX.27214&Grus grus&Common Crane  \\
MX.27527&Vanellus vanellus&Northern Lapwing  \\
MX.27610&Numenius phaeopus&Whimbrel  \\
MX.27613&Numenius arquata&Eurasian Curlew  \\
MX.27622&Tringa nebularia&Common Greenshank  \\
MX.27626&Tringa ochropus&Green Sandpiper  \\
MX.27628&Tringa glareola&Wood Sandpiper  \\
MX.27634&Actitis hypoleucos&Common Sandpiper  \\
MX.27649&Scolopax rusticola&Eurasian Woodcock  \\
MX.27665&Gallinago media&Great Snipe  \\
MX.27666&Gallinago gallinago&Common Snipe  \\
MX.27748&Larus canus&Mew Gull  \\
MX.27750&Larus argentatus&European Herring Gull  \\
MX.27753&Larus fuscus&Lesser Black-backed Gull  \\
MX.27759&Larus marinus&Great Black-backed Gull  \\
MX.27774&Larus ridibundus&Black-headed Gull  \\
MX.27908&Columba oenas&Stock Dove  \\
MX.27911&Columba palumbus&Common Wood Pigeon  \\
MX.28715&Cuculus canorus&Common Cuckoo  \\
MX.28998&Strix aluco&Tawny Owl  \\
MX.29003&Strix uralensis&Ural Owl  \\
MX.29068&Asio otus&Long-eared Owl  \\
MX.29172&Caprimulgus europaeus&European Nightjar  \\
MX.29324&Apus apus&Common Swift  \\
MX.30333&Jynx torquilla&Eurasian Wryneck  \\
MX.30443&Dendrocopos major&Great Spotted Woodpecker  \\
MX.30453&Picoides tridactylus&Eurasian Three-toed Woodpecker  \\
MX.30504&Dryocopus martius&Black Woodpecker  \\
MX.30530&Picus canus&Grey-headed Woodpecker  \\
MX.32065&Alauda arvensis&Eurasian Skylark  \\
MX.32183&Motacilla alba&White Wagtail  \\
MX.32213&Anthus pratensis&Meadow Pipit  \\
MX.32214&Anthus trivialis&Tree Pipit  \\
MX.32561&Lanius collurio&Red-backed Shrike  \\
MX.32608&Bombycilla garrulus&Bohemian Waxwing  \\
MX.32696&Troglodytes troglodytes&Eurasian Wren  \\
MX.32772&Prunella modularis&Dunnock  \\
MX.32801&Erithacus rubecula&European Robin  \\
MX.32819&Luscinia luscinia&Thrush Nightingale  \\
MX.32895&Phoenicurus phoenicurus&Common Redstart  \\
MX.32949&Saxicola rubetra&Whinchat  \\
MX.33106&Turdus merula&Common Blackbird  \\
MX.33117&Turdus pilaris&Fieldfare  \\
MX.33118&Turdus iliacus&Redwing  \\
MX.33119&Turdus philomelos&Song Thrush  \\
MX.33121&Turdus viscivorus&Mistle Thrush  \\
MX.33630&Locustella naevia&Common Grasshopper Warbler  \\
MX.33651&Acrocephalus arundinaceus&Great Reed Warbler  \\
MX.33676&Hippolais icterina&Icterine Warbler  \\
MX.33873&Phylloscopus trochilus&Willow Warbler  \\
MX.33874&Phylloscopus collybita&Common Chiffchaff  \\
MX.33878&Phylloscopus sibilatrix&Wood Warbler  \\
MX.33934&Sylvia atricapilla&Eurasian Blackcap  \\
MX.33935&Sylvia borin&Garden Warbler  \\
MX.33936&Sylvia communis&Common Whitethroat  \\
MX.33937&Sylvia curruca&Lesser Whitethroat  \\
MX.33939&Sylvia nisoria&Barred Warbler  \\
MX.33954&Regulus regulus&Goldcrest  \\
MX.33989&Muscicapa striata&Spotted Flycatcher  \\
MX.34021&Ficedula hypoleuca&European Pied Flycatcher  \\
MX.34029&Ficedula parva&Red-breasted Flycatcher  \\
MX.34505&Aegithalos caudatus&Long-tailed Tit  \\
MX.34535&Poecile montanus&Willow Tit  \\
MX.34549&Periparus ater&Coal Tit  \\
MX.34553&Lophophanes cristatus&European Crested Tit  \\
MX.34567&Parus major&Great Tit  \\
MX.34574&Cyanistes caeruleus&Eurasian Blue Tit  \\
MX.34616&Certhia familiaris&Eurasian Treecreeper  \\
MX.35146&Emberiza citrinella&Yellowhammer  \\
MX.35167&Emberiza rustica&Rustic Bunting  \\
MX.36237&Fringilla coelebs&Common Chaffinch  \\
MX.36239&Fringilla montifringilla&Brambling  \\
MX.36283&Carduelis chloris&European Greenfinch  \\
MX.36287&Carduelis spinus&Eurasian Siskin  \\
MX.36306&Carduelis carduelis&European Goldfinch  \\
MX.36310&Carduelis cannabina&Common Linnet  \\
MX.36331&Carpodacus erythrinus&Common Rosefinch  \\
MX.36356&Loxia pytyopsittacus&Parrot Crossbill  \\
MX.36358&Loxia curvirostra&Red Crossbill  \\
MX.36359&Loxia leucoptera&Two-barred Crossbill  \\
MX.36366&Pyrrhula pyrrhula&Eurasian Bullfinch  \\
MX.36368&Coccothraustes coccothraustes&Hawfinch  \\
MX.36573&Passer domesticus&House Sparrow  \\
MX.36817&Sturnus vulgaris&Common Starling  \\
MX.37090&Garrulus glandarius&Eurasian Jay  \\
MX.37122&Pica pica&Eurasian Magpie  \\
MX.37142&Corvus monedula&Western Jackdaw  \\
MX.37156&Corvus frugilegus&Rook  \\
MX.37178&Corvus corax&Northern Raven  \\
MX.73566&Corvus corone&Hooded Crow  \\
   \hline
\end{longtable}}

\end{document}